\documentclass[twocolumn]{aastex63}
\usepackage{xfrac}
\usepackage{float}
\usepackage{graphicx}
\usepackage{epstopdf}

\newcommand{\halp}{H$\alpha$}
\newcommand{\hbet}{H$\beta$} 
\newcommand{\CaII}{Ca\,{\footnotesize II}}
\newcommand{\LiI}{Li\,{\footnotesize I}}
\newcommand{\FeI}{Fe\,{\footnotesize I}}
\newcommand{\CaI}{Ca\,{\footnotesize I}}

\graphicspath{{./}{figures/}}

\received{}
\revised{}
\accepted{}
\submitjournal{ApJ}

\shorttitle{}
\shortauthors{Szab\'o et al.}

\graphicspath{{./}{figures/}}

\begin{document}

\title{A multi-epoch, multi-wavelength study of the classical FUor V1515~Cyg approaching quiescence}

%

\correspondingauthor{Zs\'ofia M. Szab\'o}
\email{szabo.zsofia@csfk.org}

\author[0000-0001-9830-3509]{Zs. M. Szab\'o}
\affiliation{Konkoly Observatory, Research Centre for Astronomy and Earth Sciences, E\"otv\"os Lor\'and Research Network (ELKH), Konkoly-Thege Mikl\'os \'ut 15-17, 1121 Budapest, Hungary}
\affiliation{CSFK, MTA Centre of Excellence, Budapest, Konkoly Thege Miklós út 15-17., H-1121, Hungary}
\affiliation{Max-Planck-Institut für Radioastronomie, Auf dem Hügel 69, D-53121 Bonn, Germany}
\affiliation{Scottish Universities Physics Alliance (SUPA), School of Physics and Astronomy, University of St Andrews, North Haugh, St Andrews, KY16 9SS, UK}
\affiliation{E\"otv\"os Lor\'and University, Department of Astronomy, P\'azm\'any P\'eter s\'et\'any 1/A, 1117 Budapest, Hungary}

\author[0000-0001-7157-6275]{\'A. K\'osp\'al}
\affiliation{Konkoly Observatory, Research Centre for Astronomy and Earth Sciences, E\"otv\"os Lor\'and Research Network (ELKH), Konkoly-Thege Mikl\'os \'ut 15-17, 1121 Budapest, Hungary}
\affiliation{CSFK, MTA Centre of Excellence, Budapest, Konkoly Thege Miklós út 15-17., H-1121, Hungary}
\affiliation{Max Planck Institute for Astronomy, K\"onigstuhl 17, D-69117 Heidelberg, Germany}
\affiliation{ELTE Eötvös Loránd University, Institute of Physics, Pázmány Péter sétány 1/A, H-1117 Budapest, Hungary}

\author[0000-0001-6015-646X]{P. \'Abrah\'am}
\affiliation{Konkoly Observatory, Research Centre for Astronomy and Earth Sciences, E\"otv\"os Lor\'and Research Network (ELKH), Konkoly-Thege Mikl\'os \'ut 15-17, 1121 Budapest, Hungary}
\affiliation{CSFK, MTA Centre of Excellence, Budapest, Konkoly Thege Miklós út 15-17., H-1121, Hungary}
\affiliation{ELTE Eötvös Loránd University, Institute of Physics, Pázmány Péter sétány 1/A, H-1117 Budapest, Hungary}

\author[0000-0003-4099-1171]{S. Park}
\affiliation{Konkoly Observatory, Research Centre for Astronomy and Earth Sciences, E\"otv\"os Lor\'and Research Network (ELKH), Konkoly-Thege Mikl\'os \'ut 15-17, 1121 Budapest, Hungary}
\affiliation{CSFK, MTA Centre of Excellence, Budapest, Konkoly Thege Miklós út 15-17., H-1121, Hungary}

\author[0000-0001-5018-3560]{M. Siwak}
\affiliation{Konkoly Observatory, Research Centre for Astronomy and Earth Sciences, E\"otv\"os Lor\'and Research Network (ELKH), Konkoly-Thege Mikl\'os \'ut 15-17, 1121 Budapest, Hungary}
\affiliation{CSFK, MTA Centre of Excellence, Budapest, Konkoly Thege Miklós út 15-17., H-1121, Hungary}

\author[0000-0003-1665-5709]{J. D. Green}
\affiliation{Space Telescope Science Institute, 3700 San Martin Dr., Baltimore, MD 21218, USA}

\author[0000-0001-5449-2467]{A. P\'al}
\affiliation{Konkoly Observatory, Research Centre for Astronomy and Earth Sciences, E\"otv\"os Lor\'and Research Network (ELKH), Konkoly-Thege Mikl\'os \'ut 15-17, 1121 Budapest, Hungary}
\affiliation{CSFK, MTA Centre of Excellence, Budapest, Konkoly Thege Miklós út 15-17., H-1121, Hungary}
\affiliation{E\"otv\"os Lor\'and University, Department of Astronomy, P\'azm\'any P\'eter s\'et\'any 1/A, 1117 Budapest, Hungary}
\affiliation{ELTE Eötvös Loránd University, Institute of Physics, Pázmány Péter sétány 1/A, H-1117 Budapest, Hungary}
\affiliation{MIT Kavli Institute for Astrophysics and Space Research, 70 Vassar Street, Cambridge, MA 02109, USA}

\author[0000-0002-0433-9656]{J. A. Acosta-Pulido}
\affiliation{Instituto de Astrofísica de Canarias, Avenida Vía Láctea, Tenerife, Spain}
\affiliation{Departamento de Astrofísica, Universidad de La Laguna, Tenerife, Spain}

\author[0000-0003-3119-2087]{J.-E. Lee}
\affiliation{School of Space Research, Kyung Hee University, 1732, Deogyeong-daero, Giheung-gu, Yongin-si, Gyeonggi-do 17104, Republic of Korea}

\author{M. Ibrahimov}
\affiliation{Institute of Astronomy, Russian Academy of Sciences, 48 Pyatnitskaya st., 119017, Moscow, Russia}

\author{K. Grankin}
\affiliation{Crimean Astrophysical Observatory, p/o Nauchny, 298409, Republic of Crimea}

\author{B. Kovács}
\affiliation{Konkoly Observatory, Research Centre for Astronomy and Earth Sciences, E\"otv\"os Lor\'and Research Network (ELKH), Konkoly-Thege Mikl\'os \'ut 15-17, 1121 Budapest, Hungary}

\author{Zs. Bora}
\affiliation{Konkoly Observatory, Research Centre for Astronomy and Earth Sciences, E\"otv\"os Lor\'and Research Network (ELKH), Konkoly-Thege Mikl\'os \'ut 15-17, 1121 Budapest, Hungary}
\affiliation{CSFK, MTA Centre of Excellence, Budapest, Konkoly Thege Miklós út 15-17., H-1121, Hungary}

\author{A. Bódi}
\affiliation{Konkoly Observatory, Research Centre for Astronomy and Earth Sciences, E\"otv\"os Lor\'and Research Network (ELKH), Konkoly-Thege Mikl\'os \'ut 15-17, 1121 Budapest, Hungary}
\affiliation{CSFK, MTA Centre of Excellence, Budapest, Konkoly Thege Miklós út 15-17., H-1121, Hungary}
\affiliation{MTA CSFK Lend\"ulet Near-Field Cosmology Research Group, 1121, Budapest, Konkoly Thege Mikl\'os \'ut 15-17, Hungary}

\author{B. Cseh}
\affiliation{Konkoly Observatory, Research Centre for Astronomy and Earth Sciences, E\"otv\"os Lor\'and Research Network (ELKH), Konkoly-Thege Mikl\'os \'ut 15-17, 1121 Budapest, Hungary}
\affiliation{CSFK, MTA Centre of Excellence, Budapest, Konkoly Thege Miklós út 15-17., H-1121, Hungary}
\affiliation{MTA-ELTE Lend{\"u}let "Momentum" Milky Way Research Group, Hungary}

\author{G. Csörnyei}
\affiliation{Konkoly Observatory, Research Centre for Astronomy and Earth Sciences, E\"otv\"os Lor\'and Research Network (ELKH), Konkoly-Thege Mikl\'os \'ut 15-17, 1121 Budapest, Hungary}
\affiliation{CSFK, MTA Centre of Excellence, Budapest, Konkoly Thege Miklós út 15-17., H-1121, Hungary}

\author{Marek Dr{\'o}{\.z}d{\.z}}
\affiliation{Mount Suhora Astronomical Observatory, Cracow Pedagogical University, ul. Podchorazych 2, 30-084 Krak{\'o}w, Poland}

\author{O. Hanyecz}
\affiliation{Konkoly Observatory, Research Centre for Astronomy and Earth Sciences, E\"otv\"os Lor\'and Research Network (ELKH), Konkoly-Thege Mikl\'os \'ut 15-17, 1121 Budapest, Hungary}
\affiliation{CSFK, MTA Centre of Excellence, Budapest, Konkoly Thege Miklós út 15-17., H-1121, Hungary}

\author{B. Ignácz}
\affiliation{Konkoly Observatory, Research Centre for Astronomy and Earth Sciences, E\"otv\"os Lor\'and Research Network (ELKH), Konkoly-Thege Mikl\'os \'ut 15-17, 1121 Budapest, Hungary}
\affiliation{CSFK, MTA Centre of Excellence, Budapest, Konkoly Thege Miklós út 15-17., H-1121, Hungary}

\author{Cs. Kalup}
\affiliation{Konkoly Observatory, Research Centre for Astronomy and Earth Sciences, E\"otv\"os Lor\'and Research Network (ELKH), Konkoly-Thege Mikl\'os \'ut 15-17, 1121 Budapest, Hungary}
\affiliation{CSFK, MTA Centre of Excellence, Budapest, Konkoly Thege Miklós út 15-17., H-1121, Hungary}

\author[0000-0002-8770-6764]{R. Könyves-Tóth}
\affiliation{Konkoly Observatory, Research Centre for Astronomy and Earth Sciences, E\"otv\"os Lor\'and Research Network (ELKH), Konkoly-Thege Mikl\'os \'ut 15-17, 1121 Budapest, Hungary}
\affiliation{CSFK, MTA Centre of Excellence, Budapest, Konkoly Thege Miklós út 15-17., H-1121, Hungary}
\affiliation{Department of Optics \& Quantum Electronics, University of Szeged, D\'om t\'er 9, Szeged, 6720, Hungary}

\author[0000-0002-8813-4884]{M. Krezinger}
\affiliation{Konkoly Observatory, Research Centre for Astronomy and Earth Sciences, E\"otv\"os Lor\'and Research Network (ELKH), Konkoly-Thege Mikl\'os \'ut 15-17, 1121 Budapest, Hungary}
\affiliation{CSFK, MTA Centre of Excellence, Budapest, Konkoly Thege Miklós út 15-17., H-1121, Hungary}
\affiliation{E\"otv\"os Lor\'and University, Department of Astronomy, P\'azm\'any P\'eter s\'et\'any 1/A, 1117 Budapest, Hungary}

\author{L. Kriskovics}
\affiliation{Konkoly Observatory, Research Centre for Astronomy and Earth Sciences, E\"otv\"os Lor\'and Research Network (ELKH), Konkoly-Thege Mikl\'os \'ut 15-17, 1121 Budapest, Hungary}
\affiliation{CSFK, MTA Centre of Excellence, Budapest, Konkoly Thege Miklós út 15-17., H-1121, Hungary}
\affiliation{ELTE Eötvös Loránd University, Institute of Physics, Pázmány Péter sétány 1/A, H-1117 Budapest, Hungary}

\author[0000-0002-6293-9940]{Waldemar Og{\l}oza}
\affiliation{Mount Suhora Astronomical Observatory, Cracow Pedagogical University, ul. Podchorazych 2, 30-084 Krak{\'o}w, Poland} 

\author{A. Ordasi}
\affiliation{Konkoly Observatory, Research Centre for Astronomy and Earth Sciences, E\"otv\"os Lor\'and Research Network (ELKH), Konkoly-Thege Mikl\'os \'ut 15-17, 1121 Budapest, Hungary}
\affiliation{CSFK, MTA Centre of Excellence, Budapest, Konkoly Thege Miklós út 15-17., H-1121, Hungary}

\author[0000-0003-0926-3950]{K. Sárneczky}
\affiliation{Konkoly Observatory, Research Centre for Astronomy and Earth Sciences, E\"otv\"os Lor\'and Research Network (ELKH), Konkoly-Thege Mikl\'os \'ut 15-17, 1121 Budapest, Hungary}
\affiliation{CSFK, MTA Centre of Excellence, Budapest, Konkoly Thege Miklós út 15-17., H-1121, Hungary}

\author[0000-0002-3658-2175]{B. Seli}
\affiliation{Konkoly Observatory, Research Centre for Astronomy and Earth Sciences, E\"otv\"os Lor\'and Research Network (ELKH), Konkoly-Thege Mikl\'os \'ut 15-17, 1121 Budapest, Hungary}
\affiliation{CSFK, MTA Centre of Excellence, Budapest, Konkoly Thege Miklós út 15-17., H-1121, Hungary}
\affiliation{E\"otv\"os Lor\'and University, Department of Astronomy, P\'azm\'any P\'eter s\'et\'any 1/A, 1117 Budapest, Hungary}

\author[0000-0002-1698-605X]{R. Szakáts}
\affiliation{Konkoly Observatory, Research Centre for Astronomy and Earth Sciences, E\"otv\"os Lor\'and Research Network (ELKH), Konkoly-Thege Mikl\'os \'ut 15-17, 1121 Budapest, Hungary}
\affiliation{CSFK, MTA Centre of Excellence, Budapest, Konkoly Thege Miklós út 15-17., H-1121, Hungary}

\author{Á. Sódor}
\affiliation{Konkoly Observatory, Research Centre for Astronomy and Earth Sciences, E\"otv\"os Lor\'and Research Network (ELKH), Konkoly-Thege Mikl\'os \'ut 15-17, 1121 Budapest, Hungary}
\affiliation{CSFK, MTA Centre of Excellence, Budapest, Konkoly Thege Miklós út 15-17., H-1121, Hungary}
\affiliation{MTA CSFK Lend\"ulet Near-Field Cosmology Research Group, 1121, Budapest, Konkoly Thege Mikl\'os \'ut 15-17, Hungary}

\author{A. Szing}
\affiliation{Konkoly Observatory, Research Centre for Astronomy and Earth Sciences, E\"otv\"os Lor\'and Research Network (ELKH), Konkoly-Thege Mikl\'os \'ut 15-17, 1121 Budapest, Hungary}
\affiliation{CSFK, MTA Centre of Excellence, Budapest, Konkoly Thege Miklós út 15-17., H-1121, Hungary}

\author[0000-0002-6471-8607]{K. Vida}
\affiliation{Konkoly Observatory, Research Centre for Astronomy and Earth Sciences, E\"otv\"os Lor\'and Research Network (ELKH), Konkoly-Thege Mikl\'os \'ut 15-17, 1121 Budapest, Hungary}
\affiliation{CSFK, MTA Centre of Excellence, Budapest, Konkoly Thege Miklós út 15-17., H-1121, Hungary}

\author[0000-0001-8764-7832]{J. Vinkó}
\affiliation{Konkoly Observatory, Research Centre for Astronomy and Earth Sciences, E\"otv\"os Lor\'and Research Network (ELKH), Konkoly-Thege Mikl\'os \'ut 15-17, 1121 Budapest, Hungary}
\affiliation{CSFK, MTA Centre of Excellence, Budapest, Konkoly Thege Miklós út 15-17., H-1121, Hungary}
\affiliation{ELTE Eötvös Loránd University, Institute of Physics, Pázmány Péter sétány 1/A, H-1117 Budapest, Hungary}
\begin{abstract}
Historically, FU~Orionis-type stars are low-mass, pre-main sequence stars. The members of this class experience powerful accretion outbursts and remain in an enhanced accretion state for decades or centuries. V1515~Cyg, a classical FUor, started brightening in the 1940s and reached its peak brightness in the late 1970s. Following a sudden decrease in brightness it stayed in a minimum state for a few months, then started a brightening for several years.
We present results of our ground-based photometric monitoring complemented with optical/NIR spectroscopic monitoring.
Our light curves show a long-term fading with strong variability on weekly and monthly time scales.
The optical spectra show P~Cygni profiles and broad blue-shifted absorption lines, common properties of FUors. However, V1515~Cyg lacks the P~Cygni profile in the Ca\,{\footnotesize II} 8498 \AA{} line, a part of the Ca infrared triplet (IRT), formed by an outflowing wind, suggesting that the absorbing gas in the wind is optically thin.
The newly obtained near-infrared spectrum shows the strengthening of the CO bandhead and the FeH molecular band, indicating that the disk has become cooler since the last spectroscopic observation in 2015. 
The current luminosity of the accretion disk dropped from the peak value of 138\,$L_{\odot}$ to about 45\,$L_{\odot}$, suggesting that the long-term fading is also partly caused by the dropping of the accretion rate.  

\end{abstract}


 \section{Introduction} \label{sec:intro}


A rather small, but spectacular class of the low-mass pre-main sequence stars are the FU~Orionis type objects (briefly FUors). The group was defined by \citet{herbig1977} based on the properties of the first few outbursting sources observed in the optical, today known as classical FUors. However, these and the other optically visible sources represent only part of a more complex phenomenon that is conspicuous also at infrared wavelengths. FUors are characterized by their enormous inner disks brightness increase, which is caused by the enhanced accretion from the disk onto the protostar. Enhanced accretion is most likely caused by disk instabilities, and this stage can last for several decades, or likely even centuries \citep{paczynski1976,lin1985,kenyon-and-hartmann1988,kenyon-and-hartmann1991,bell1995,turner1997,audard2014,kadam2020}. 
FUors show very similar optical spectra with the properties of F or G supergiants with broad absorption lines, P~Cygni profile of H$\alpha$, wind components, and strong Li\,{\footnotesize I} 6707\,\AA{} absorption in the optical regime. 

This paper is the second in a series started by \citet{szabo2021}, with the aim of characterising past and current evolutionary stages of the so called classical FUors by means of archival and new photometric as well as spectroscopic observations. The target of the present study, V1515 Cyg, was the third discovered FUor. Unlike the first two classical FUors (FU~Orionis and V1057~Cyg), V1515~Cyg did not reach the photometric maximum brightness within a year, but this process lasted significantly longer: it started in the 1940's and it reached the peak (13.7~mag in the $B$-band) at the end of the 1970's \citep{landolt1977}. 
In 1980 a significant, yet only temporary fading occurred, which was the sharpest decline yet observed during a FUor outburst \citep{kolotilov1981,kolotilov1983,KH1991}. V1515~Cyg faded 1.5 mag in the $B$-band within a few months, but after reaching the minimum at the end of 1980, the source started to brighten again \citep{clarke2005}. The fading observed in V1515~Cyg after reaching the light curve maximum is commonly observed among the classical FUors: both FU~Ori and V1057~Cyg showed similar dips after reaching their maximum \citep{kolotilov1997,KH1991,kenyon-and-hartmann1991}.

Unlike V1057~Cyg, where the pre-outburst spectrum showed properties of a classical T Tauri-type star (CTTS; \citealt{herbig1977,herbig2003}), no spectroscopic observations were carried out prior to the brightening of V1515~Cyg.
The spectra obtained during the first fading event in the 1980's by \citet{kolotilov1983} showed absorption lines typical for high-luminosity G2--G5 star and despite the fading, the spectrum of V1515~Cyg has not fundamentally changed since the observations obtained around 1974--1975 by \citet{herbig1977}. All features common for FUors like broad H$\alpha$ line having P~Cyg profile, strong Li~{\footnotesize I}~6707\,\AA{} as well as strong Na~{\footnotesize D} lines, were present. 
The presence of blueshifted absorption in the H$\alpha$ and H$\beta$ lines are common signatures for the outflowing wind in these objects. In the case of V1515~Cyg the blueshifted features were interpreted as an expanding envelope by \citet{kolotilov1983}.
Using photometric data, \citet{KH1991} proposed to explain the 1980 fading by a sudden dust grain condensation event in the outflowing wind. They proposed that the slow recovery from the minimum was most likely caused by the gradual dispersal (expansion) of a newly formed dust cloud. 

Several years ago, we started a photometric monitoring program of a few classical FUors at the Piszkéstető Mountain Station of Konkoly Observatory in Hungary, including V1515~Cyg. After combination with other public-domain photometric data of this source, we realized that about 15 years ago the source started a long-term fading trend. Given that the last detailed photometric analysis of V1515~Cyg was presented by \citet{clarke2005}, while the last optical and infrared spectral analysis were performed by \citet{KH1991}, \citet{agra-amboage2014} and \citet{connelley2018}, we decided to increase the cadence of our photometric monitoring and obtain new spectroscopic observations to properly characterise this fading process.
Although the FUor phenomenon can occur throughout the YSO evolution \citep[see e.g.,][]{kenyon&hartmann1996,fischer2022}, it is still an open question how an eruption ends, and how the properties of a system undergoing an eruption are evolving.
Considering the long-term fading trend, V1515~Cyg could easily be the first FUor going back to the quiescent stage, therefore, it is crucial to gather as much data as possible to better understand an important phenomenon in the evolution of young stars.


In Section \ref{sec:obs} we briefly summarize our observations and the data reduction. We analyse the data in Section \ref{sec:res} and we discuss obtained results in Section \ref{sec:discussion}. Lastly, we summarize our findings in Section \ref{sec:conclusions}.

\section{Observations and data reduction} 
\label{sec:obs}

\subsection{Ground-based optical and near-infrared photometry}
\label{sec:photometry}

The bulk of photometric observations was obtained at Piszkéstető Observatory between 2005 and 2021 in $B$, $V$, $R_{\rm C}$, $I_{\rm C}$, $g'$, $r'$, and $i'$ filters. Three telescopes, each having somewhat different optical system, were used. Between 2005 October and 2011 May we observed the star with the 1\,m Ritchey-Chr\'etien-Coudé (RCC) telescope, equipped with a 1300$\times$1340 pixel Roper Scientific VersArray: 1300B CCD camera (pixel scale: 0$\farcs$306). In August 2011 -- December 2019 we observed the star with the 60/90~cm Schmidt telescope (pixel scale: 1$\farcs$027) and from May 2020 we use the Astro Systeme Austria AZ800 alt-azimuth direct drive 80-cm Ritchey-Chr\'etien (RC80) telescope (pixel scale 0$\farcs$55). 

In 2006 we also observed V1515~Cyg at Mt.~Maidanak observatory by means of the 100 and the 60~cm Zeiss telescopes, both equipped with the same IMG1001E 1k$\times$1k CCD camera and $BVR_CI_C$ filters set: the larger reflector was used during 9 nights between 17 -- 27 June, while the smaller during 35 nights between 4 July -- 28 August. Due to different focal lengths, the larger telescope offered 6.35, while the smaller 11.74~arcmin  field of view, with the pixel scale of 0.372 and 0.688~arcsec~pix$^{-1}$, respectively. Typically 4 images in $B$, 2 -- 3 images in $V$, and single images in $R_CI_C$ filters per night were obtained. After standard reduction steps on bias, dark and flat field, we extracted aperture photometry of V1515~Cyg using three comparison stars for the 100~cm, and five comparison stars for the 60~cm telescope. The results from separate images obtained during the same night were averaged. This resulted in typical uncertainty of 0.04 -- 0.05~mag in $B$, 0.01 -- 0.02~mag in $V$, and 0.01~mag in $R_CI_C$ filters.

In July -- September, 2021, we observed the star at the Mt. Suhora Observatory of the Cracow Pedagogical University (Poland). We used the 60\,cm Carl-Zeiss telescope equipped with an Apogee AltaU47 camera, giving 1$\farcs$116 pixel scale and $19\farcm0\times19\farcm0$ field of view. Johnson $BV$ and Sloan $g'r'i'$ filters were used.

Occasionally we also used other telescopes: on 2006 July 19 and 2012 October 14 we obtained $B$, $V$, and $R_{\rm J}$ data with the IAC80 telescope of the Instituto de Astrof\'\i{}sica de Canarias located at Teide Observatory (Canary Islands, Spain) (0$\farcs$537 pixel scale). 
The data reduction for the RCC, Schmidt, RC80, Mt. Suhora and IAC80 observations was carried out in the same way as described in detail by \citet{szabo2021}. We show our photometric results in Table \ref{tab:phot} in Appendix A and in Figs.~\ref{fig:lc} and \ref{fig:lc_piszkes}. The typical uncertainty of our measurements is 0.02--0.03\,mag in $B$ and 0.01\,mag in all other filters. 

We obtained near-infrared images in the $J$, $H$, and $K_{\rm s}$ bands at six epochs between 2006 July 15 and 2012 October 13 using the 1.52\,m Telescopio Carlos Sanchez (TCS) at the Teide Observatory. Lastly, we used the NOTCam instrument on the NOT on 2021 August 9. Because of the brightness of our target in the infrared, we used a 5 mm diameter pupil mask intended for very bright objects to diminish the telescope aperture, which gave about 10\% transmission. The detailed data reduction was carried out the same way as for V1057~Cyg using TCS as described in \citet{szabo2021}. In the same paper we also provided further details about the above instruments. The typical photometric uncertainties are of 0.01--0.03\,mag. We present the $JHK_s$ photometry in Table~\ref{tab:phot} in Appendix A.
We also observed V1515~Cyg with the $2.56$\,m Nordic Optical Telescope (NOT) located at the Roque de los Muchachos Observatory (Canary Islands; Plan ID 63-401, PI: Zs.~M.~Szabó) We used the NOTCam instrument in wide-field imaging mode on 2021 April 25. The instrument consists of a 1024$\times$1024 pixel HgCdTe Rockwell Science Center `HAWAII'. The field of view was  $4^{\prime}\times 4^{\prime}$ with a pixel scale of 0$\farcs$234.

We supplemented our ground-based optical photometry with publicly available data from the ASAS-SN \citep{shappee2014, kochanek2017} and ZTF surveys \citep{masci2018}, which were reduced by dedicated pipelines. We also used the DASCH database \citep{dasch2012} for our work. The first ASAS-SN data were obtained on 2015 February 24, while the latest on 2022 January 4, whereas the ZTF observations range from 2018 March 28, to 2021 November 3. The DASCH data ranges between 1944 and 1989, and seasonal averages were used for plotting purposes in Fig.~\ref{fig:lc}. We also use the 1982 -- 2003 Mt.~Maidanak $UBVR$ data set, already analysed in detail by \citet{clarke2005}.

\subsection{Space-based optical and infrared photometry}

We used the Transiting Exoplanet Survey Satellite \citep[{\it TESS,}][]{ricker2015},~observed V1515~Cyg between 2019 July 18 and September 11 (Sectors~14 and 15), and then between 2021 July 20 and August 20 (Sector~41). The photometry was extracted from the full-field images using 1.5~pix aperture radius in the same way as described in \citet{szabo2021}. In the final step, the instrumental {\it TESS} magnitudes were aligned to our Schmidt- (direct) and RC80-telescope (transformed) measurements in the $I_C$ band, and the results are shown in Fig.~\ref{fig:lc_TESS}. The seven breaks visible in the light curve in 2019, and the single break in 2021, are caused either by the regular data transfers to the ground or by scattered Earth and Moon light, preventing accurate sky subtraction -- these affected data were discarded from the presented plots and from further analyses.

On 2018 August 29, V1515~Cyg was observed with the Stratospheric Observatory for Infrared Astronomy (SOFIA; \citealt{young12}) using the Faint Object infraRed CAmera for the SOFIA Telescope (FORCAST; \citealt{herter13}). We obtained mid-infrared imaging using the F056, F077, F088, F111, F242, F315, and F348 filters with effective wavelengths of 5.6, 7.7, 8.8, 11.1, 24.2, 31.5, and 34.8$\,\mu$m, respectively (Plan ID 06\textunderscore0062, PI: J.~D.~Green). Total on-source integration times were between and 45\,s and 87\,s depending on the filter. The images were processed using the SOFIA pipeline and retrieved as Level 2 data products from the SOFIA Science Archive as ingested into the IRSA database. We obtained aperture photometry on the images with a series of apertures with increasing radii and selected an aperture for the final photometry where this curve of growth flattened. The resulting fluxes are: 
$F_{5.6}=0.486\pm0.071$\,Jy, $F_{7.7}=0.502\pm0.065$\,Jy, $F_{8.8}=0.531\pm0.069$\,Jy, $F_{11.1}=0.981\pm0.069$\,Jy, $F_{24.2}=2.42\pm0.13$\,Jy, $F_{31.5}=2.06\pm0.16$\,Jy, $F_{34.8}=2.21\pm0.18$\,Jy.
We also used data from the Wide-field Infrared Survey Explorer (\textit{WISE}, \citealt{wright2010}). In Tab.~\ref{tab:phot_wise} in Appendix A we summarize the saturation corrected WISE data for V1515~Cyg shown in Fig.~\ref{fig:lc}. 
\subsection{Spectroscopy}
\label{sec:spectroscopy}

We obtained a new optical spectrum of V1515~Cyg with the high-resolution FIbre-fed Echelle Spectrograph (FIES) instrument on the NOT on 2021 May 29. We used a fibre with an entrance aperture of $2\farcs5$  which provided a spectral resolution $R$=25\,000 between 370 -- 900\,nm. The final spectrum is composed of three individual spectra, each with 2400 s exposure time. We reduced the spectra using the FIEStool software.
We also observed V1515~Cyg with the Bohyunsan Optical Echelle Spectrograph \citep[BOES;][]{kim2002} installed on the 1.8\,m telescope at the Bohyunsan Optical Astronomy Observatory (BOAO). BOES provides $R$=30\,000 (using a 300\,$\mu$m fiber) in the wavelength range $\sim$4000--9000\,\AA{}. We binned the spectra with 2$\times$2 pixels to increase the S/N. We took four spectra of V1515 Cyg between 2015 and 2018, for the exact dates, see the spectroscopic log of observations in Tab.~\ref{spec_dates}.
We reduced these spectra in a standard way within {\sc IRAF}: after bias and flatfield corrections, the ThAr lamp spectrum was used for wavelength calibration, and continuum fitting was performed using the \texttt{continuum} task. 
Finally, heliocentric velocity correction was applied by the \texttt{rvcorrect} task and the published radial velocity of V1515~Cyg \citep[$-$15\,km\,s$^{-1}$;][]{hartmann2004}. Unlike the usual observing method (and our 2021 NOT spectrum) with several exposures, the BOES spectra were taken with one, long exposure. This made the spectrum sensitive for cosmic rays with no means to remove them using the usual procedures.

As no telluric standard stars were observed either for FIES or BOES, we performed the telluric correction using the molecfit software \citep{smette2015,kausch2015} by fitting the telluric absorption bands of O$_2$ and H$_2$O. This generally provided good correction except for the deepest lines where the detected signal was close to zero.

\begin{deluxetable}{ccccc}
\label{spec_dates}
\tabletypesize{\scriptsize} 
\tablecaption{Log of Spectroscopic Observations \label{tbl_obs_log}}
\tablewidth{0pt}
\tablehead{
\colhead{Telescope} & \colhead{Instrument} & \colhead{Spectral} & \colhead{Observation} & \colhead{Exp. time} \\[-3mm]
&  & \colhead{Resolution} & \colhead{Date [UT]} & \colhead{[sec]}}
\startdata
BOAO & BOES & 30,000 & 2015 Dec 26 & 3600 \\
$\cdots$ & $\cdots$ & $\cdots$ & 2017 May 29 & 3600 \\
$\cdots$ & $\cdots$ & $\cdots$ & 2018 Oct 07 & 600 \\
$\cdots$ & $\cdots$ & $\cdots$ & 2018 Dec 19 & 3600 \\
NOT & FIES & 25,000 & 2021 May 29  & 2400 $\times$ 3 \\
NOT & NOTCam & 2500 & 2021 Aug 09  & 400 $\times$ 4 \\
\enddata
\end{deluxetable}


On 2021 August 9, we used the NOTCam on the NOT to obtain a new near-infrared spectrum of V1515~Cyg and Iot~Cyg (telluric standard, A5~V) in the $JHK_s$ bands (1.25--2.2\,$\mu$m). We used the low-resolution camera mode ($R$=2500) with ABBA dither positions, and exposure times ranged from 60 to 100 seconds (Tab.~\ref{tbl_obs_log}). For each image, flat-fielding, bad pixel removal, sky subtraction, aperture tracing, and wavelength calibration steps were performed within {\sc IRAF}. 
A Xenon lamp spectrum was used for the wavelength calibration. The Hydrogen absorption lines in Iot~Cyg were removed by Gaussian fitting and then the spectrum of V1515~Cyg was divided by the normalized spectrum of Iot~Cyg for telluric correction. 
Flux calibration was performed by applying the accretion disk model (Sec.~\ref{sec:accdisc}). It is based on the NOT $JHK_s$ photometry taken on the same night, but the photometry was taken in broadband filters, hence gives fluxes only for the effective wavelengths of these filters. We therefore use the SED model (desribed in Sec.~\ref{res:sed}) fitted to the photometric points to interpolate the fluxes for any given wavelength needed for the flux calibration of the spectrum.

\section{Results and Analysis} 
\label{sec:res}

\subsection{Light Curves}
\label{sec:lc}

\begin{figure*}
\includegraphics[width=\textwidth]{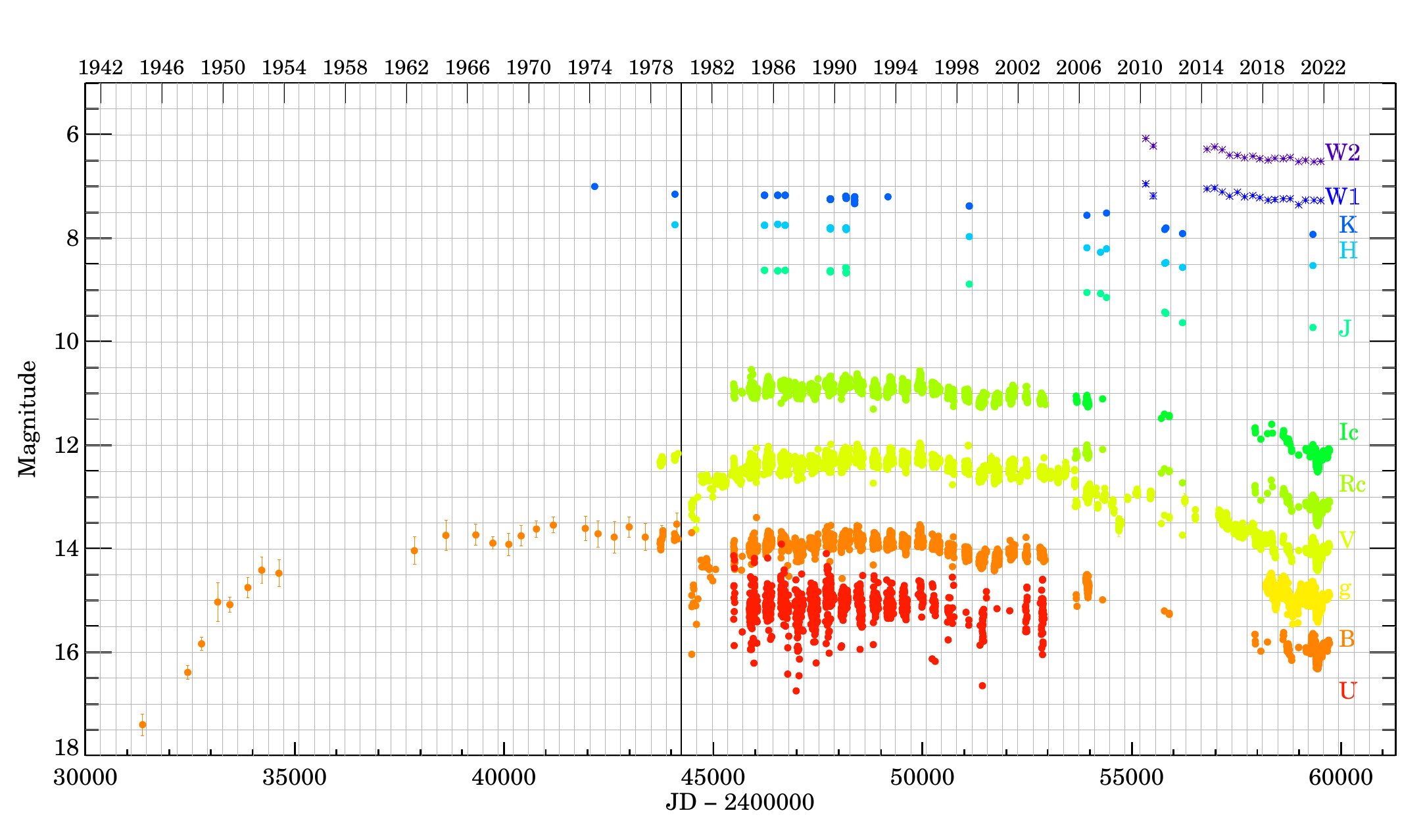}
\caption{Long-term optical and near-infrared light curves of V1515~Cyg. The vertical solid black line marks the fading event occured in 1980. We complemented our work with data from the literature prior to 2006 and it is summarized in Tab.~\ref{tbl_phot_obs_log}.}
\label{fig:lc}
\end{figure*}

In order to study the brightness variations of V1515~Cyg in the long-term, we constructed light curves with data gathered from the literature \citep{herbig1977,landolt1977,kolotilov1983,KH1991,kenyon-and-hartmann1991,herbst1999,clarke2005,dasch2012} and we complemented the archival data with results from our monitoring. We present the long-term photometric light curves of V1515~Cyg in Fig.~\ref{fig:lc}.
In order to stay consistent between numerous data sets and to simplify further analyses, we transformed our RC80, Mt. Suhora, and ZTF Sloan $r'i'$ magnitudes into Johnson--Cousins $R_{\rm C}I_{\rm C}$ system by constructing the SED for each day observations and interpolate for the effective wavelengths of the $R_C$ and $I_C$ filters using a spline function in the log-log space. The resulting interpolated fluxes were transformed to magnitudes with the zero points of the $R_CI_C$ filters.
We stress that this approach gave more consistent results 
than the transformation formulas between Johnson and Sloan systems given by \citet{jordi2006}. Any remaining offsets between overlapping data sets in optical bands were then removed by constant shifts, as listed in Tab.~\ref{tab:shifts} in the Appendix. We give a brief summary of the different data sets used for Fig.~\ref{fig:lc} in Tab.~\ref{tbl_phot_obs_log}.

As mentioned in Section \ref{sec:intro}, the outburst of V1515~Cyg was unlike those any other classical FUors: it brightened for $\sim$30 years, before reaching the peak brightness in the late 1970's \citep{landolt1975,landolt1977}, then in 1980 it started a quick fading event, reaching the minimum brightness, and re-brighten again for several years \citep{kolotilov1981,kolotilov1983,KH1991,clarke2005}.
Between 1984--2004 the light curve showed a plateau shape \citep{clarke2005}.
However, a significant long-term fading of V1515~Cyg started around 2005--2006, with a sudden drop in the brightness ($\sim$1 mag in the $BVR$ filters), a trend that is currently still ongoing with the mean rate of 0.085~mag~yr$^{-1}$ in the $V$-band. We note, that both the plateau and the long-term fading are visible in the optical $BVR$, the near-infrared $JHK$, and the \textit{WISE} light curves also confirm the fading trend at 3.4 and 4.6 $\mu$m. 

The small amplitude light changes were thoroughly analysed by \citet{clarke2005} between the 1980's and early 2000's. 
However, our careful look into the same Mt.~Maidanak data set revealed a feature unnoticed earlier: between 1984 -- 2003 the small-scale variations visible on top of the major light curve plateau predominantly manifest themselves in the form of peaks. 
At first glance, they resemble accretion bursts typical for magnetospheric accretion in CTTS, however, the $UBVR$ light curves show very similar variability amplitudes. In combination with the characteristic time scales of these brightness fluctuations ($\sim 10-30$~d) this strongly suggests that all these variations most likely originated in the inner protoplanetary disk. 

The INTEGRAL $V$-band data are sparse, thus the peak-like events could not be seen clearly. Instead, these data revealed at least one well-defined dip at $JD\approx2454700$ (2008 August) -- perhaps a signature of gradual disk envelope cavity closure by increased dust condensation as the outburst is weakening. The dips are in fact more numerous in the data obtained in 2010s and 2020s.

\begin{deluxetable}{ccc}
\tabletypesize{\footnotesize} 
\tablecaption{Summary of the photometric data used for Fig.~\ref{fig:lc}\label{tbl_phot_obs_log}}
\tablewidth{0pt}
\tablehead{
\colhead{Date} & \colhead{Filters} & \colhead{Ref.}}
\startdata
1974 -- 1991    &   $JHK$     & \citet{molinari1993} \\
1975, 1976      &   $V$       & \citet{herbig1977} \\    
1976            &   $UBV$     & \citet{landolt1977} \\
1978--1983      &   $UBV$     & \citet{kolotilov1983} \\
1983--1995      &   $V$       & \citet{herbst1999} \\
1983--2005      &   $UBVR$    & \citet{clarke2005} \\
1982--1991      &   $V$   & \citet{KH1991} \\
1985--1991      &   $JHK$   & \citet{KH1991} \\
1985 -- 1998    &   $K$       & \citet{kolotilov1990} \\
2005--2011      &   $BVRI$    & This work \\
2006, 2012      &   $JHK_s$   & This work \\
2011-2019       &   $BVRI$    & This work \\
2019--2021      &   $BVr'i'$  & This work \\
2021            &   $JHK_s$   & This work \\
\enddata
\end{deluxetable}

\begin{figure*}
\includegraphics[width=\textwidth]{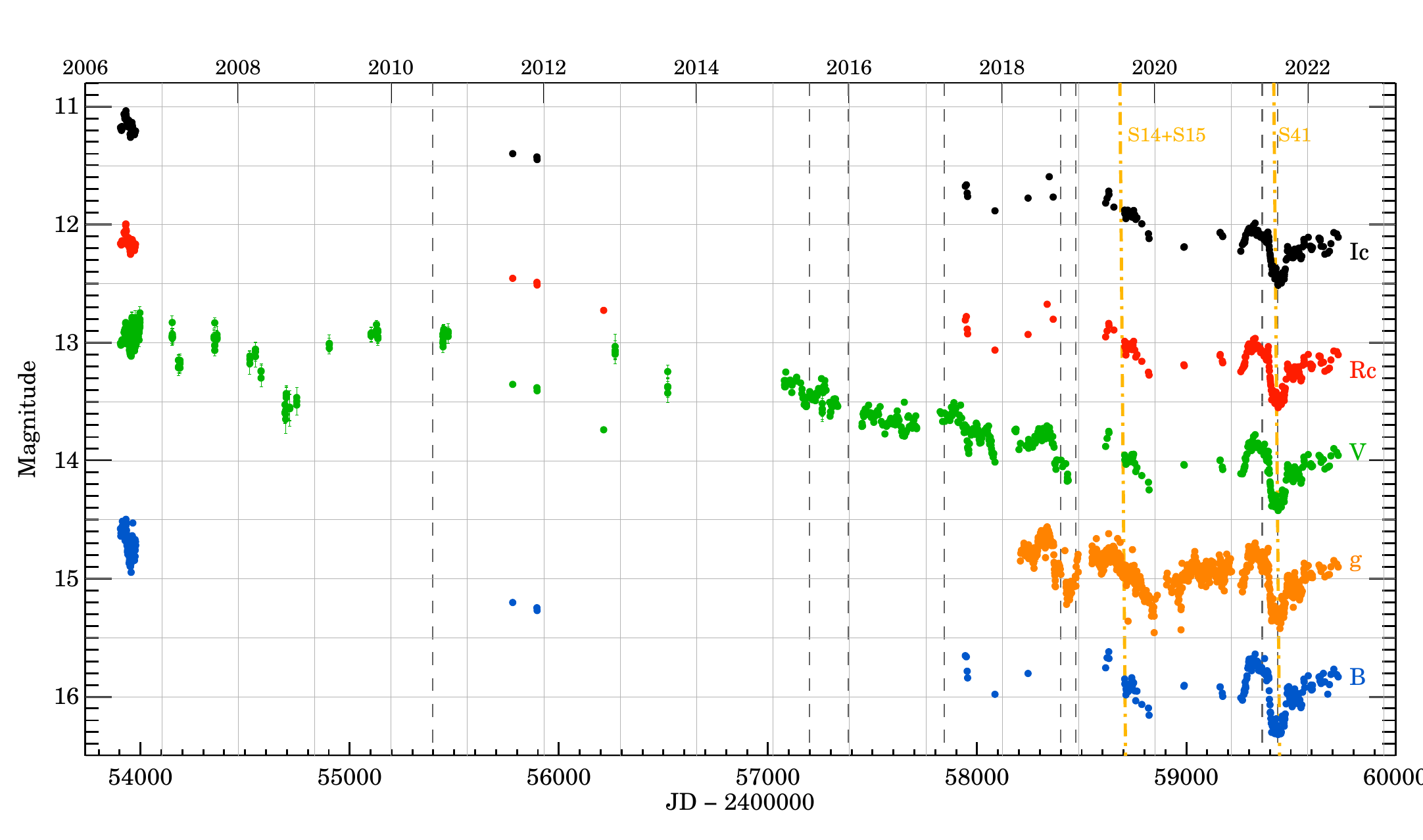}
\caption{Optical light curves of V1515~Cyg between 2006 and 2022. Most of the $BVR_CI_C$ data were obtained at the Piszkéstető Observatory, while the rest of the data taken in the $V$- and $g$-band are from the ASAS-SN database. The vertical dashed dark grey lines mark the BOAO and NOT optical/IR observations between 2012 and 2021 as well as the 2015 spectrum of V1515~Cyg by \citet{connelley2018}. Yellow dash-dotted lines indicated the \textit{TESS} Sector 14, 15 and 41 observations of the source.}
\label{fig:lc_piszkes}
\end{figure*}

\begin{figure*}
\includegraphics[width=0.5\linewidth]{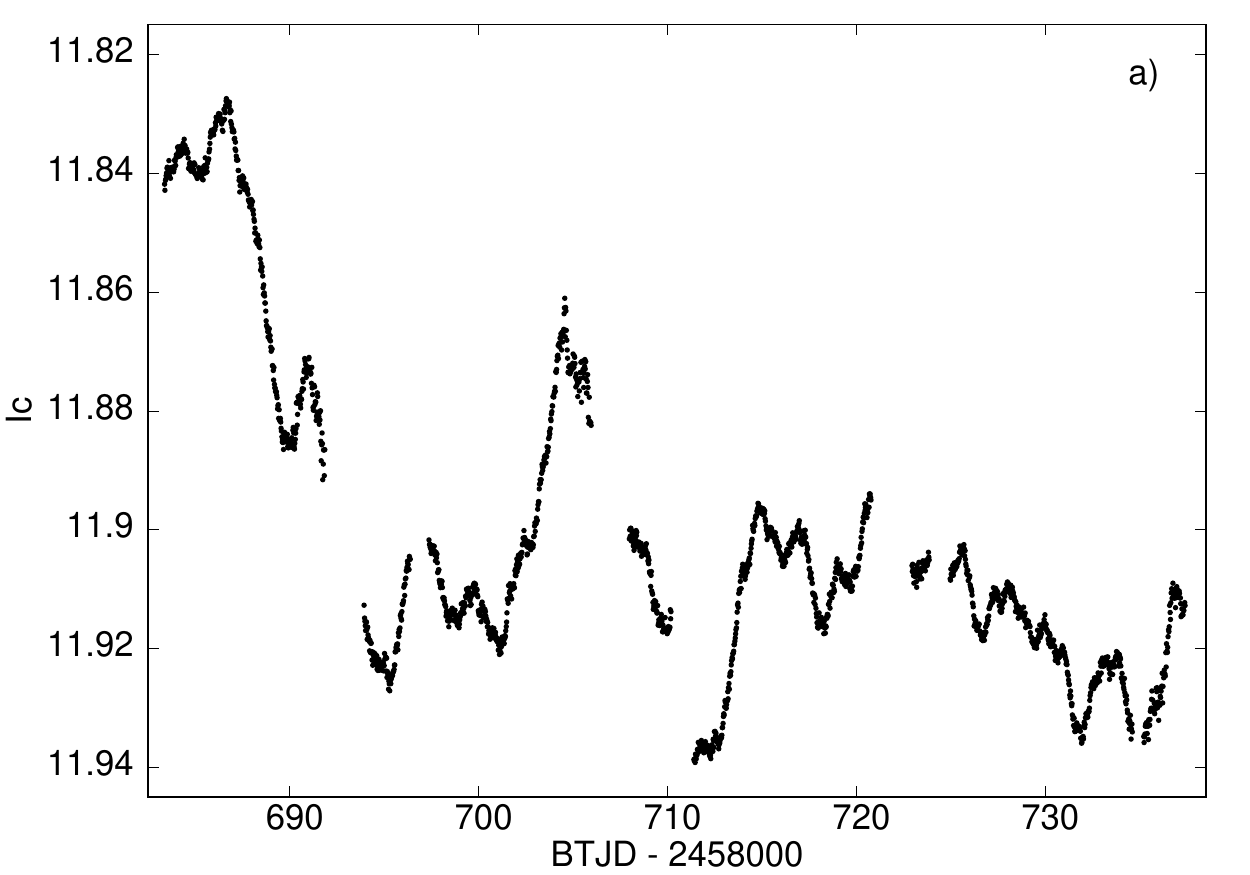}
\includegraphics[width=0.5\linewidth]{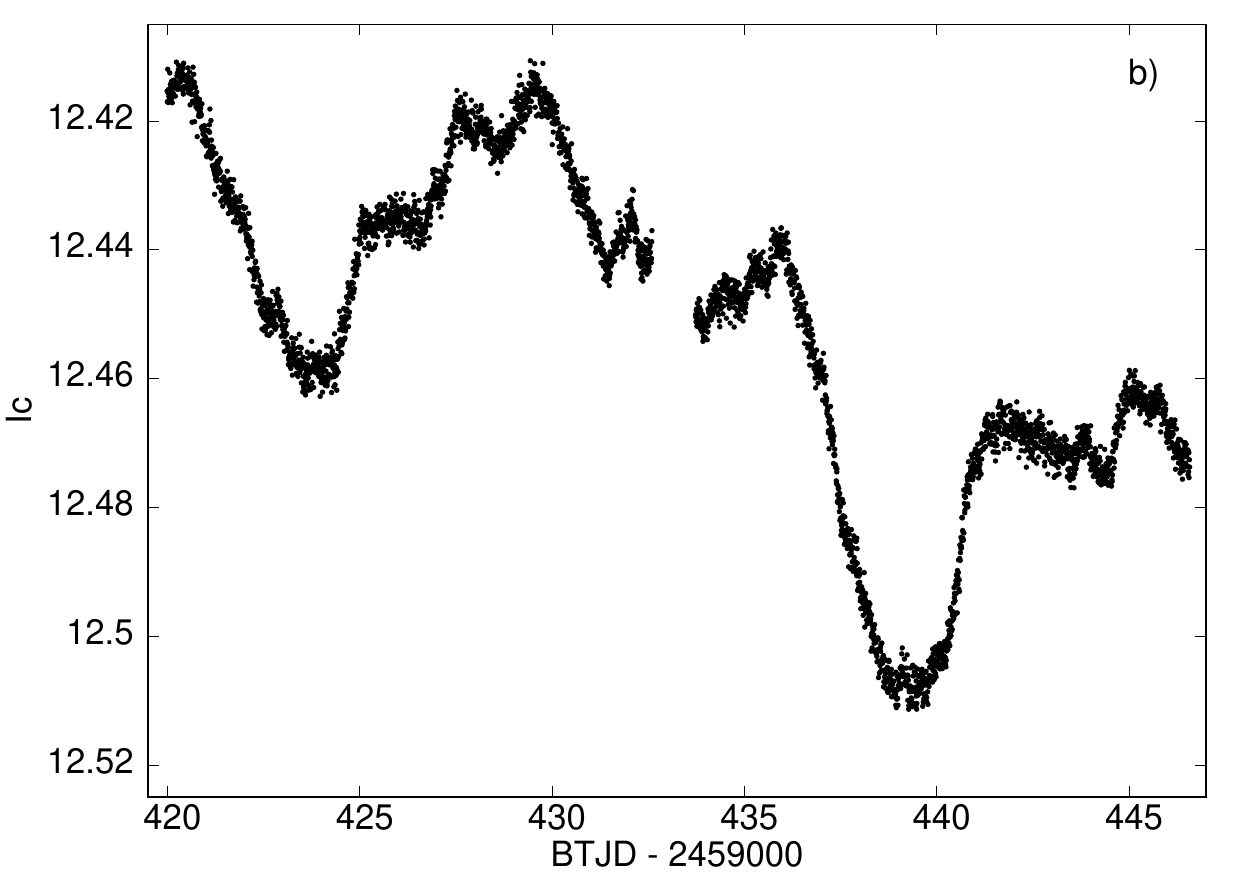}
\caption{Light curves of V1515~Cyg gathered by {\it TESS} in 2019 and 2021, aligned to our $I_C$ light curves.}
\label{fig:lc_TESS}
\end{figure*}

\subsection{Frequency Analysis}
\label{sec:freqanalysis}

In order to search for periodic and quasi-periodic oscillations (QPOs) in the ground-based data, we utilised the $V$- and $g$-band light curves, providing both the best temporal coverage and showing variability present in other optical bands. 
In the first step, all outlier points were removed.
As the ASAS-SN survey gradually switched to the $g$-band, we re-scaled the amplitude of these variations to match those observed in the $V$-band by the multiplicative factor of 0.95. We determined this value during the time period, when simultaneous $V$- and $g$-band data were obtained, then, we aligned the scaled $g$-band light curve to the $V$-band by a constant shift.

In the first step we searched for long-term quasi-periodic variations in the 1982--2021 dataset superimposed on the major light curve plateau. 
For this purpose we subtracted the major long-term trend from the light curve by fitting a 2nd order polynomial. Although the Lomb-Scargle \citep{zechmeister2009} diagram computed from the entire light curve reveals a forest of peaks, none of them represent stable period nor quasi-period. 
For this reason, in the second approach we decided to split the light curve based on the dominant variability morphology and analyse each part separately, i.e.~the first part obtained before 2005 ($JD=2453500$), dominated by peaks, and the second part obtained after, which is dominated by dips. 
No significant quasi-periodicity was found, except for the three consecutive dip-like 350--370\,d variations visible directly in the light curve between $JD=2458000-2459000$ (2017--2020).

In the third step we decided to analyse each well-sampled observing season separately. Prior to this analysis, any trends visible during certain seasons were removed by using low order (1--2) polynomials to enhance the shorter changes. We obtained that in the 1987 data set there is a significant 13.91\,d QPO with false alarm probability of 4$\times$10$^{-5}$. This is similar to the 13.89\,d period found by \citet{clarke2005} in the same data. In addition, we found a significant 13.06\,d peak with false alarm probability of 10$^{-3}$ in the 2003 data set. 
During the other seasons, the light curve usually exhibited only two consecutive oscillations of a similar pattern, quickly disappearing and giving way to light changes with different patterns. 

In order to study the small-scale variability unavailable from the ground, we utilised the {\it TESS} light curves obtained in 2019 and 2021. 
We performed frequency $f$ analysis using the procedure of \citet{ruc08}, where the amplitude $a_f$ errors are estimated by means of the bootstrap sampling technique. Prior to the analysis, the general downward trends in the brightness visible during both in 2019 and 2021 (Fig.~\ref{fig:lc_TESS}a,b) were removed by linear fits.
Subsequently, the residual magnitudes were transformed to normalized flux units. We found that in both seasons the amplitude-frequency spectrum exhibits a Brownian random-walk nature (Fig.~\ref{fig:freq_TESS}a,b), in which the amplitudes scale with the frequency as $a_f\sim f^{-1}$ \citep{press1978}. Only in the 2021 data set there is a single peak at $f=0.439$~c~d$^{-1}$, which appears to represent a significant 2.28\,d QPO. 

\subsection{Wavelet Analysis}
\label{sec:waveletanalysis}

It is known that Fourier analysis cannot trace temporal changes nor localize in time finite oscillatory packages such as those present in our data. As it was for the first time proven by \citet{ruc08,Rucinski2010}, the wavelet analysis can be successfully applied to extract such information from the evenly-sampled space-based light curves of vigorously accreting CTTS and Herbig~Ae stars.
In particular, the Morlet wavelet allows us to separate individual oscillatory packages and directly characterise their temporal changes. 
Although the variability observed in the above mentioned classes of young stars is caused by the changing visibility, number, and the lifetime of the numerous hot spots resulting from the (most often) unstable or moderately stable magnetospheric accretion \citep{blinova2016}, or in some particular cases the pulsed accretion 
\citep{tofflemire2017,kospal2018, tofflemire2019,Fiorellino2022}, it was already theoretically proven that the hot spot behavior actually reflects the physical conditions in the inner disks of CTTS (see \citealt{blinova2016} and references within). 
In the case of the FUors, where the disk radiation is dominating over the stellar one, and the magnetosphere is likely totally crushed by the enhanced disk-plasma pressure so that hot spots typical for CTTS cannot form, effects brought by these inner disk instabilities can be directly observed in the visual light only in the objects with residual traces of the envelope \citep{siwak2013, green2013b, baek2015, siwak2018, siwak2020}.

\begin{figure*}
\includegraphics[width=0.5\linewidth]{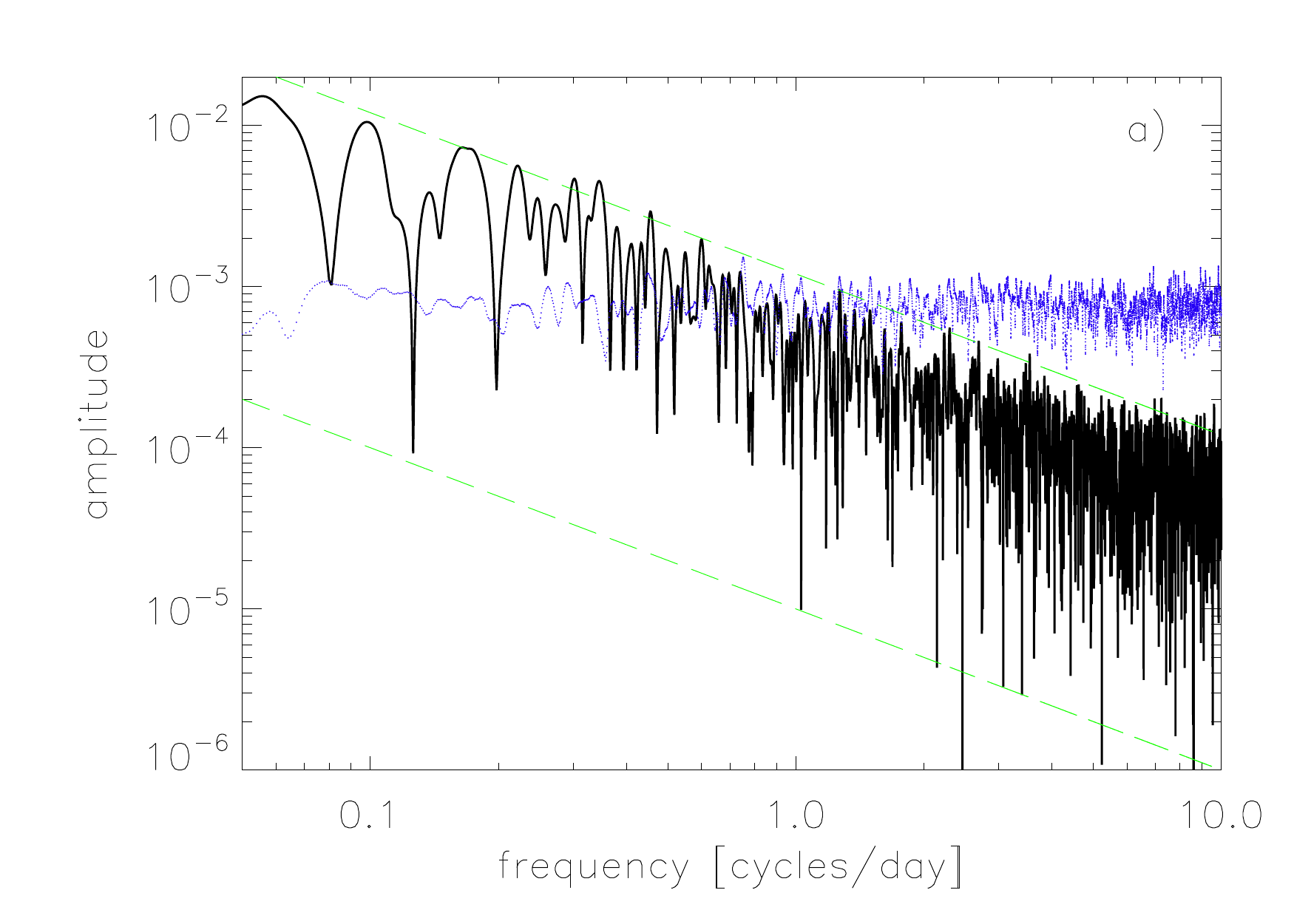}
\includegraphics[width=0.5\linewidth]{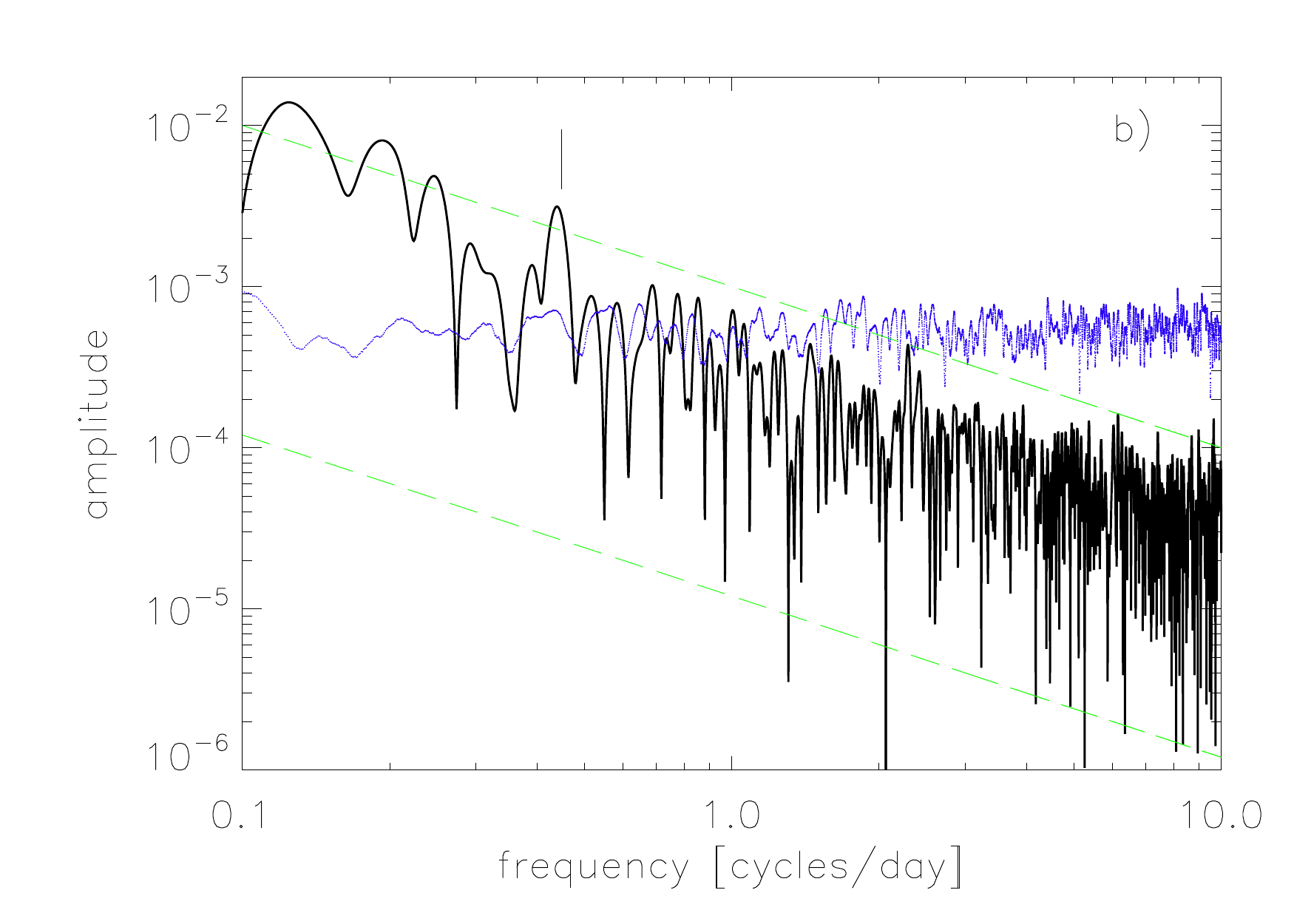}\\ 
\includegraphics[width=0.5\linewidth]{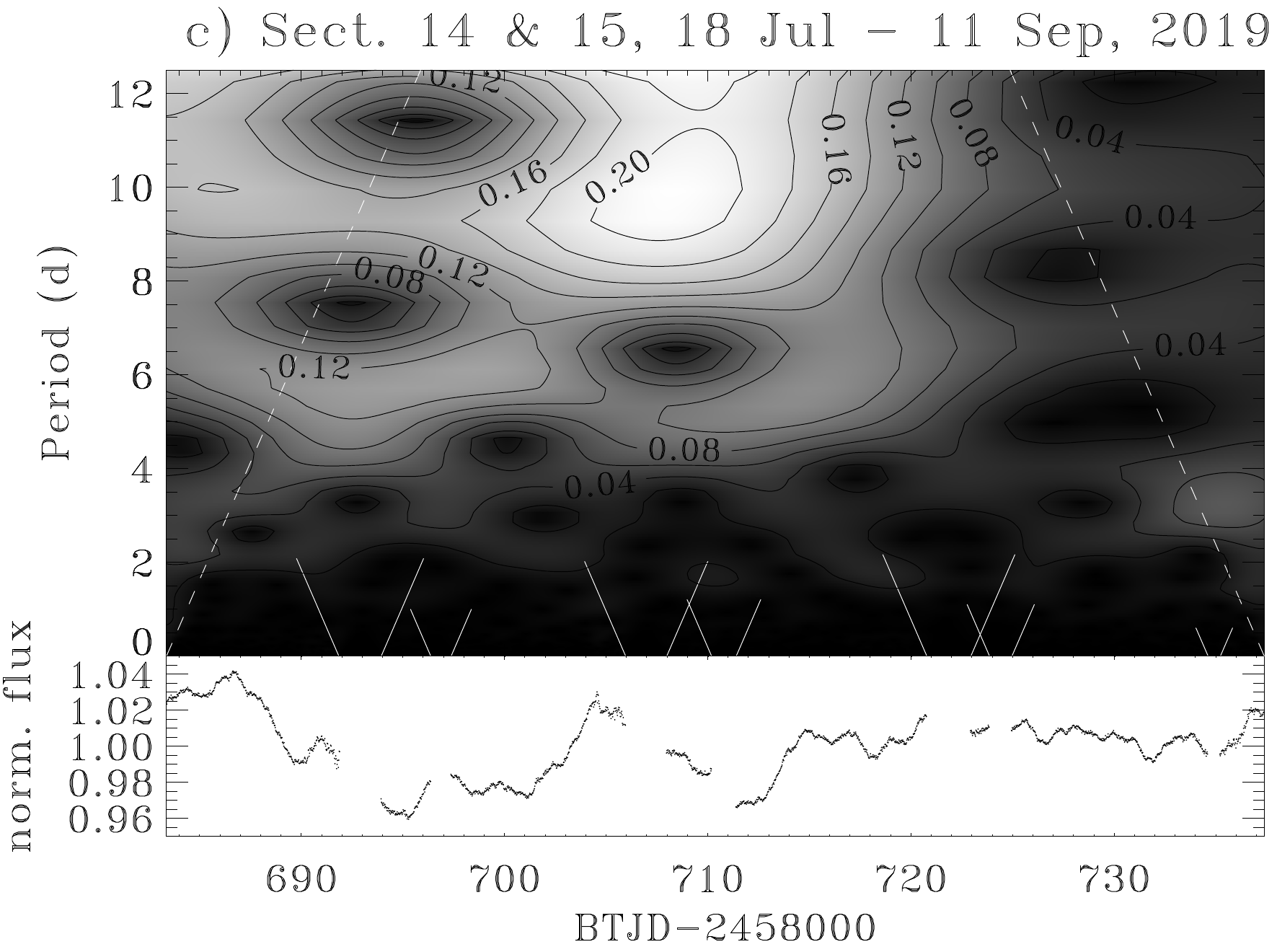}
\includegraphics[width=0.5\linewidth]{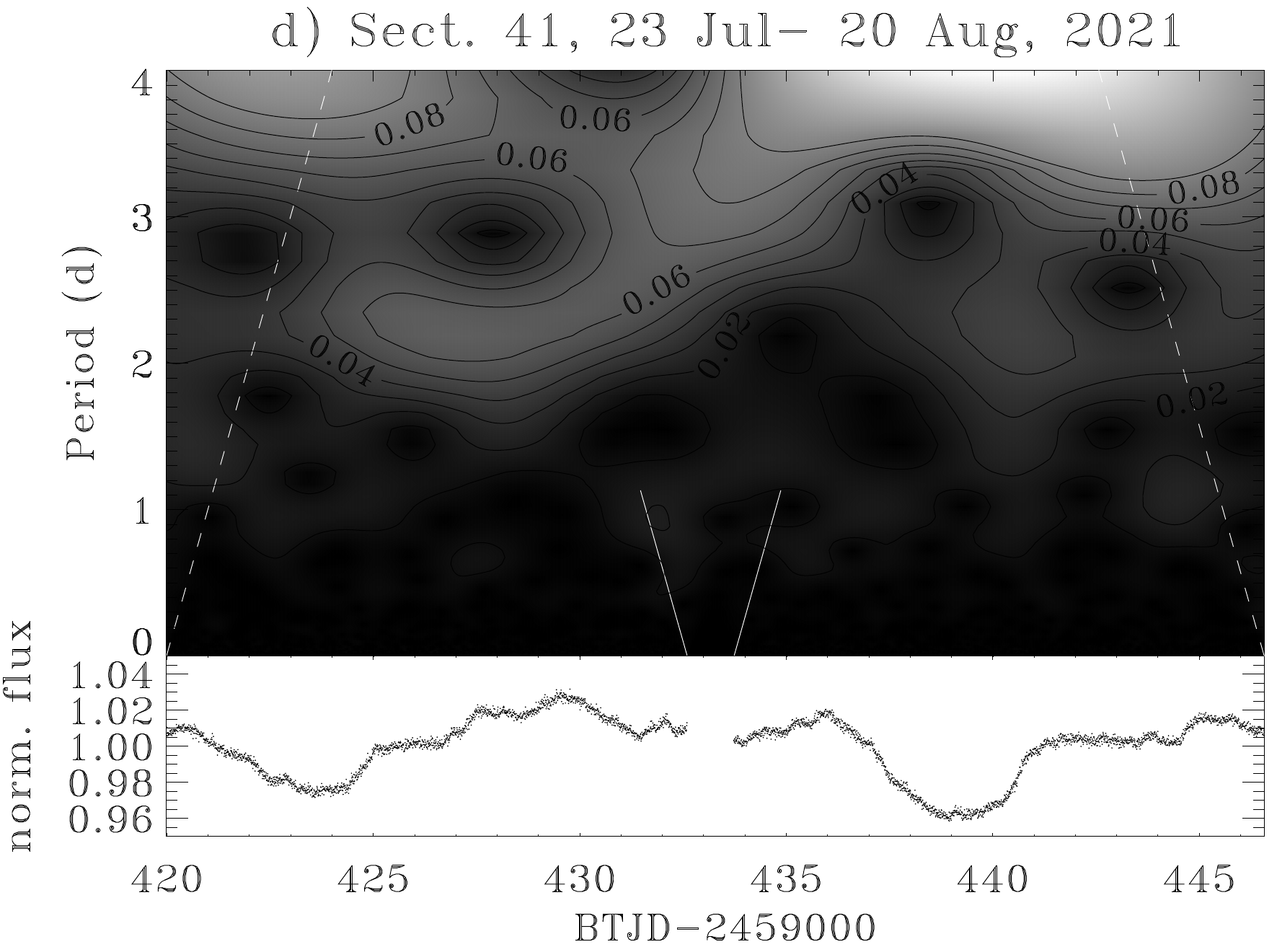}\\ 
\includegraphics[width=0.5\linewidth]{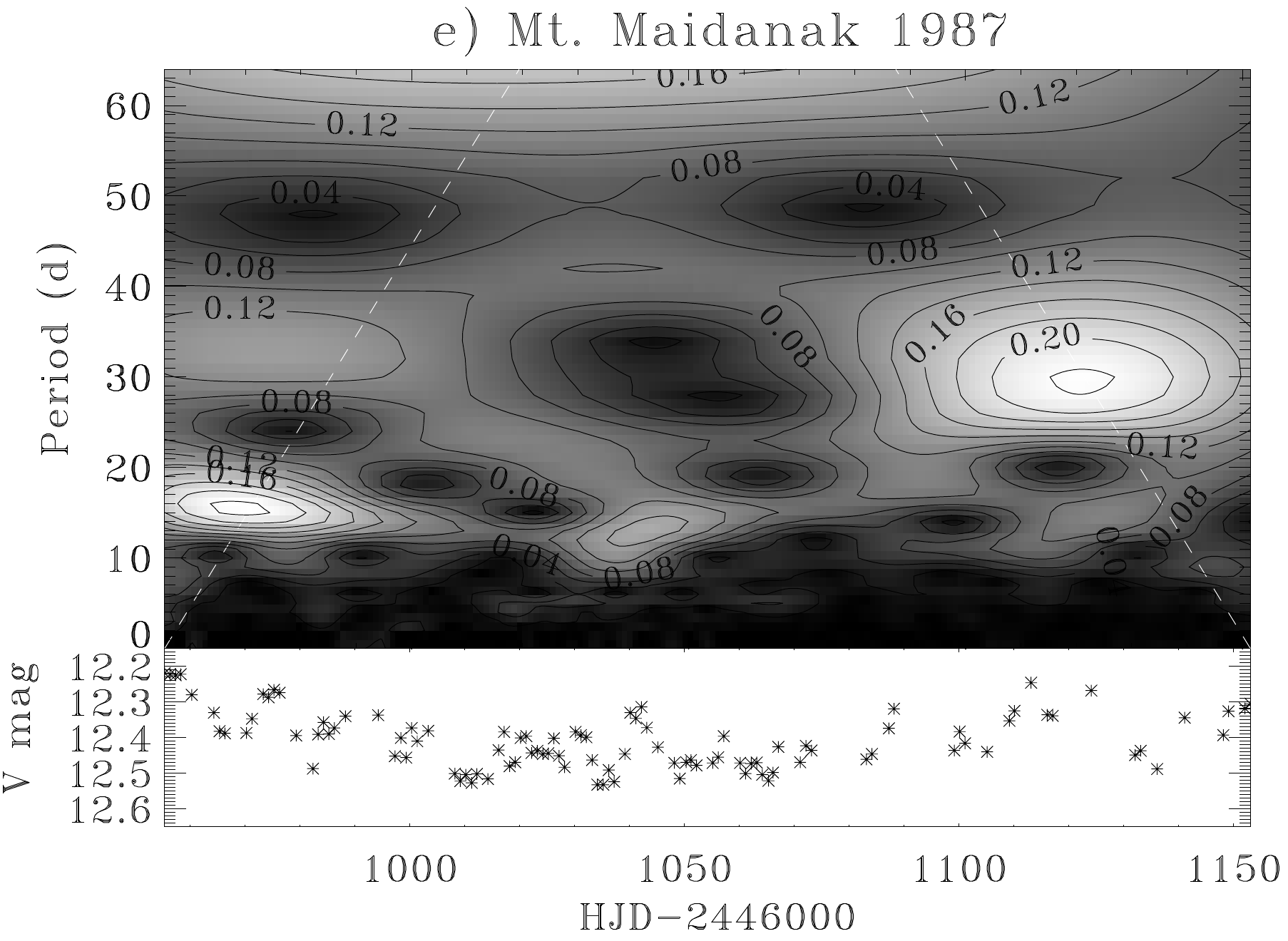}
\includegraphics[width=0.5\linewidth]{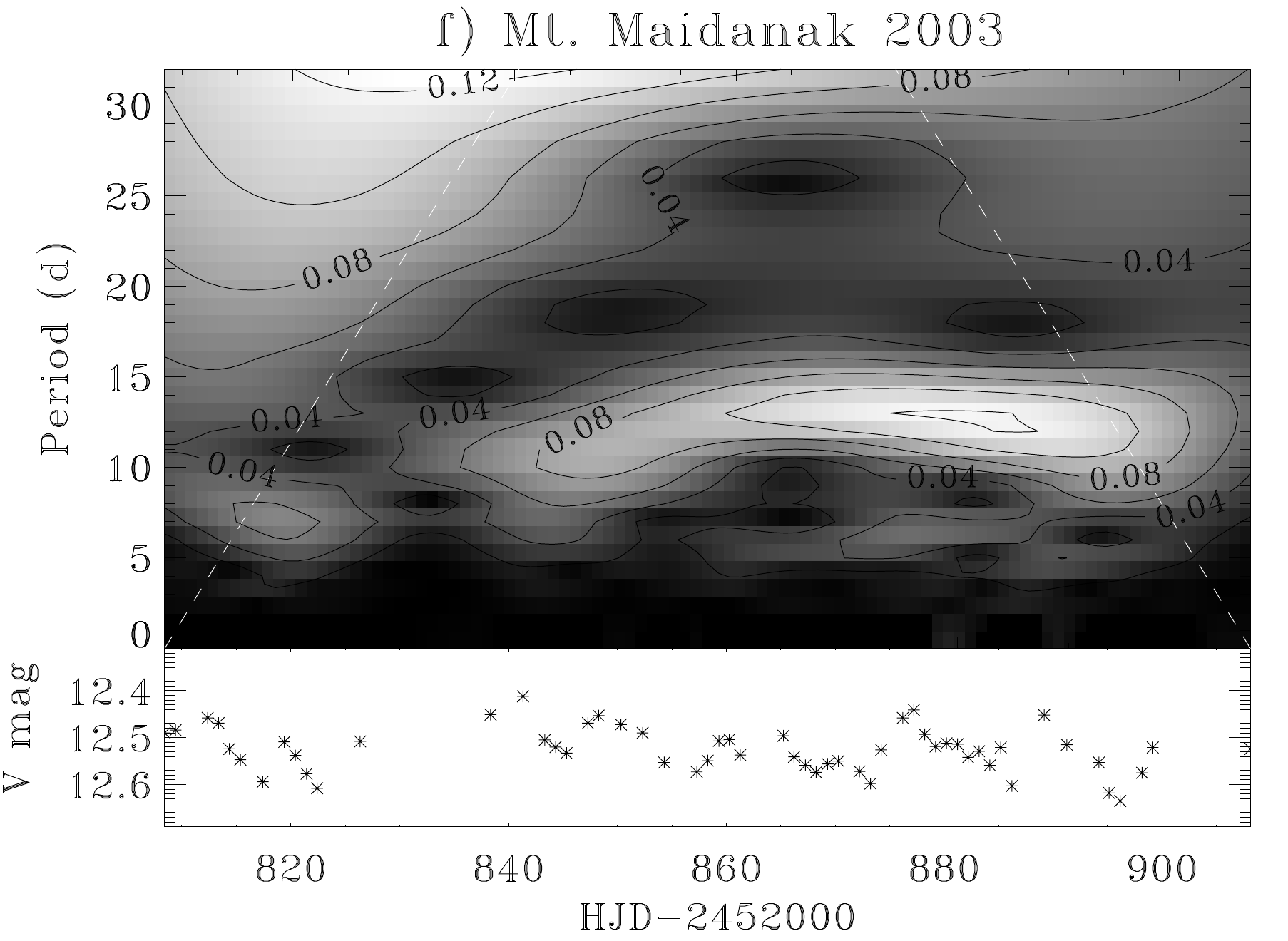}
\caption{Upper panels: Fourier spectra (black line) calculated from the de-trended and transformed to normalised flux units 2019 (Sectors 14 \&15) and 2021 (Sector 41) {\it TESS} light curves. The stochastic nature ($a_f \sim f^{-1}$) of these oscillations is indicated by the two parallel green dashed lines. The amplitude errors are represented by blue dots. Middle panels: respective Morlet wavelet spectra for {\it TESS}. Bottom panels: Morlet wavelet spectra showing the 13.1-13.9\,d QPO observed in 1987 and 2003. The original Johnson $V$ data are shown below.
In all wavelet spectra, the major edge effects lie outside of the white dashed lines. The secondary effects caused by interruptions in {\it TESS} data acquisition are marked by white solid lines.
}
\label{fig:freq_TESS}
\end{figure*}

We obtained that the Morlet-wavelet spectrum of the 2019 {\it TESS} data does reveal only a single, weak signature of QPO drifting from $\sim$6 to $\sim$5.5\,d between $BTJD-2458000=690-720$ (Fig.~\ref{fig:freq_TESS}c). 
In accord with the visual examination of the light curve, we note random changes, usually showing only one cycle or two self-similar oscillatory cycles. 
The shortest variations of this kind have a period of 1.5\,d, but most of them are longer than 2\,d. We have to stress that this conclusion is unfortunately disturbed by the seven interruptions in the {\it TESS} data acquisition, strongly affecting the short periodic part of the spectrum, as indicated by the solid white lines in Fig.~\ref{fig:freq_TESS}c,d. 
Moreover, there is a non-zero chance that any of these affected regions may propagate towards the longer periods inside the open cone(s), which can be imaginatively extended as indicated by the white continuous lines.

The shortest significant variations in the 2021 {\it TESS} wavelet spectrum are of 0.8--0.9\,d and are mostly localised in the middle of the data set. The spectrum does also reveal that the 2.28\,d peak obtained in Sec.~\ref{sec:freqanalysis} is unstable, and is seen only between $BTJD-2459000\approx425-430$, i.e.~it lasted for barely two oscillatory cycles (Fig.~\ref{fig:freq_TESS}d), it re-appeared around $BTJD-2459000\approx441$, but it is not well defined in the light curve itself. 
However, together with the former case it can produce an excess of power in the frequency spectrum, which leads to the spurious detection. 

Similarly to \citet{szabo2021}, we also applied the weighted wavelet Z-transform (WWZ, \citealt{foster1996}), designed for analysis of unevenly sampled time series, and available within the {\sc Vartools} package \citep{hartman2016}. In general, the analysis did not show any oscillatory packages other than those directly inferred from the best-sampled Mt.~Maidanak light curves themselves, i.e.~typically consisting of two self-similar oscillations (see in Sec.~\ref{sec:freqanalysis}).
We show only the results for the 1987 and 2003 seasons, although instead of the WWZ, we decided to present the ordinary Morlet wavelet spectra. We stress that the spectra were computed from the original data interpolated into evenly-spaced time grid with 1\,day step, which is equal to the median spacing of the Mt.~Maidanak original data (Fig.~\ref{fig:freq_TESS}e,f). The signatures of the 13.1\,d QPO are evidently best defined for the 2003 data set, but we stress that the detection efficiency is a function of the data sampling, as shown in the associated bottom panels. For that reason, the 13.9\,d QPO from the 1987 season is less securely defined, as the maximum amplitude appears before the first dashed line, denoting edge effects, and then the amplitude and the period appear to change.

\subsection{The Curious Case of the 2021 Phenomenon}
\label{sec:cc}

\begin{figure*}
\includegraphics[width=1.0\textwidth]{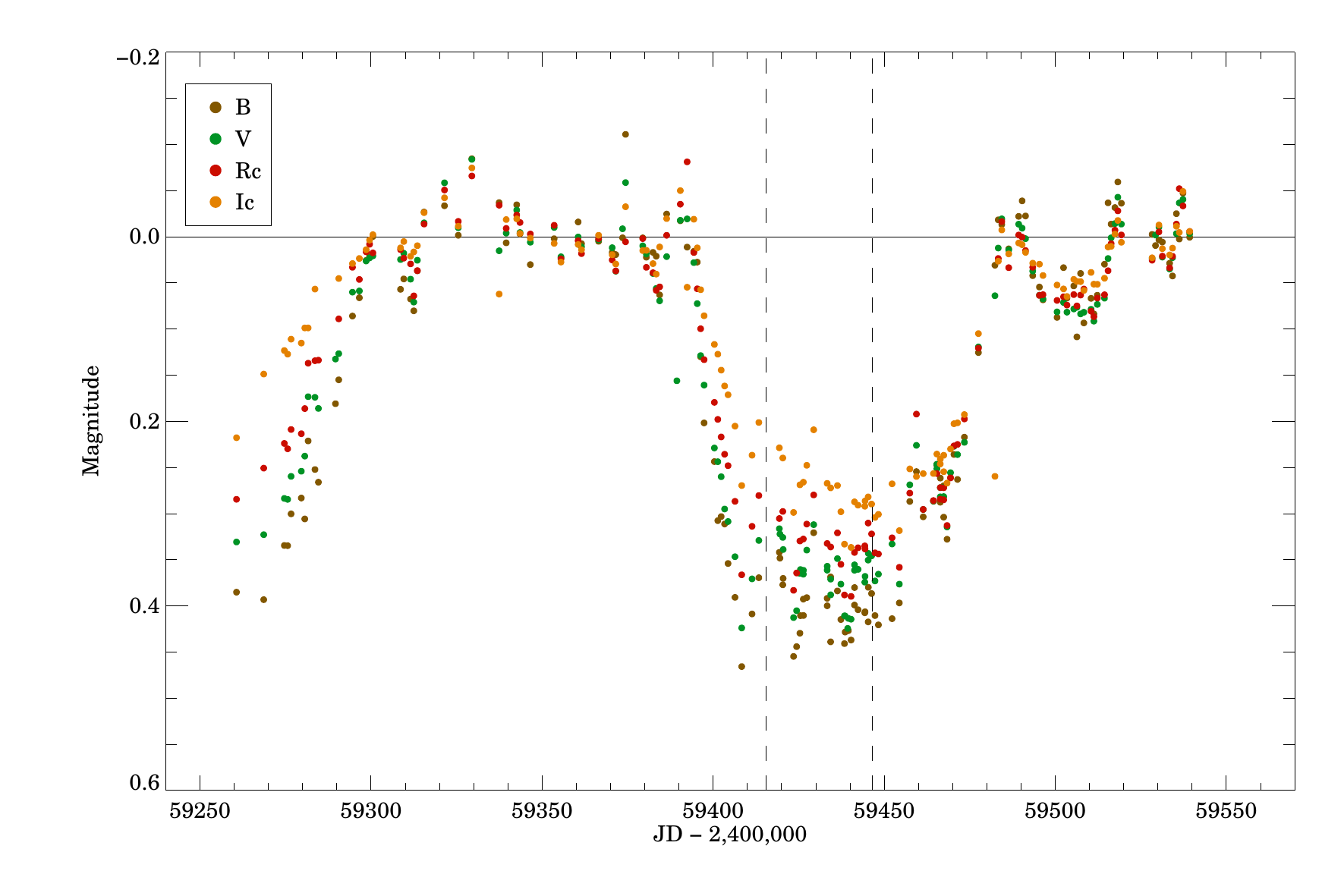}\\
\includegraphics[width=1.0\textwidth]{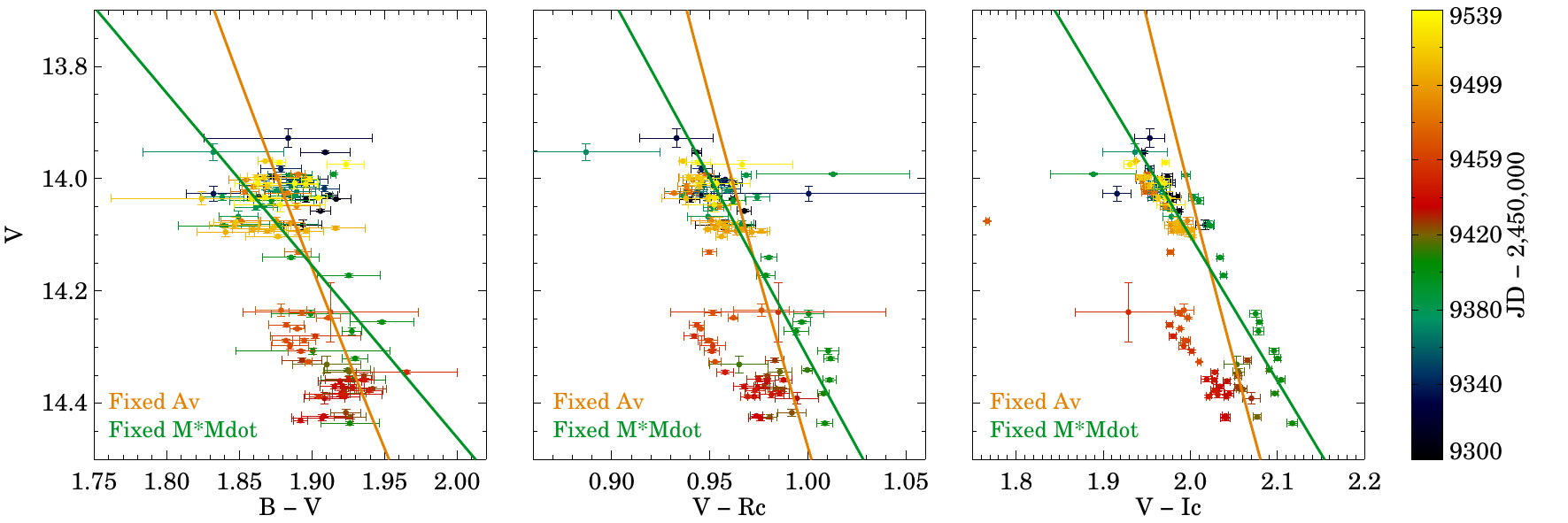}
\caption{Upper panel: detrended 2021 data indicated in different filters. {\it TESS} coverage is indicated by two parallel dashed lines. Bottom panel: color-magnitude diagrams for the detrended light curve. 
Evolutionary paths obtained from accretion disk modelling (Sec.~\ref{sec:accdisc}) are indicated by lines of different colors.}
\label{fig:lc_2021}
\end{figure*}

The data coverage from ground-based telescopes was exceptional in 2021.
The early 2021 season data showed brightening trend by 0.2 -- 0.4\,mag, which stabilised for about 1 -- 1.5 month, then (in June 2021) it faded quickly
and stayed nearly-constant for about one month, exactly during the {\it TESS} observations in July -- August. In early September, the brightness started to rise and stabilized on the level expected from the long-term fading trend.

Taking advantage of the excellent coverage, we analysed this phenomenon in detail as follows. First we detrended the long-term fading trend by linear fit to the high brightness level, as indicated in the upper panel of Fig.~\ref{fig:lc_2021}. One can notice that brightness evolution in different filters occurred in slightly different way. This is better emphasized on color-magnitude diagrams, as shown in the bottom panels of Fig.~\ref{fig:lc_2021}. 
It is evident that the brightness drop followed the extinction path. Then, in the light curve minimum the colors became bluer, which suggests scattering by small grains, like in UXor stars \citep[see e.g.,][]{abraham2018}. Finally during the rising branch, all color indices remained almost unchanged, i.e.~the brightness went to the upper level almost on a grey path.

We performed the same analysis for the 2019 and 2020 observing seasons, but conclusive results could not be reached due to the coarser sampling. Therefore, we only focused on the 2021 data set, since this is the most useful one to draw conclusions thanks to the good data sampling. \\
We discuss in detail the possible physical interpretation of the  dimming events of V1515~Cyg in comparison with the other classical FUors and the classical T~Tauri star (CTTS) RW~Aur in Sec.~\ref{sec:dimming}.
\subsection{Spectroscopy}
\label{sec:spectroscopy}
We monitored V1515~Cyg from 2012 to 2021 with optical and near-infrared spectrographs and obtained six spectra. We detected P~Cygni profiles, broad blue-shifted absorption lines, and metallic lines in absorption.
For the identification of the lines the NIST Atomic Spectra Database\footnote{https://physics.nist.gov/PhysRefData/ASD/lines\_form.html} was used. 


\subsubsection{Optical Spectroscopy}
\label{sec:optical_spectroscopy}

\begin{figure*}
    \centering
    \includegraphics[width=0.45\textwidth]{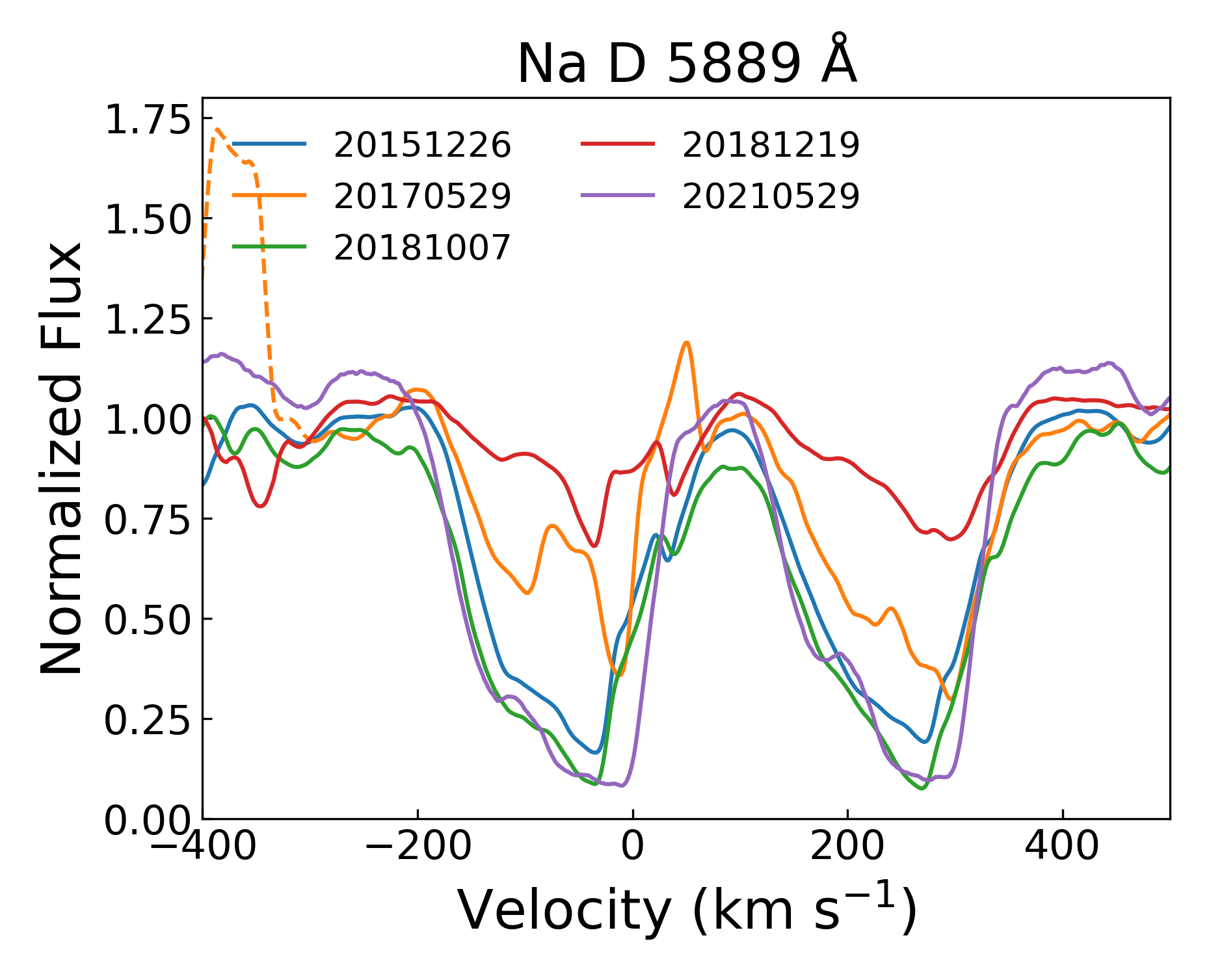}
    \includegraphics[width=0.45\textwidth]{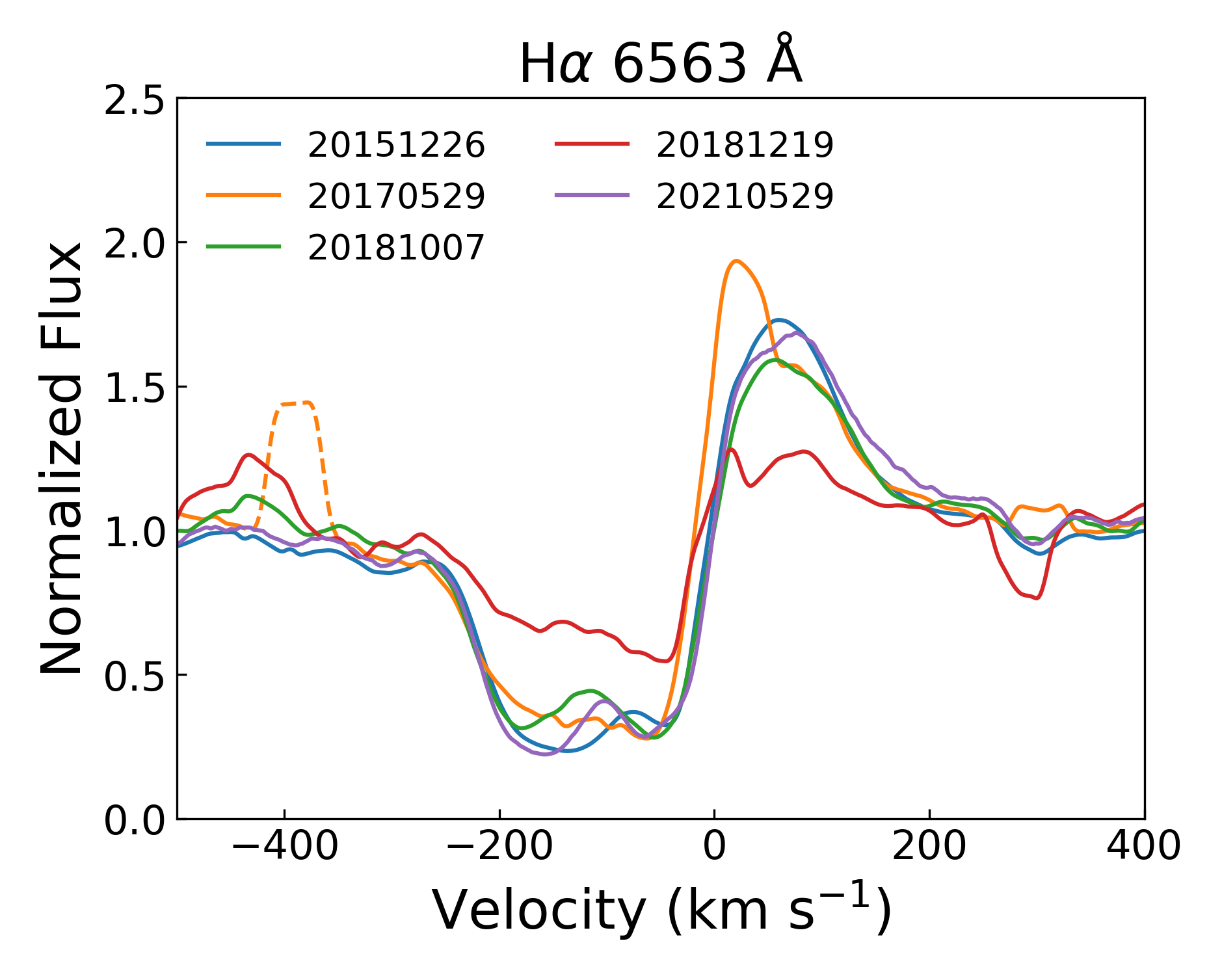}
    \includegraphics[width=0.45\textwidth]{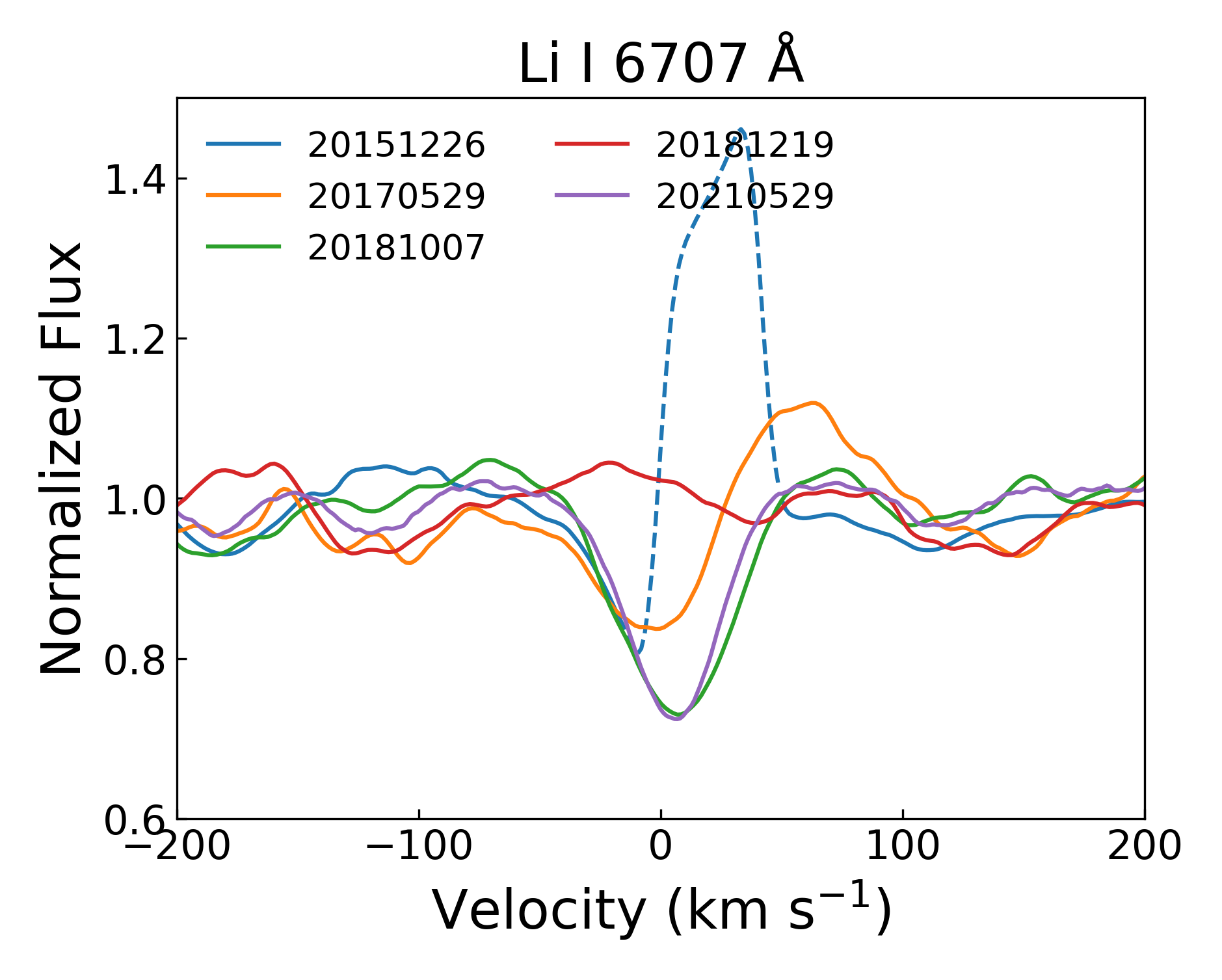}
    \includegraphics[width=0.45\textwidth]{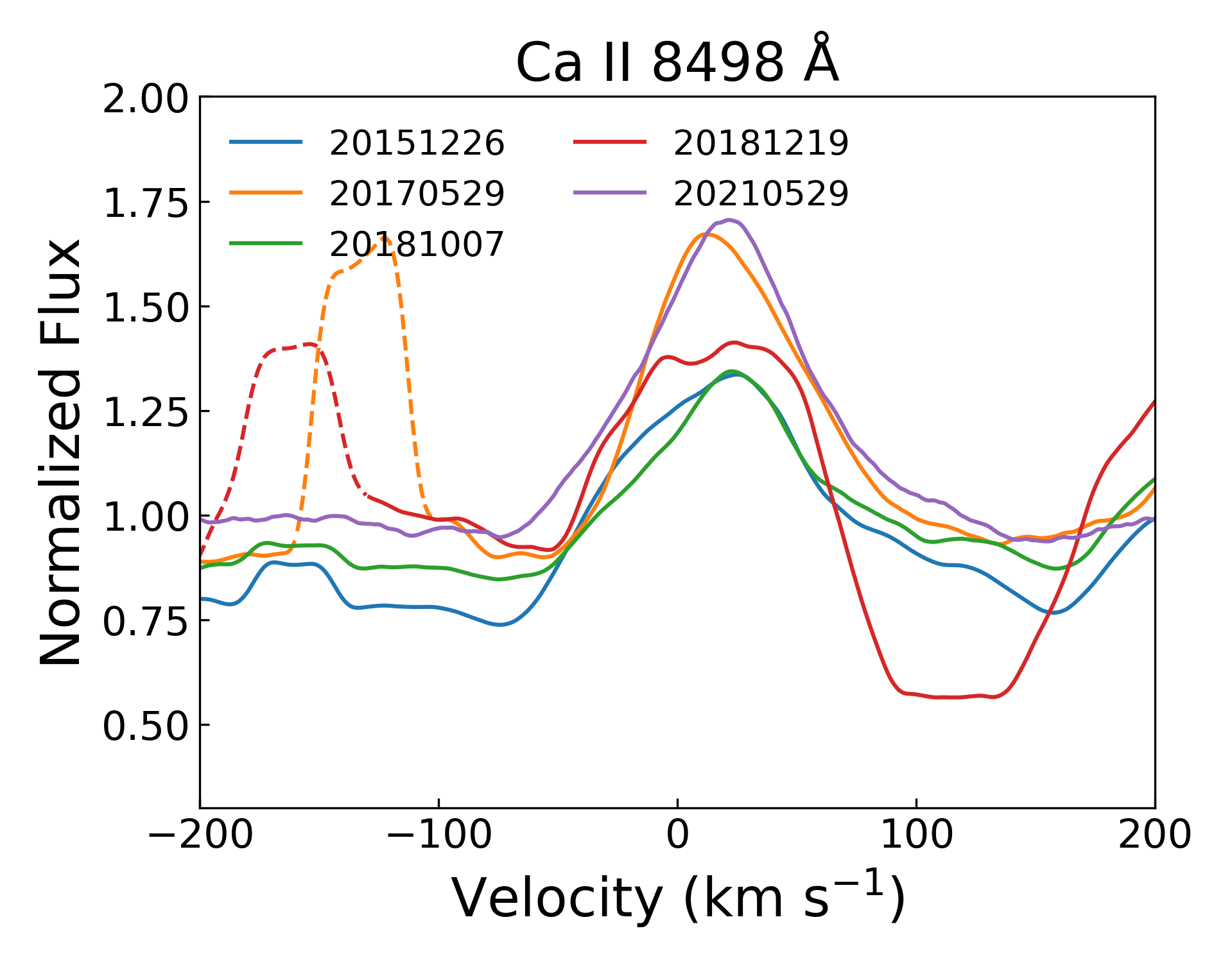}
    \caption{Time evolution of some characteristic lines in the optical spectra of V1515~Cyg. To show each line profile clearly, we smoothed them with a 30 pixel moving window. 
    The \LiI\ 6707\,\AA{} line observed in 2015 December 26 is affected by an artifact, which is indicated with a dashed line in the third panel. The artifacts on the other lines are also presented with dashed lines.
    }
    \label{fig_oplot}
\end{figure*}

Classical FUors share several common spectroscopic features in their optical spectra, such as the P~Cygni profile of \halp, strongly blue-shifted absorption lines like the Li\,{\footnotesize I} 6707\,\AA{} absorption, and the fact that their spectral type is wavelength-dependent \citep{kenyon&hartmann1996, audard2014}. 
Some of these spectroscopic characteristics are also seen in our observations, presented in the upper panel of Fig.~\ref{fig_oplot}. Apart from 2018 December spectrum, all the other spectra show remarkably similar profiles within the measurement uncertainties. 
The 2018 December spectrum in Fig.~\ref{fig_oplot} is different in the way that the absorption profile was not as deep as in all the other epochs. 
We note that for plotting purposes in this section, we used our NOT/FIES spectrum obtained in 2021 May.

\begin{figure}
    \centering
    \includegraphics[width=\columnwidth]{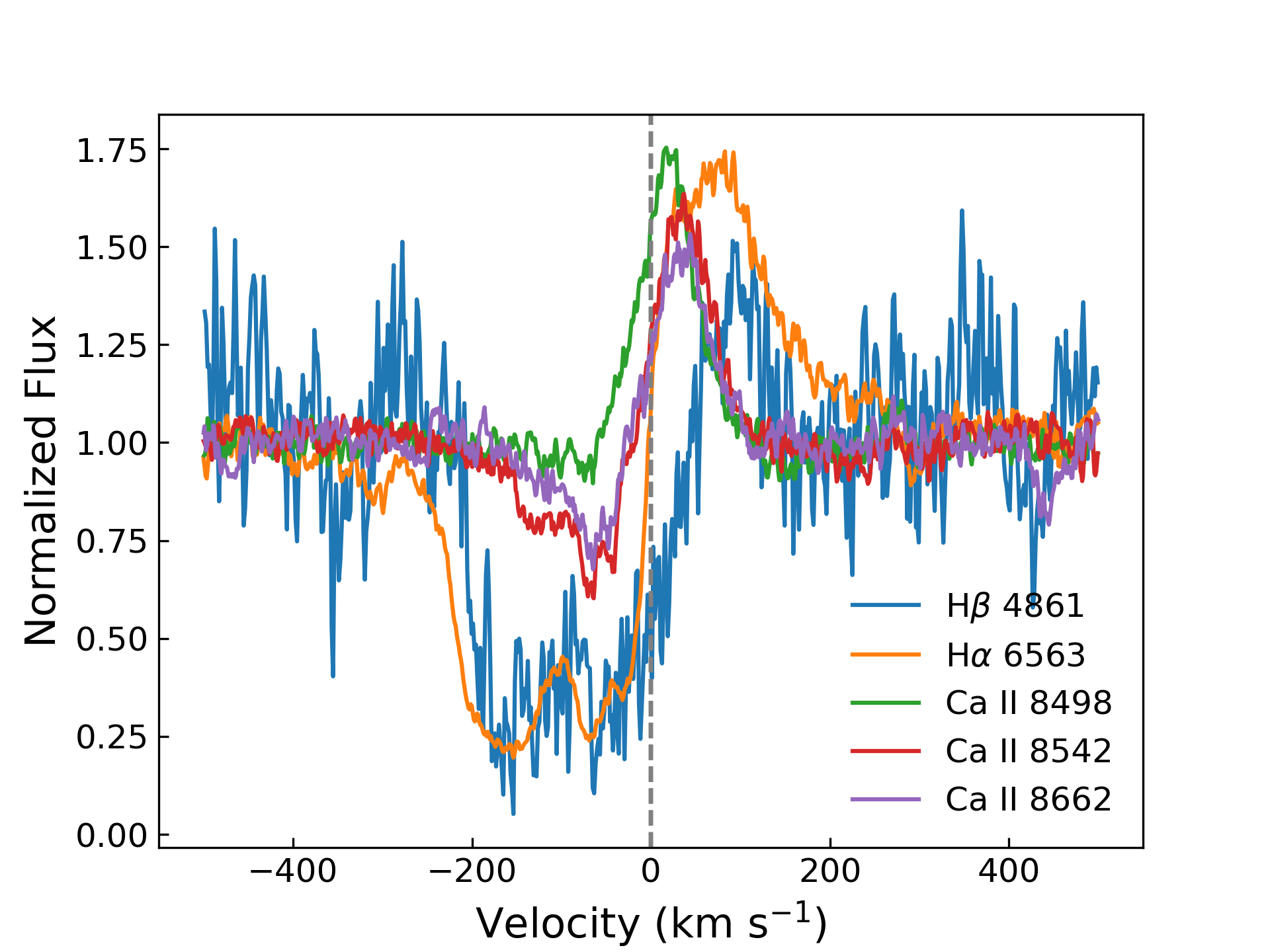}
    \caption{Comparison of the profiles of lines with P~Cygni shape in V1515~Cyg in 2021 May.}
    \label{fig_pcyg_oplot}
\end{figure}

We observed several P~Cygni profiles, specifically in \halp, \hbet, and two lines of the Ca infrared triplet (IRT), \CaII~8542, and \CaII~8662 \AA{} lines, but interestingly  not in the \CaII~8498~\AA{} line. Fig.~\ref{fig_pcyg_oplot} shows observed P~Cygni profiles in the 2021 spectrum.
In the case of the \halp\ line, the previously published optical spectrum of V1515~Cyg presented by \citet{herbig1977} and \citet{KH1991} showed a pure absorption profile, but we observed \halp\ with a clear P~Cygni profile in all our observations taken between 2015 and 2021 (middle panel in Fig.~\ref{fig_oplot}), including a clear emission component. 
\halp\ and \hbet\ show higher velocity of both blue-shifted absorption ($\sim -$300\,km\,s$^{-1}$) and red-shifted emission ($\sim$100\,km\,s$^{-1}$) components compared to the \CaII\ lines ($\sim -$200\,km\,s$^{-1}$ and $\sim$50\,km\,s$^{-1}$). 

\begin{figure*}
    \centering
    \includegraphics[width=0.32\textwidth]{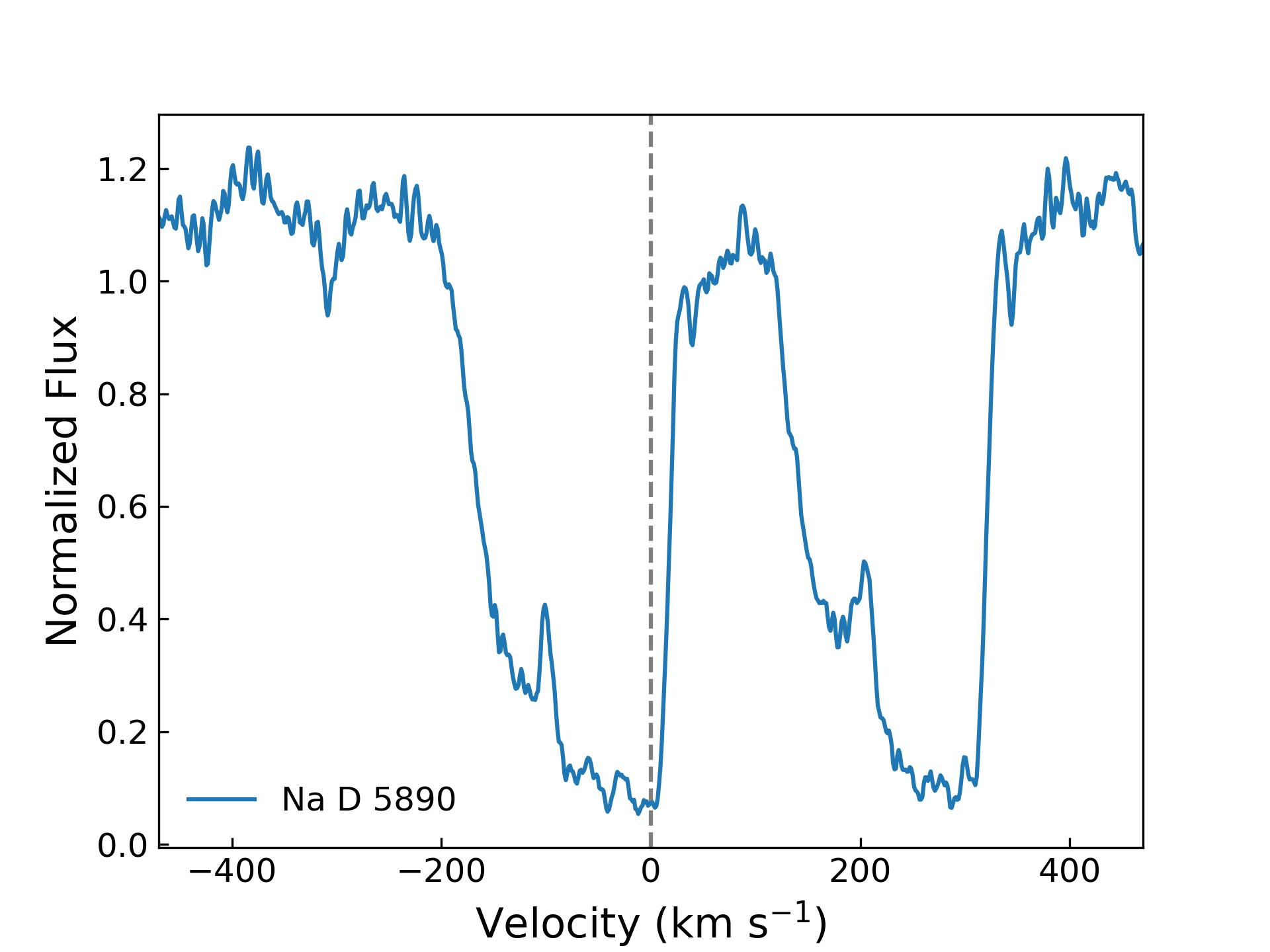}
    \includegraphics[width=0.32\textwidth]{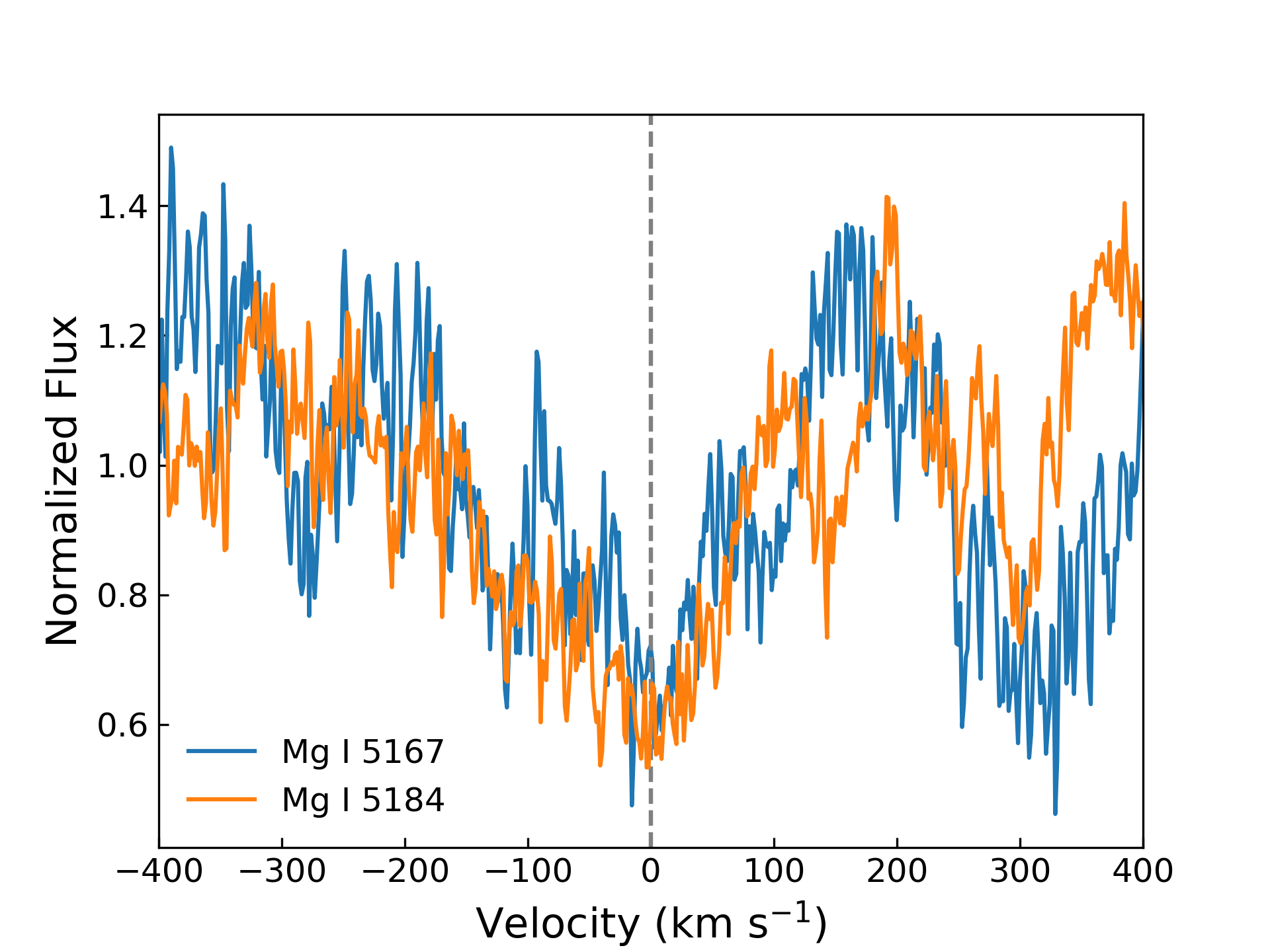}
    \includegraphics[width=0.32\textwidth]{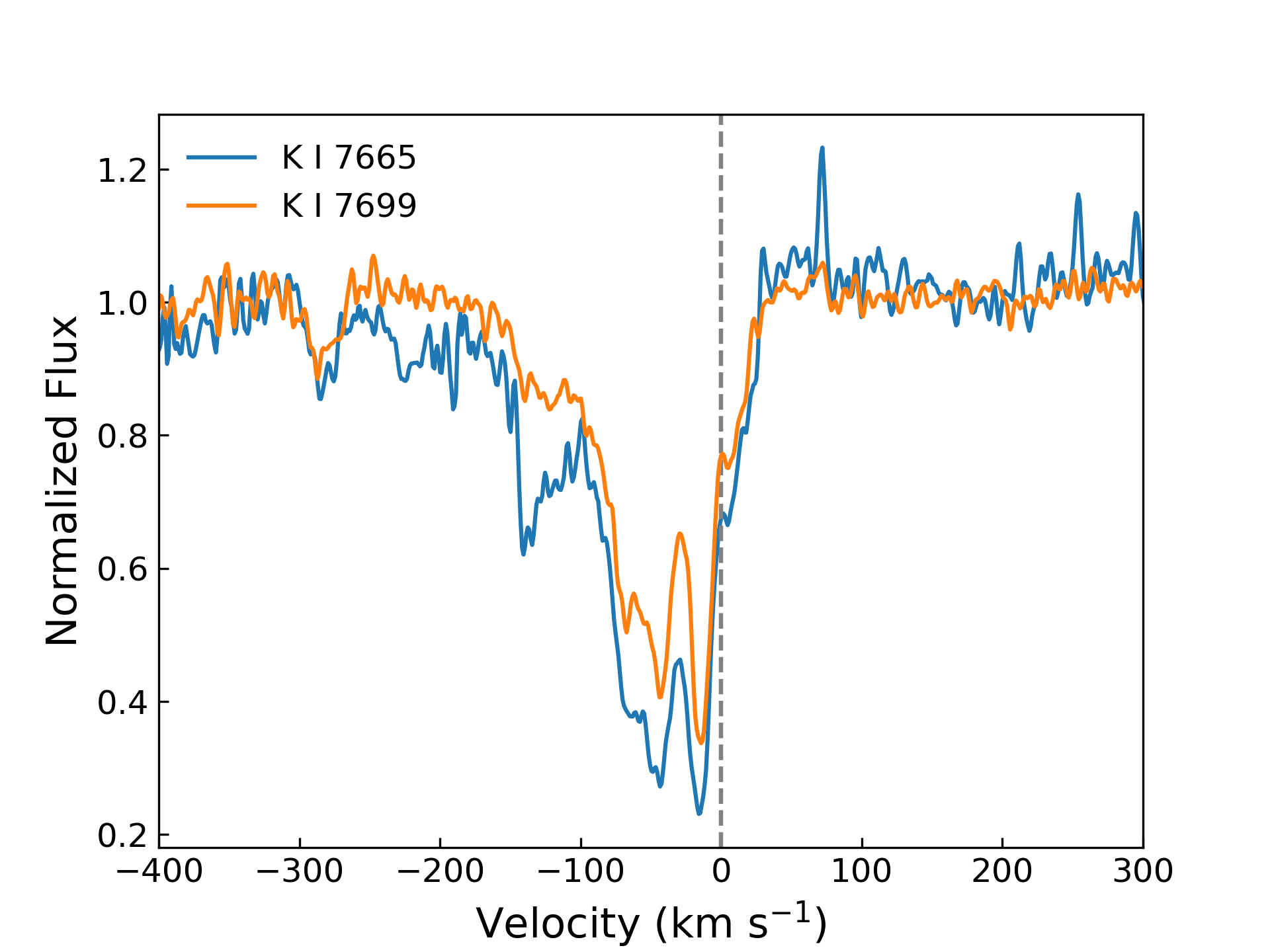}
    \caption{Broad blue-shifted absorption lines of V1515 Cyg in 2021 May: Na\,{\footnotesize I}\,D doublet (5889 and 5895\,\AA{}), Mg\,{\footnotesize I} 5167and 5184\,\AA{}, and K\,{\footnotesize I}  (7664 and 7698\,\AA{}) doublet. \label{fig_broad_oplot}} 
\end{figure*}

\begin{figure}
    \centering
    \includegraphics[width=\columnwidth]{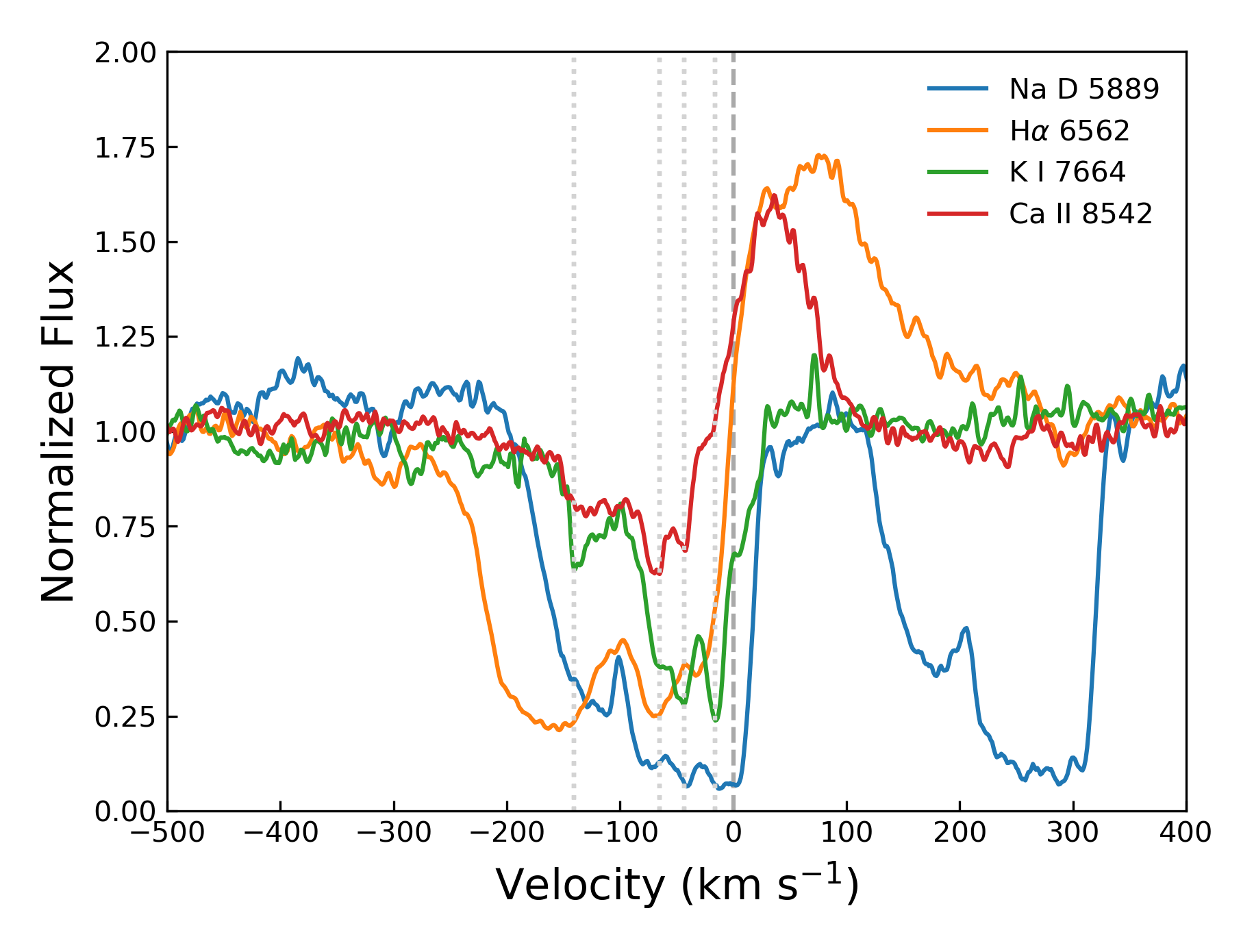}
    \caption{Multiple velocity components observed in V1515~Cyg in 2021 May: Na\,D 5889, \halp~6562, K~{\scriptsize I} 7664, and \CaII~8542\,\AA{}. The dashed line indicates zero velocity, and dotted lines mark blue-shifted velocity components of about --141, --65, --44, and --16\,km\,s$^{-1}$. The spectra are smoothed by 5 pixels to show line profiles clearly.
    \label{fig_pcyg_velocity}}
\end{figure}

Wind signatures in the form of broad blue-shifted absorption lines were also observed in the Na\,{\footnotesize I}\,D 5889 and 5895\,\AA{}, Mg\,{\footnotesize I} 5167 and 5184\,\AA{}, and K\,{\footnotesize I} doublet lines (7664 and 7698\,\AA{}) (Fig.~\ref{fig_broad_oplot}). 
The highest velocity component extended up to about 250\,km s$^{-1}$, similar to the blue-shifted absorption component of \halp\ and \hbet. 
Moreover, some lines show multiple velocity components, namely the Na\,{\footnotesize I}\,D 5889, \halp\ 6562, K\,{\footnotesize I} 7664, and \CaII\ 8542\,\AA{} lines (Fig.~\ref{fig_pcyg_velocity}). These lines share at least three velocity components among around $-$141, $-$65, $-$44, and $-$16\,km\,s$^{-1}$, implying that the lines might form in the same region. Here we note that jet tracer lines, such as [S\,{\footnotesize II}], [N\,{\footnotesize II}], and [O\,{\footnotesize III}] were not detected in our multi-epoch spectra as opposed to V1057~Cyg \citep[][]{szabo2021}.

\begin{figure}
    \centering
    \includegraphics[width=\columnwidth]{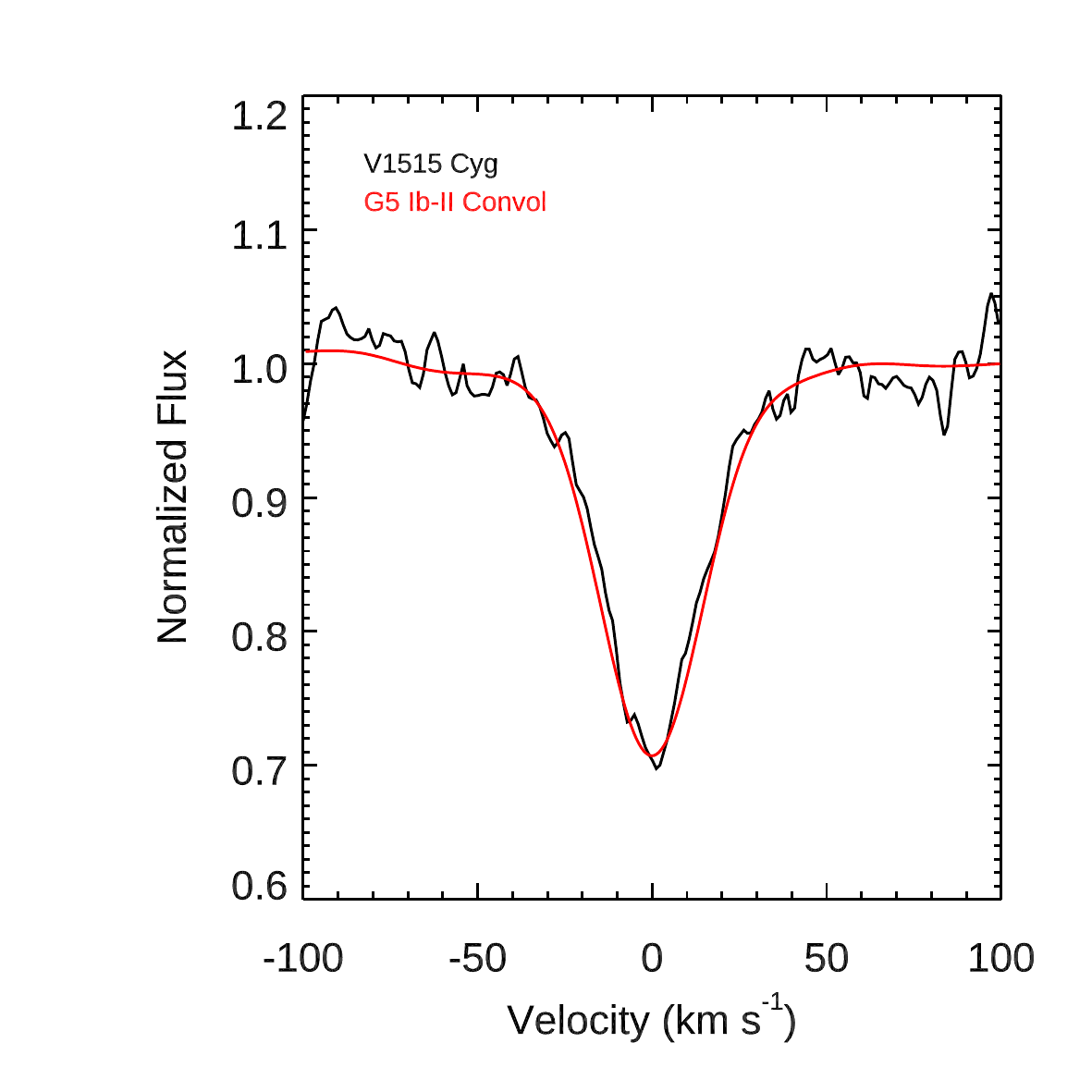}
    \caption{Averaged metallic absorption profiles of V1515~Cyg in 2021 May (black) and standard star (red) convolved with a rotational velocity of 15\,km\,s$^{-1}$.}
    \label{fig_vsini}
\end{figure}

We also observed numerous metallic absorption lines during our observations. These lines include various Fe\,{\footnotesize I} lines between 5446 and 7511\,\AA{}, Ca\,{\footnotesize I} 6122--6471\,\AA{}, Ti\,{\footnotesize I} 5965, 7344\,\AA{}, Ni\,{\footnotesize I} 6108, 6767\,\AA{}, and the Li\,{\footnotesize I} 6707\,\AA{} line (see some of these lines in Fig.~\ref{fig_comp_atomic} in Section~\ref{sec:comp_spec}).
In order to estimate the rotational velocity ($v$sin$i$), we convolved the stellar spectrum with a Gaussian profile to account for the rotational broadening of the lines and compared their spectra to V1515~Cyg.
For this analysis, we adopted the standard stellar spectra from F8 to G8 supergiants used in \citet{park2020}, which are observed with the same observing setup as V1515~Cyg.
First, we averaged four spectral lines (\FeI~6141.7, \FeI~6411.6, \CaI~6439.1, and \FeI~7511.0\,\AA{}) by normalizing each peak intensity for V1515~Cyg and standard stars. Then, the stellar spectrum was convolved with a Gaussian to account for various rotational broadening from 0 to 30\,km\,s$^{-1}$ with a 1\,km\,s$^{-1}$ step.
The best fit was found by ${\chi}^2$ minimization, and the best fit rotational velocity is 15\,km\,s$^{-1}$ found using G5~Ib-II type stellar template (Fig.~\ref{fig_vsini}).
The obtained rotational velocity is similar to the previously found $\sim$20\,km\,s$^{-1}$ \citep{hartmann1985} and consistent with low disk inclination \citep{gramajo2014, hillenbrand2019, milliner2019}. Furthermore, the best fit spectral type (G5~Ib-II) is also consistent with G supergiants \citep{KH1991}. 


\subsubsection{Near-infrared Spectroscopy} \label{sec:spec_NIR}
\begin{figure*}
    \centering
    \includegraphics[width=\textwidth]{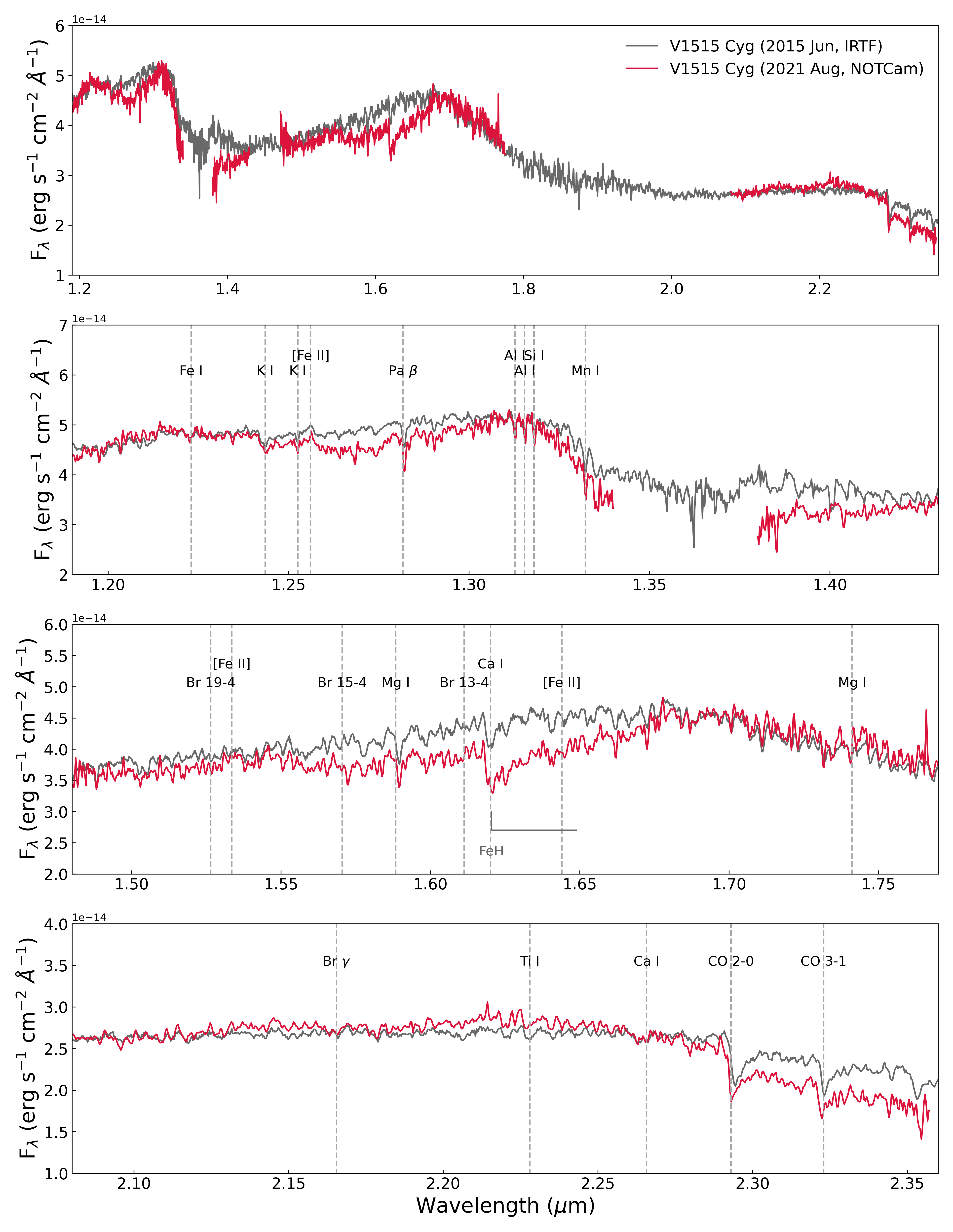}
    \caption{The NIR $J$, $H$, and $K$ spectrum of V1515~Cyg. Gray and red colors present the spectra observed with IRTF \citep{connelley2018} and NOTCam, respectively. First panel shows the full $JHK$ spectrum, second the $J$, third the $H$, then the $K$-band spectrum.}
    \label{fig_notcam}
    
\end{figure*}

Our recent NIR spectrum taken in 2021 shows several absorption lines: Pa$\beta$, Al\,{\footnotesize I}, Mg\,{\footnotesize I}, FeH molecular bands, and strong CO absorption bandhead features are present, similarly like in the previous spectra of V1515~Cyg taken by \citet{connelley2018}.
In Fig.~\ref{fig_notcam} we compare their 2015 June spectrum and our spectrum from 2021 August: in the first panel we show the full $JHK$ spectra, the next three panels show the $J$, $H$, and $K$-band spectrum respectively.
Although the NIR photometric observations are insufficient since 2012, V1515~Cyg has almost the same brightness in 2012 and 2020 observations. There is no significant flux change of the $JHK_{s}$ spectrum in 2015 and 2020.
However, our observation shows interesting changes in some spectral features: the FeH ($>$ 1.62~$\mu$m) and CO ($>$ 2.29~$\mu$m) molecular bands became stronger.
Both the CO rovibrational and the FeH molecular bands are stronger at lower temperatures \citep{wallace2001, bieging2002, hargreaves2010}, implying that V1515~Cyg became cooler than observed in the previous observation in 2015 by \citet{connelley2018}.    

The Pa$\beta$ equivalent widths (EWs) became stronger in our recent NOTCam (1.33\,$\pm$\,0.53\,\AA{}) observation compared to the IRTF (1.03\,$\pm$\,0.09\,\AA{}) observations of \citet{connelley2018}. However the uncertainties are higher for the new observations, therefore, this change may not be significant. 
The usual scaling relations between the fluxes of the hydrogen emission lines were established for T~Tauri stars \citep[see e.g.,][]{alcala2017}. These relations work because the accretion in all of these stars happen according to the magnetospheric accretion scenario \citep[see e.g.,][]{hartmann2016}. This scenario cannot be applied to FUors, as indicated by their lack of emission lines. The absorption lines are associated with an active accretion disk, and their formation is explained by temperature inversion (hot midplane, colder disk atmosphere) \citep[see e.g.,][]{kenyon&hartmann1996}. In V1515~Cyg we see the Pa$\beta$ line in absorption, as most of the lines observed in the spectrum of FUors in general. Therefore, the hydrogen lines cannot be used for the mass accretion calculation unlike in T~Tauri stars.
\subsection{Accretion Disk Modeling}
\label{res:accdisc}
In this section we model the inner part of the system with a steady, optically thick and geometrically thin viscous accretion disk. We already applied this method for a number of young eruptive stars \citep{kospal2016,kospal2017a,abraham2018,kun2019,szegedi-elek2020,szabo2021} to separate the effects of variable accretion, extinction and study their long term evolution quantitatively.
The mass-accretion rate in our model is constant in the radial direction \citep[Eq.~1 in ][]{kospal2016}. We neglect any contribution from the star itself, assuming that all optical and near-infrared emission in the outburst originates from the hot accretion disk. We calculated synthetic SEDs of the  disk by integrating the blackbody emission of concentric annuli between the stellar radius and $R_{\rm out}$. The inner radius of the disk mainly influences the emission in the optical bands. In practice, we usually cannot discriminate between a small stellar radius with higher line-of-sight extinction and a larger radius with lower extinction, using only broad-band optical photometry. In the subsequent modeling we prescribe that the average of our $A_V$ values obtained between 2012 and 2021 must comply with the $A_V$=3.5$\pm$0.4 mag proposed by \citet{connelley2018} based on an infrared spectrum taken in 2015. This constraint sets $R_{\rm in}$ = 1.4~R$_{\odot}$. The outer radius of the accretion disk can be estimated by fitting the mid-IR emission of the disk. We fixed $R_{\rm out} = 0.45$~au, which matches the $L$- and $M$-band observations of V1515~Cyg from 1989 \citep[][]{kenyon-and-hartmann1991}. The inclination of the disk is also an important input parameter of our accretion disk model. Most studies agree that V1515~Cyg is seen close to pole-on \citep[see e.g.,][]{gramajo2014,milliner2019}. In the following we adopt $i$=10$^{\circ}$.

\begin{figure}
\includegraphics[width=\columnwidth]{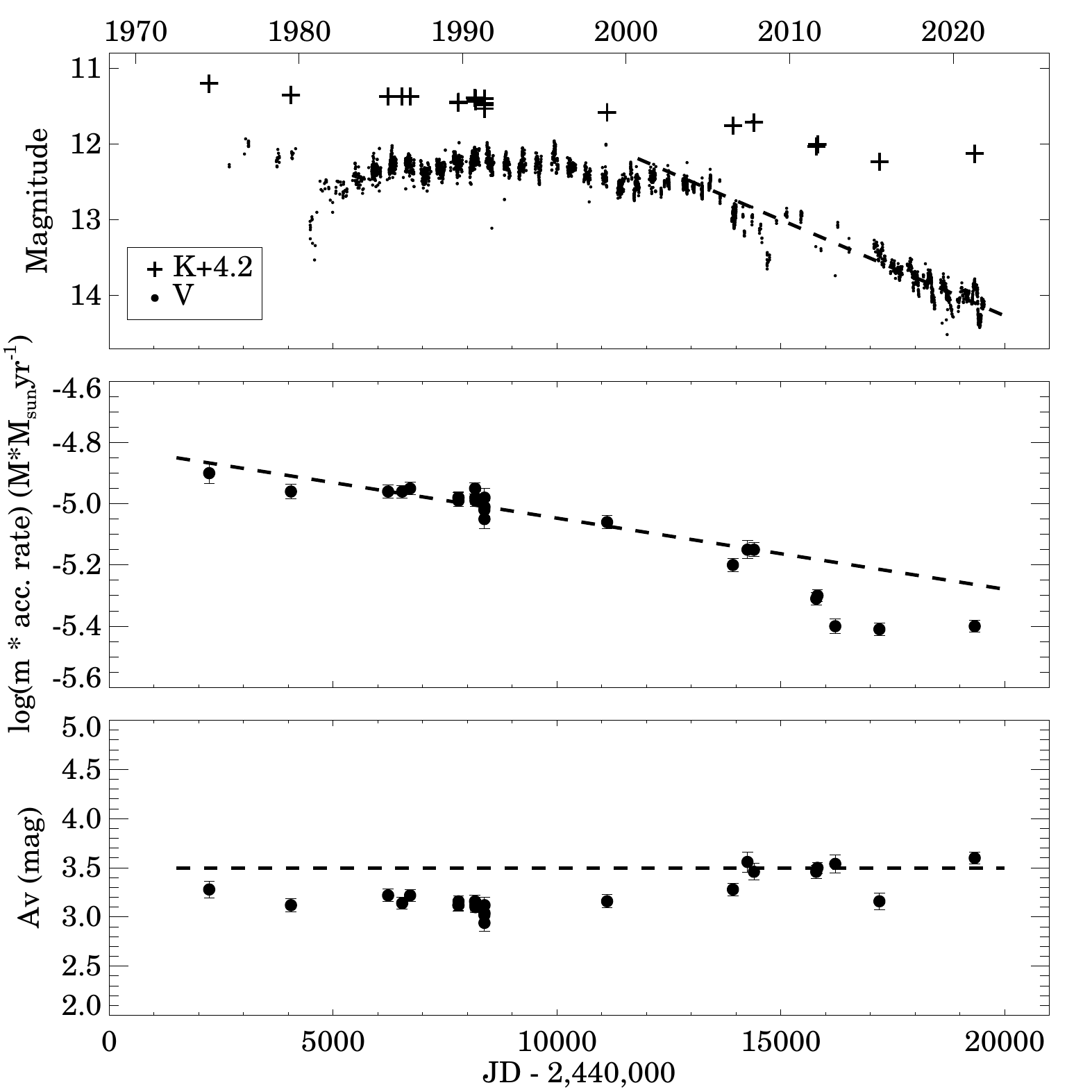}
\caption{Optical $V$-band and near-infrared $K$-band light curves of V1515~Cyg (upper panel), temporal evolution of the accretion rate (middle panel) and extinction toward the star (bottom panel) derived from our accretion disk modeling described in Sect.~\ref{sec:accdisc}.}
\label{fig:accdisk}
\end{figure}

Our models have only two free parameters: the product of the accretion rate and the stellar mass $M\dot{M}$, as well as the line-of-sight extinction $A_V$. Disk SEDs were calculated for a large range of $M\dot{M}$ values, and at each step the fluxes were reddened using a grid of $A_V$ values assuming the standard extinction law from \citet{cardelli1989} with $R_V = 3.1$. Finding the best $M\dot{M}$ -- $A_V$ combination was performed with ${\chi}^2$ minimization, by taking into account all quasi-simultaneous flux values between 0.4 and 2.5$\,\mu$m. The formal uncertainties of the data points were set to a homogeneous 5\% of the measured flux value, which also accounted for possible differences among photometric systems. The model fits reproduced the measurements well, with low reduced $\chi^2$ values. The resulting temporal evolution of the accretion rate and extinction values, together with the optical $V$-band and the near-infrared $K$-band light curve, are plotted in Fig.~\ref{fig:accdisk}, where the $K$-band light curve is shifted by $+$4.2 mag for plotting purposes.
We discuss these results further in the next subsection and in the Sec.~\ref{sec:discussion}.

\subsection{Spectral Energy Distribution}
\label{res:sed}

\begin{figure}
\includegraphics[width=\columnwidth]{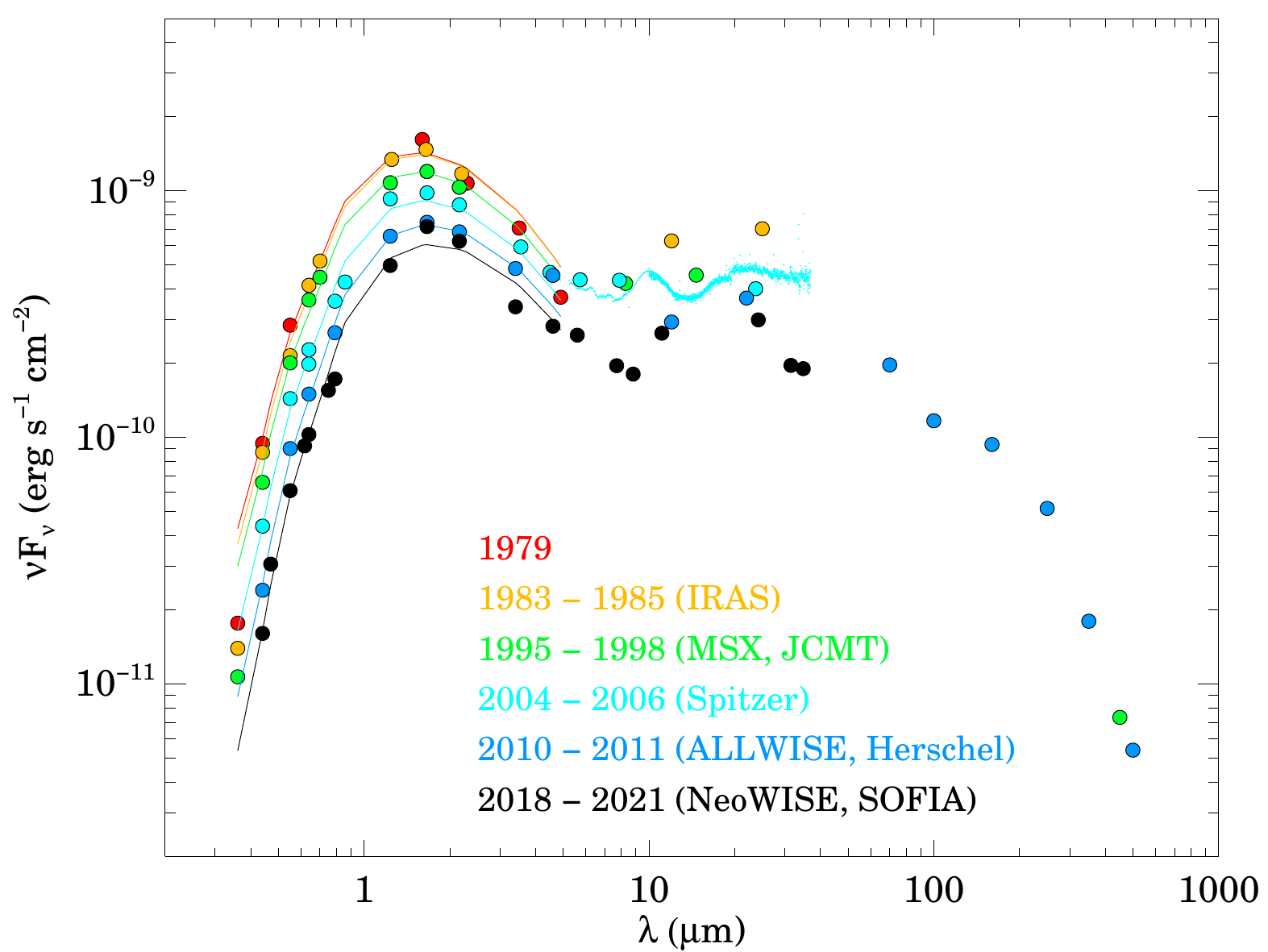}
\caption{
Spectral energy distribution of V1515~Cyg at different representative epochs. The data points are from Fig.~\ref{fig:lc}, as well as from space-based (\textit{IRAS}, \textit{Spitzer},  \textit{WISE}, \textit{Herschel}) and airborne (SOFIA) missions, as indicated in the legend. Solid curves show the results of our accretion disk models for the individual epochs.
}
\label{fig:sed}
\end{figure}

Fig.~\ref{fig:sed} displays the SED of V1515~Cyg at several representative phases of its outburst. The optical and infrared points up to ${\lambda}{\sim}5$ $\mu$m were taken from Fig.~\ref{fig:lc}, while photometry at longer wavelengths are used from different space-borne missions, such as the \textit{IRAS} \citep[][]{iras1984,iras-catalogue1994}, \textit{MSX} \citep[][]{egan2003}, \textit{Spitzer}\footnote{Data were taken from "A Spitzer Legacy Survey of the Cygnus-X Complex" catalogue, available at the IRSA database \url{https://irsa.ipac.caltech.edu/data/SPITZER/Cygnus-X/}} \citep[][]{spitzer-cygnus2007}, \textit{Herschel} \citep[][]{green2013} and the \textit{WISE} \citep[][]{wright2010}, and we also used data from the airborne \textit{SOFIA} \citep[][]{young12} missions. 
The data points from the Herschel and the AllWISE catalogs were obtained within a year, thus we combined the two data sets into one SED. In the far-infrared domain (${\lambda}>25$ $\mu$m) the IRAS and ISO measurements were apparently contaminated by some extended emission in their large beams, thus were discarded from the plot. Contamination was confirmed by archival 70 $\mu$m and 160 $\mu$m images by the Herschel Space Telescope, which show that V1515~Cyg is located at the edge of a dust cloud, on a structured far-infrared background.

There are several properties of the SED of V1515~Cyg, which are discussed below in Sec.~\ref{sec:accdisc}

\section{Discussion} \label{sec:discussion}

\subsection{Dimming events in V1515~Cyg}
\label{sec:dimming}

In Sec.~\ref{sec:cc}, we described the interesting dimming event of V1515~Cyg observed in 2021. In the following we investigate the possible physical cause/interpretation of such dimming events. 
\citet{kenyon-and-hartmann1991} and \citet{kolotilov&kenyon1997} assumed that the occultation events were due to dust formation in an expanding shell in the wind. However, \citet{clarke2005} considered that in the case of V1515~Cyg the occultation events are interactions between the wind and the surrounding infalling envelope. They proposed that, for instance, V1057~Cyg has a strong wind and was able to clear the circumstellar material in response to the outburst. The wind in V1515~Cyg has always been weaker, therefore it has never been able to completely clear the line of sight of dusty material, which can explain the variations since the beginning of the outburst. The dust lifted by the wind could cause this variability. In contrast, FU~Ori was able to clear a large cavity, hence its light curve is free from this kind of variability \citep[][]{clarke2005}. This could also explain why there is no correlation between the photometric and the spectroscopic variation suggested by \citet{clarke2005}.
We tried to draw an analogy with the T~Tauri star, RW~Aur, where multiple, similar dimming events were observed with similarly good data sampling. \citet{koutoulaki2019} suggested that the deep minima are most likely caused by a layer of dust obscuring the inner 0.05 -- 0.1\,au of the system, while the accretion rate did not change significantly  \citep[see also other studies, e.g.,][and references therein]{petrov2015,facchini2016}. However, there are many obvious differences between the two systems. Firstly, that although V1515~Cyg is approaching the quiescence phase, it is still in outburst, hence exhibiting a more enhanced accretion rate compared to RW~Aur. Therefore, the optical photometry of V1515~Cyg is still dominated by the accretion disk (and the star is negligible), as is usual in FUors in outburst. Moreover, in V1515~Cyg we see the continuous decrease of the accretion rate despite the dimming events. Another difference is the that RW~Aur is a binary, while V1515~Cyg seems to be a single star.

In order to understand the observed dimming event in 2021, additional, similarly well sampled multi-filter light curves, ideally complemented by spectroscopic monitoring, are needed to look for similarities or differences between these two kinds of dimming events.


\subsection{Accretion Disk Modeling and SED}
\label{sec:accdisc}

Fig.~\ref{fig:accdisk} shows the temporal evolution of the accretion rate and the extinction together with the optical $V$-band and the near-infrared $K$-band light curves of V1515~Cyg. The long-term evolution of the accretion rate until 2010 is approximately consistent with an exponential decay, with an e-folding time of 20600 days ($\sim$56\,years). The dashed line in Fig.~\ref{fig:accdisk} was fitted to the $V$-band data after 2003 in order to determine the e-folding time of the long-term optical fading, corresponding to approximately 12\,years.

The peak and current $M\dot{M}$ values are 1.29$\times$10$^{-5}$ and 4.17$\times$10$^{-6}$~M$_{\odot}^2$yr$^{-1}$, respectively. Assuming a stellar mass of 0.3\,M$_{\odot}$, the real accretion is higher by a factor of about 3. The $A_V$ curve is rather constant: a weak gradual decrease is hinted before 2000, which might be responsible for the brightening of the system during this period, in spite of the decreasing accretion rates. The current luminosity of the accretion disk is about 45\,$L_{\odot}$, dropped from a peak value of 138\,$L_{\odot}$. 



The SED of the source has several properties which we discuss further below.
The SED is dominated by a peak at optical-near-NIR wavelengths. The clear excess above a blackbody-like extrapolation of this peak to infrared wavelengths implies the presence of a significant amount of circumstellar matter (the total circumstellar mass, based on IRAM observations, is 0.04\,M$_{\odot}$ \citep{feher2017}). 
The spectral range around 10 $\mu$m and 18 $\mu$m exhibits broad emission features, which is attributed to submicrometer size silicate grains \citep[see also e.g.,][]{green2006}. The strength of the short wavelength peak monotonically decreased with time, reflecting the evolution of the hot inner accretion disk modeled in Sec.~\ref{res:accdisc}. 
A temporal descrease of the mid-infrared flux is also suggested, although the data coverage is less complete than in the optical.

At far-infrared and submillimeter wavelengths the \textit{Herschel} observations may represent the emission of the circumstellar disk, with no clear sign of a massive cold circumstellar envelope. There is no information on the temporal evolution of the system during the outburst at these wavelengths. The apparent consistency between the 450 $\mu$m JCMT/SCUBA \citep[][]{sandell2001} single dish measurement from 1998 \citep{sw2001} and the 500$\,\mu$m  \textit{Herschel/SPIRE} observations from 2011 suggests that the flux change at these long wavelengths is not significant. 

Due to the dispersion of the data points, the color evolution can be consistent with evolution along either constant accretion or constant extinction path, in general the color evolution is consistent with the evolution along constant accretion or constant extinction path, obtained from our accretion disk model in Sec.~\ref{res:accdisc} and Sec.~\ref{sec:accdisc}. Considering the small variability amplitude and the photometric uncertainty, it is not possible to decide which is the dominant mechanism. As the global long-term color evolution is accounted by the modeling, we do not present here the global color-magnitude diagrams but refer the reader to Sec.~\ref{res:accdisc} and Sec.~\ref{sec:accdisc} for the general results.


\subsection{Comparison between V1515~Cyg and V1057~Cyg} \label{sec:comp_spec}

In this section we give a comparison between V1515~Cyg and V1057~Cyg. First we compare their light curves, then their spectroscopic properties using our previous study by \citet{szabo2021} with the caveat that the spectra for the two sources were taken in different stages of their outburst evolution. Both sources are considered classical FUors \citep{herbig1977, herbig2003, connelley2018}, and we were able to observe similar properties in the spectra of both targets, comparing them in the following.

\emph{Similarities.}
A similar aspect of both sources can be found in their light curves: both objects underwent a sharp dip event after reaching the peak brightness. In the case of V1515~Cyg this happened right after reaching the light maximum in 1980, recovering and then showing a plateau stage until the early 2000's (Fig.~\ref{fig:lc}). However, the light curve of V1057~Cyg is different, this source also experienced a similar, sharp dip, around 1995 after an exponential fading trend. Similarly, just like in V1515~Cyg, V1057~Cyg also recovered from this event and showed a plateau stage afterwards \citep[called 'second plateau'][]{szabo2021}. Interestingly, FU~Ori also showed a similar event, followed by a long-term plateau stage \citep[see e.g.,][]{hartmann2016}.

\begin{figure*}
    \centering
    \includegraphics[width=\textwidth]{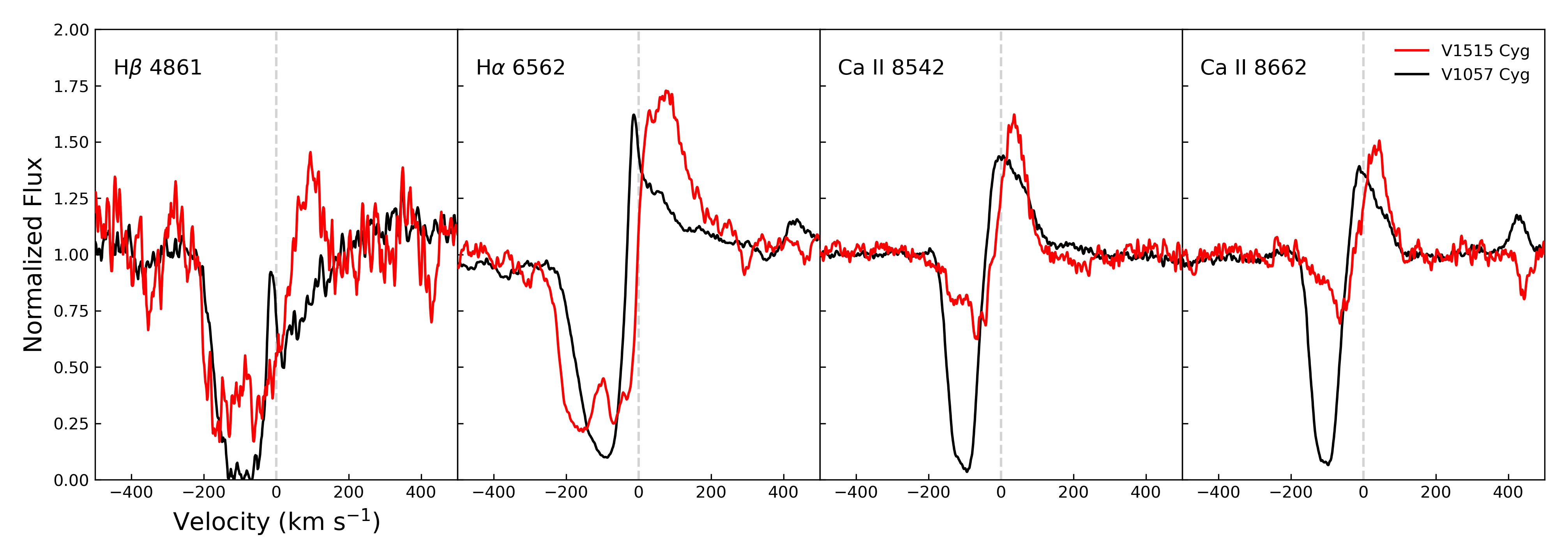}
    \caption{P Cygni profiles of \hbet\ 4861, \halp\ 6562, \CaII\ 8542, and \CaII\ 8662\,\AA{}.}
    \label{fig_comp_pcyg}
\end{figure*}

\begin{figure*}
    \centering
    \includegraphics[width=\textwidth]{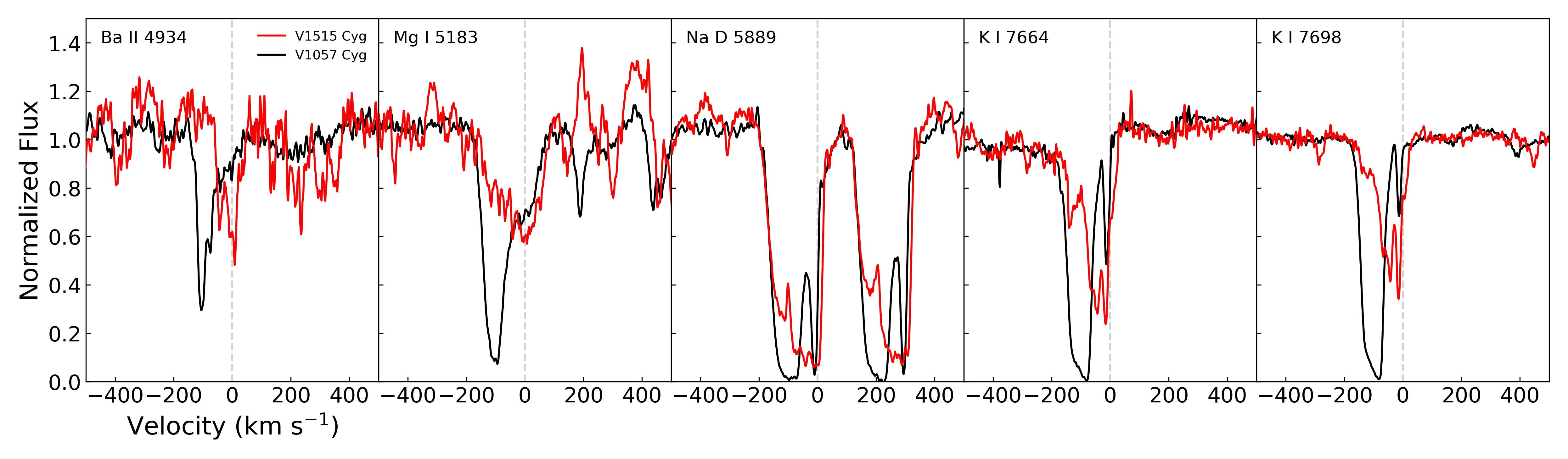}
    \caption{Broad blue-shifted absorption lines: Ba\,{\footnotesize II} 4934, M\,{\footnotesize I} 5183, Na\,D 5889/5895, K\,{\footnotesize I} 7664, and K\,{\footnotesize I} 7698\,\AA{}. \label{fig_comp_broad}}
\end{figure*}

In order to compare two FUors homogeneously, we used the latest observation of our NOT/FIES optical spectra for both.
Our targets show four Balmer lines from \halp\ to H$\delta$ in our observations with broad blue-shifted absorption profiles. Among these four lines, relatively clear (higher S/N) \halp\ and \hbet\ are presented in Fig.~\ref{fig_comp_pcyg}. \halp\ and \hbet\ show P~Cygni profiles with broad blue-shifted absorption component and red-shifted emission profile in both targets. 
The two FUors show P~Cygni profiles of the \halp, \CaII~8542, and 8662\,\AA{} lines as shown in Fig.~\ref{fig_comp_pcyg}.
In addition, both FUors show blue-shifted absorption line profiles formed by an outflowing wind in several neutral and singly ionized metallic lines (Fig.~\ref{fig_comp_broad}). However, V1057~Cyg shows higher velocity components of shell features than V1515 Cyg \citep[e.g., Ba\,{\scriptsize II} 4934~\AA{} and Fig. 11 in ][]{szabo2021}.

\emph{Differences.}
Even though we see similarities in their light curves, perhaps the most striking difference between the two classical FUors is the brightening phase presented in their light curves. 
The outburst of V1057~Cyg was an abrupt one, happening within less than a year \citep[see Fig.~1 in][]{szabo2021}. The outburst of V1515~Cyg was different from the beginning compared to the other classical FUors, taking about four decades to reach the maximum brightness. This difference suggests that overall the outburst mechanisms are different.

Interestingly, in the optical spectrum of V1515~Cyg there is a lack of P~Cygni profile in the Ca\,{\footnotesize II} 8498\,\AA{} line compared to V1057~Cyg. In our previous study on V1057~Cyg, we found that all three lines of the Ca\,{\footnotesize II} IRT showed clear P~Cygni profiles, whereas in the case of V1515~Cyg, we found P~Cygni profiles only for the Ca\,{\footnotesize II} 8542 and Ca\,{\footnotesize II} 8662\,\AA{} lines. The strength/depth of the absorption is also weaker than those observed in V1057~Cyg \citep{herbig2003,szabo2021}.
Several works investigated the relation between the Ca\,{\footnotesize II} IRT lines in T~Taur stars and found that if the lines are optically thin the ratio of the three lines is 1:9:5, and in the optically thick case it is 1:1:1 \citep[see e.g.,][]{herbig1980,hamman-persson1989,hamann-persson1990,azevedo2006}. 
This means that the absorption in the 8498\,\AA{} line would be 9 times weaker than in the 8542\,\AA{} line.
The fact, that the absorption component is not detected (i.e., remains below our detection threshold) in the 8498\,\AA{} line suggests that the gas in the outflowing wind causing the absorption component is probably optically thin. 

\begin{figure*}
    \centering
    \includegraphics[width=\textwidth]{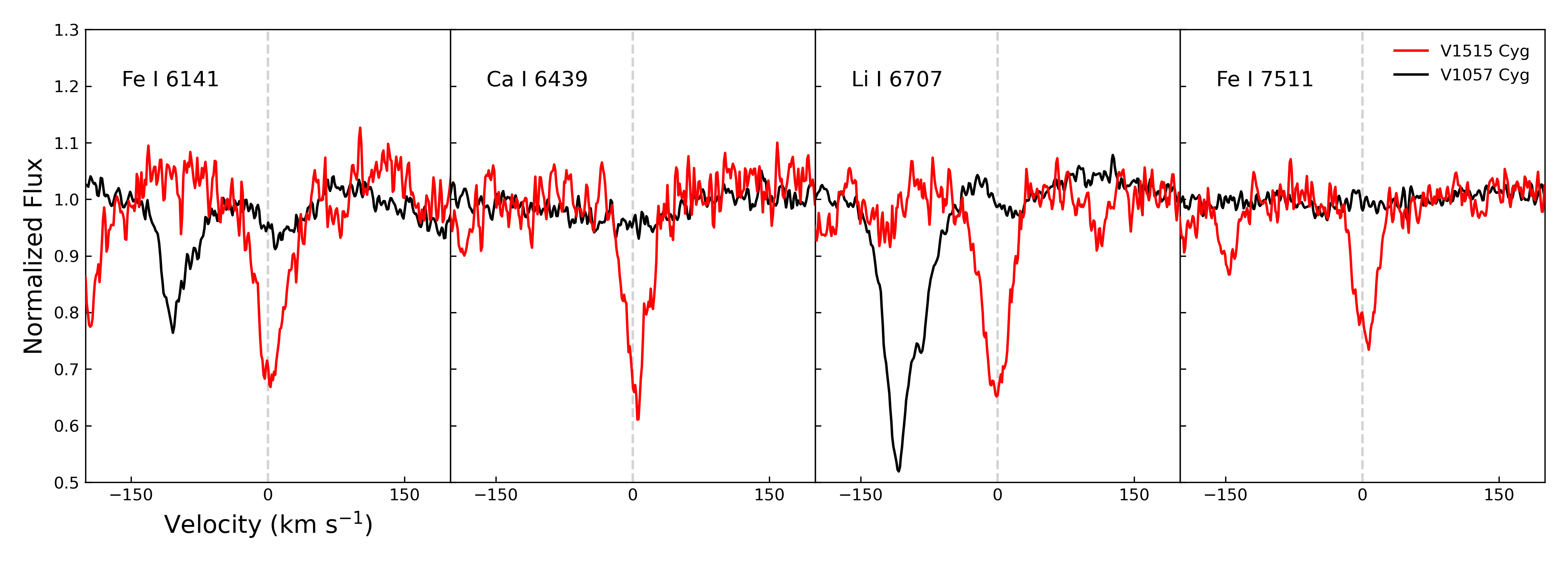}
    \caption{Atomic absorption lines of Fe\,{\footnotesize I} 6141, Ca\,{\footnotesize I} 6439, Li\,{\footnotesize I} 6707, and Fe\,{\footnotesize I} 7511\,\AA{}.}
    \label{fig_comp_atomic}
\end{figure*}

Since the strength of the blue-shifted absorption component in a P~Cygni profile is formed by an outflowing wind in the system, in the case of V1515~Cyg, the weaker strength could be an indication of the weaker outflowing wind in the structure.
Compared to V1057~Cyg where the Ca IRT lines clearly varied with time, in V1515 Cyg, we did not observe strong temporal variation of the lines either. The strength of the P~Cygni profiles varies over time, however, they stay within the same velocity range. 
In addition, no shell lines nor forbidden emission lines were observed in V1515~Cyg as opposed to V1057~Cyg \citep{szabo2021}.
Fig.~\ref{fig_comp_atomic} shows the comparison of atomic metallic lines in both FUors. Unlike V1057~Cyg, a number of atomic metallic lines are observed in the spectrum of V1515~Cyg that are not present in V1057~Cyg.

In the multi-epoch spectrum of V1515~Cyg, there is no evidence of jet tracer lines, such as [S\,{\footnotesize II}], [N\,{\footnotesize II}], and [O\,{\footnotesize III}], as opposed to V1057~Cyg \citep[][]{szabo2021}. The absence of these lines further  supports the idea of a weaker, uncollimated optically thin wind in this system.



\begin{figure*}
    \centering
    \includegraphics[width=\textwidth]{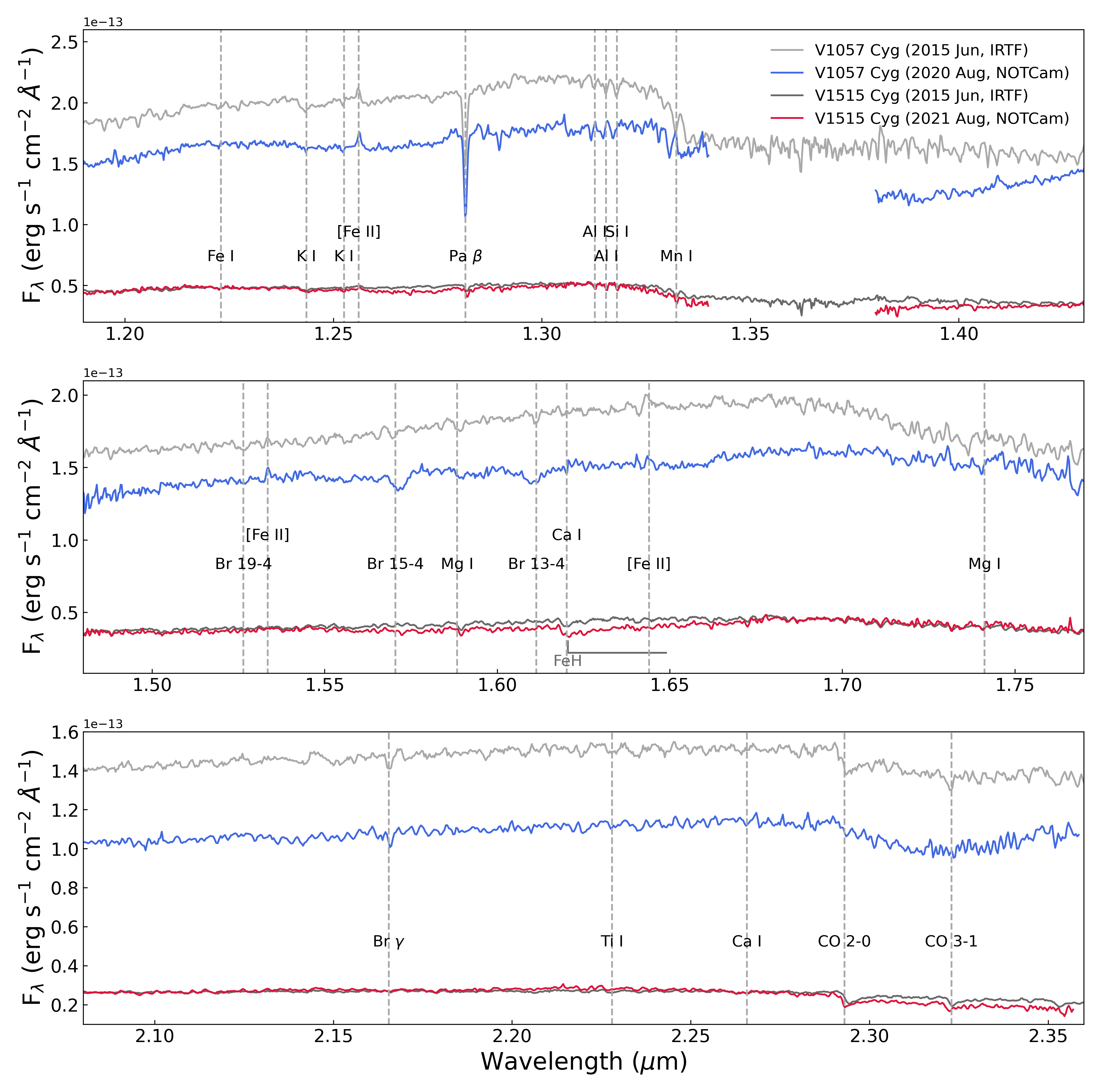}
    \caption{The comparison of the NIR spectra of V1515~Cyg and V1057~Cyg. Light gray and blue lines indicate V1057~Cyg observed with IRTF and NOTCam, respectively. Dark gray and red lines present V1515~Cyg observed with IRTF and NOTCam, respectively. \label{fig_ir_spec}}
\end{figure*}

\begin{figure}
    \centering
    \includegraphics[width=\columnwidth]{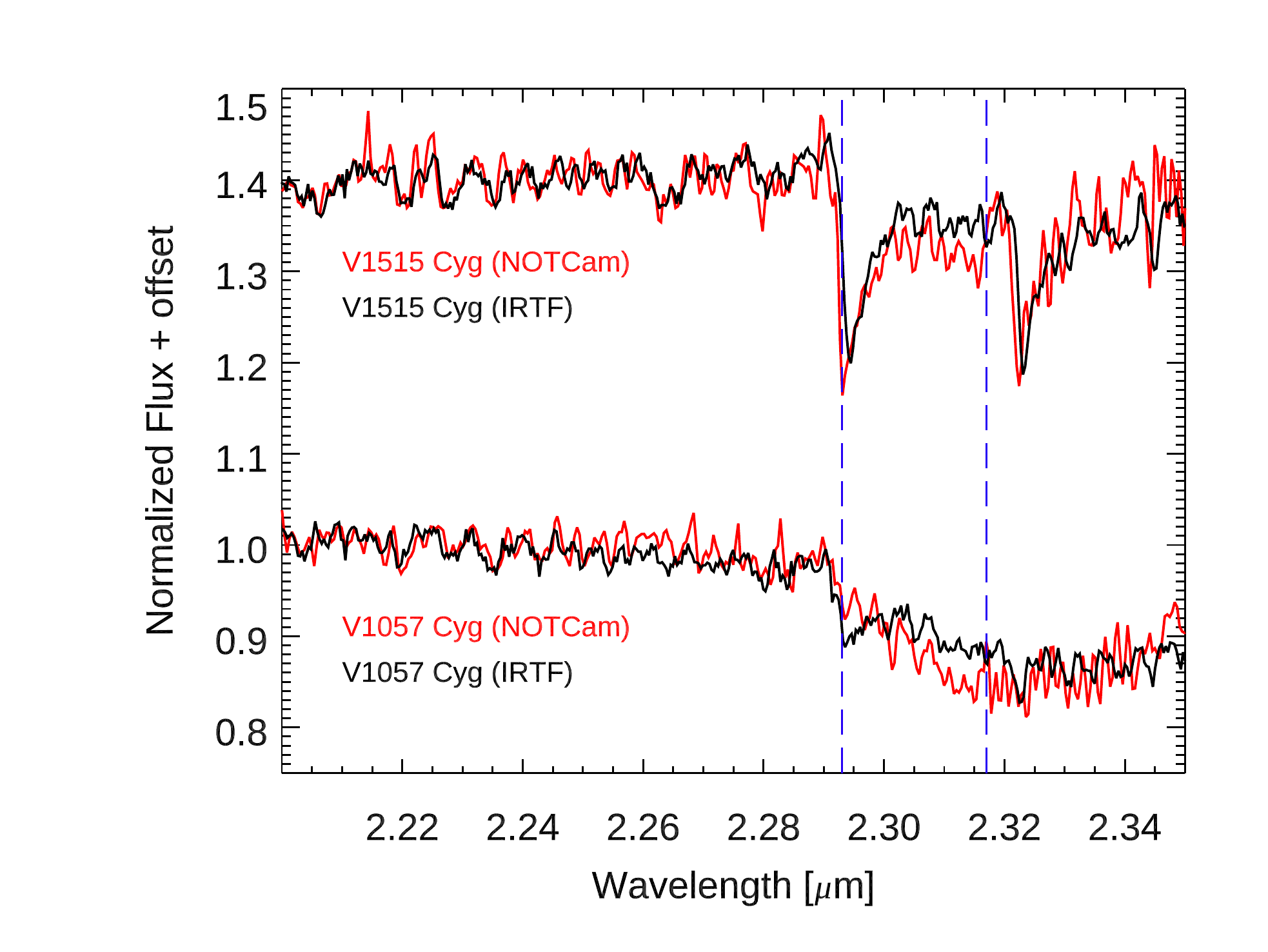}
    \caption{CO bandhead of V1515 Cyg (upper spectra) and V1057 Cyg (lower spectra). Dashed lines indicate the integration range for EW.}
    \label{fig_co_ew}
\end{figure}

\begin{figure}
    \centering
    \includegraphics[width=0.45\textwidth]{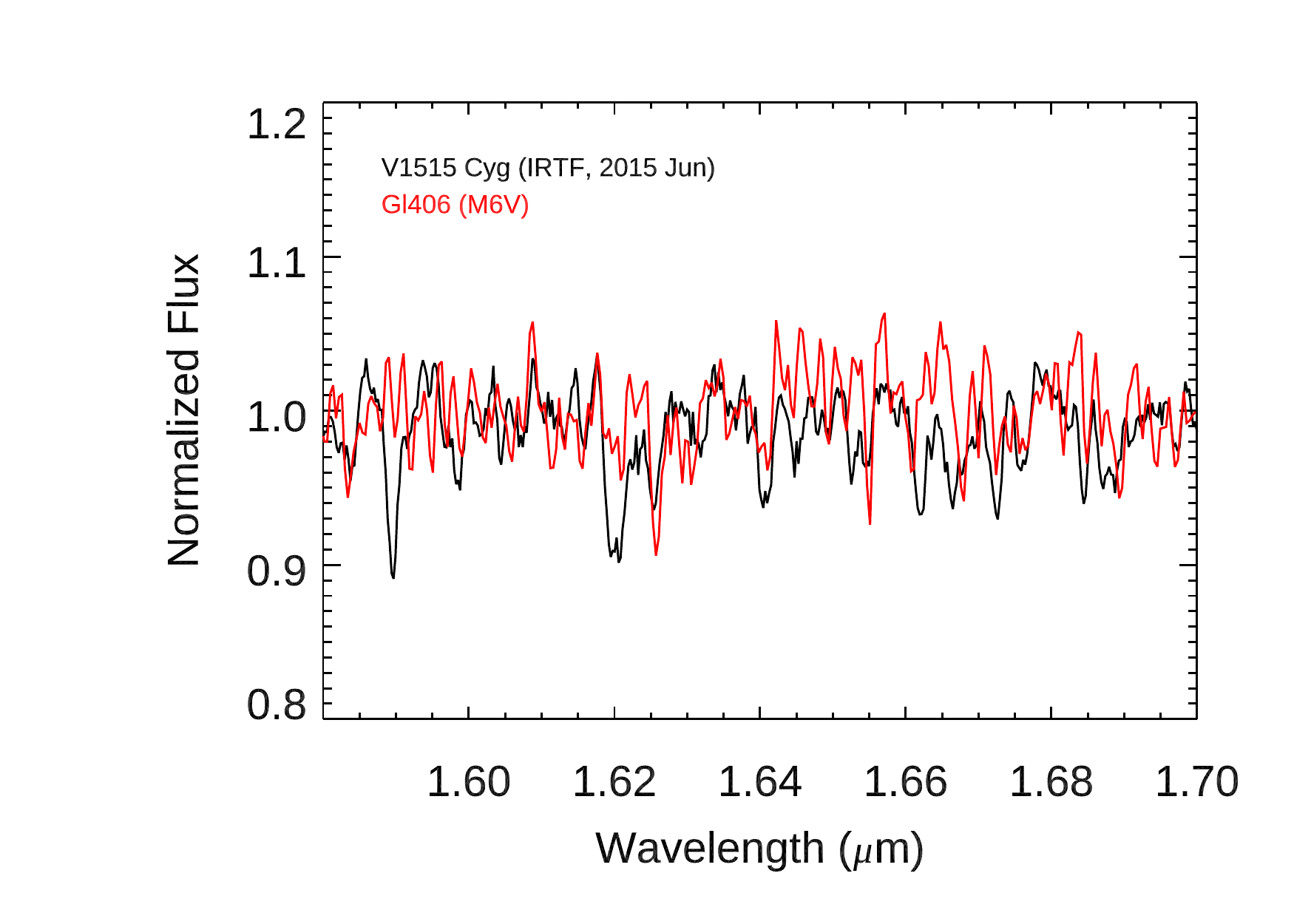}\\
    \includegraphics[width=0.45\textwidth]{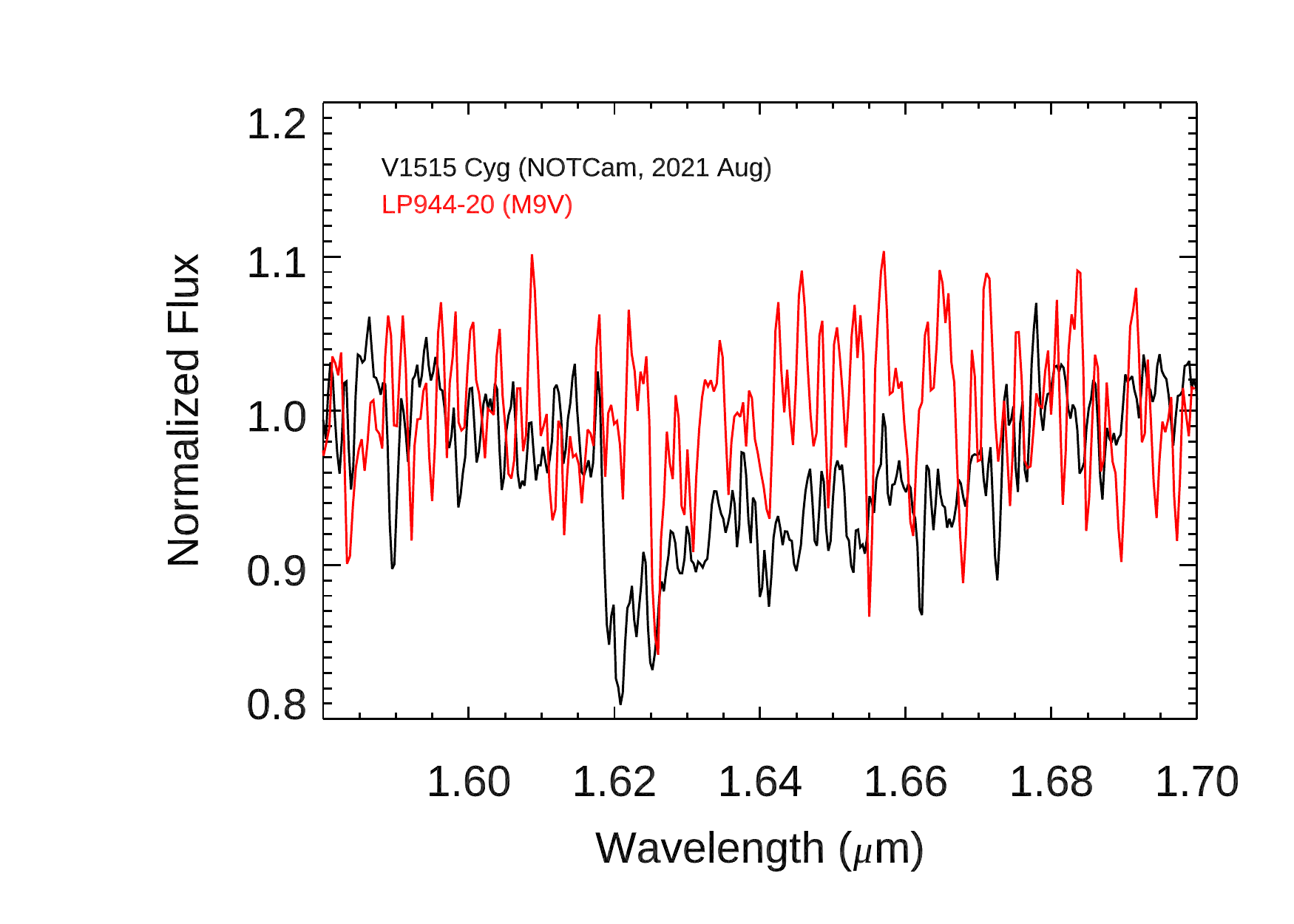}
    \caption{FeH features around 1.62\,$\mu$m. Black and red lines present V1515 Cyg and standard stellar spectrum \citep{rayner2009}. The IRTF and NOTCam spectra are well-matched with M6 and M9 dwarfs, respectively.
}
    \label{fig:FeH_comp}
\end{figure}

Fig.~\ref{fig_ir_spec} shows the comparison of the NIR spectra between V1515~Cyg (dark gray and blue) and V1057~Cyg (light gray and red). Overall, the flux of V1515~Cyg is lower than the one of V1057~Cyg, and the flux change between 2015 and 2020/21 is also smaller in V1515~Cyg. Most of the observed lines are the same as those of V1057~Cyg in the NIR spectra.
To compare spectral changes in V1515~Cyg itself and the two FUors, we measured the EW of the CO overtone bandhead features (2.293 -- 2.317\,$\mu$m) with the same method described in detail in \citet{szabo2021}.
Fig.~\ref{fig_co_ew} shows the CO bandhead of V1515~Cyg (upper spectra) and V1057~Cyg (lower spectra), and the measured EW is listed in Table~\ref{tbl_co}.
In the case of V1057~Cyg, the CO 2--0 bandhead feature became weaker with the decreasing brightness, and it is interpreted as the decreasing mass accretion rate \citep{szabo2021}. On the other hand, in the case of V1515~Cyg, the bandhead features almost remain the same strength  (Fig.~\ref{fig_co_ew}), similarly to the monitoring results of the optical spectra (Fig.~\ref{fig_oplot}).
In our observation, the EW of the CO increases, which is mostly affected by the lower temperature rovibrational transitions ($\sim$2.31\,$\mu$m) in Fig.~\ref{fig_co_ew}. This result is consistent with the presence of the stronger FeH in our observation (Fig.~\ref{fig_notcam}).
We estimated the gas temperature by comparing the FeH feature at 1.62\,$\mu$m in V1515~Cyg and those of standard stars \citep{rayner2009} between M4 and L5 spectral types (Fig.~\ref{fig:FeH_comp}). Both the target and the stellar spectrum were normalized. 
We found the best fit by eye due to the blended molecular nature of this feature. The previous spectrum taken in 2015 by \citet{connelley2018} was well-matched with an M6 -- M7 dwarf spectrum \citep[T$_{\rm eff}$$\sim$2700\,K (GJ~406);][]{kuznetsov2019}, while the 2021 spectrum is well-matched with a later type dwarf of M9 -- M9.5 \citep[T$_{\rm eff}$$\sim$2500\,K (LHS~2924);][]{gagne2015}.
Together with the decreasing temperature from the FeH, and the increased EW of the CO bandhead in the 2021 observation (see Sec~\ref{sec:spec_NIR}), we conclude that the disk of V1515~Cyg became cooler since 2015, further suggesting that the source is approaching a quiescence phase.

An accurate comparison is still difficult to make between the two sources. The two system have very different light curves, indicating that overall different  energetics were present in the early stages of the outburst \citep[see Fig.~1 in][]{szabo2021}. While a long term fading trend is present in both cases, the decline is currently much faster and drastic in V1515~Cyg. As shown, the current mass accretion rate of V1515~Cyg is much lower (4.17$\times$10$^{-6}$\,M$_{\odot}^2$yr$^{-1}$) than in V1057~Cyg (10$^{-4}$\,M$_{\odot}^2$yr$^{-1}$), which can justify the lower mass ejection rate and the optically thin outflow in the former system. Overall, the weaker wind in V1515~Cyg further suggests that the outburst mechanisms are quite different in the two sources.
It is important to mention that different geometries can also play a role in differences between systems, however the inclination of the two sources are similar as both systems are seen close to face-on \citep[see e.g.,][]{gramajo2014,milliner2019}.

\begin{deluxetable}{llll}
\tabletypesize{\scriptsize}
\tablecaption{EW of CO overtone bandhead \label{tbl_co}}
\tablewidth{0pt}
\tablehead{
\colhead{Target} & \colhead{Observation Date} & \colhead{EW} & \colhead{Reference} \\[-3mm]
\colhead{} & \colhead{[UT]} & \colhead{[\AA{]}} & \colhead{}}
\startdata
V1515~Cyg & 2015 June$^{a}$ & 16.25 $\pm$ 0.42 & This work \\
V1515~Cyg & 2021 August & 23.02 $\pm$ 0.66 & This work \\
V1057~Cyg & 2015 June$^{a}$ & 22.75 $\pm$ 0.36 & \citet{szabo2021} \\
V1057~Cyg & 2020 August & 27.03 $\pm$ 0.45 & \citet{szabo2021} \\
\enddata
\tablenotetext{a}{Spectrum from \citet{connelley2018}}
\label{tab:co}
\end{deluxetable}

\section{Conclusions} \label{sec:conclusions}

In this paper we performed a multi-epoch, multi-wavelength study of the classical FU~Orionis-type star V1515~Cyg 
and we arrived to the following conclusions: 
\begin{itemize}
    \item Our long-term photometric monitoring of V1515~Cyg supplemented with data obtained at Mt.~Maidanak and data from publicly available archives such as INTEGRAL, ASAS-SN, and ZTF show a long-term fading trend which started around 2006. The fading is not monotonous: after the maximum brightness (1984--2005), the inner disk apparently experienced several obscuration-like events. 
    The long-term evolution of the source's optical brightness after 2003 is approximately consistent with an exponential decay, with an e-folding time of approximately 12\,years.
     
    \item The Mt.~Maidanak photometry enabled us to investigate the light curve changes occurring in the timescale of 5--100\,d during the maximum. Comparisons of these multi colour light curves suggest that the majority of these light curve changes did originate from the inner disk. We confirmed the 13.9\,d QPO found in the 1987 data set by \citet{clarke2005} and we found a similar peak of 13.06\,d QPO in the 2003 data set. Although observed in two seasons only, these values are surprisingly similar to the persistent 14\,d modulation of spectral wind features found in FU~Orionis \citep{herbig2003,powell2012}. Time-series spectroscopy can decide about the possible similarities in V1515~Cyg in the future.
    
    \item Although {\it TESS} enabled investigation of short-periodic, down to $\sim$20~min oscillations, the 2019 and 2021 light curves do not show rapidly occurring light variations on such time scales. The frequency spectra exhibit the random-walk character, being of the same type as the one found in our previous study of V1057~Cyg \citep[][]{szabo2021}. It is plausible that the dimming event studied in great detail during the 2021 season could be similar to what was observed in the CTTS RW~Aur, however, there are many differences between the two system. Additional, similarly well sampled seasons are required to investigate and compare this phenomena further in the future.
    
    \item Our accretion disk modelling reveals that the current luminosity of the accretion disk dropped from the peak value of 138\,$L_{\odot}$ to about 45\,$L_{\odot}$. The peak and current $M\dot{M}$ values are 1.29$\times$10$^{-5}$ and 4.17$\times$10$^{-6}$~M$_{\odot}^2$yr$^{-1}$, respectively. This indicates that the long-term fading is also partly caused by the dropping of the accretion rate.  
    The optical and near-infrared peak present in the SED indicates the presence of significant circumstellar matter surrounding the source. The long-term evolution of the accretion rate until 2010 is approximately consistent with an exponential decay, with an e-folding time of $\sim$56\,years.

    \item We performed optical spectroscopic monitoring of the source between 2015--2021 and detected several absorption lines, which show small variations over time. Several high-velocity wind features were also observed in the form of absorption components of P~Cygni profiles which similarly show small variations. 
    The source lacks the P~Cygni profile in the Ca\,{\footnotesize II} 8498\,\AA{} line, part of the Ca infrared triplet (IRT), compared to V1057~Cyg and other classical FUors \citep{szabo2021}. The strength of the blue-shifted absorption component in this P~Cygni profile is formed by an outflowing wind. The weaker strength in this case could be an indication of the weaker, optically thin outflowing wind in the structure. 
    
    \item During our spectroscopic monitoring despite the pure emission of the Ca\,{\footnotesize II} 8498\,\AA{} line, we did not observe any other emission, nor forbidden emission lines in the spectra of V1515~Cyg.
    The absence of certain forbidden emission lines indicate that there are no jets/outflows present while these were detected in V1057~Cyg \citep{szabo2021}.
    
    \item For our study, we obtained a new NIR spectrum of V1515~Cyg in 2021, and compared it with the one obtained in 2015 by \citet{connelley2018}. In the new one, we found that both  the CO overtone bandhead features and the FeH molecular bands became stronger since the previous observation.
    Together with the decreasing temperature from the FeH, and the increased EW of the CO bandhead in the 2021 observation, we conclude that the disk of V1515~Cyg became cooler by about 200\,K since 2015, further suggesting that the source is approaching the quiescent phase.
    
    \item Despite the long-term fading of V1515~Cyg, our results from the spectroscopic monitoring and analysis still resemble the properties those of a classical/bona fide FUor.

\end{itemize}
The continuation of the photometric monitoring supplemented with additional spectroscopic snapshot observations can be crucial for our understanding of the late evolution of FUor outbursts and how they eventually end.
This study is the second part of a series with the aim of revisiting the first few classical FUors after the famous 1936 eruption of FU~Orionis, the prototype source.
Our study highlights the importance of scrutinizing these objects which ultimately can help us
to better understand the accretion processes in the early stages of Sun-like star formation.


    

\section*{Acknowledgements}
We kindly thank Michael Connelley for handing us over the 2015 IRTF spectra of V1515~Cyg in order to carry out a more accurate analysis.
This project has received funding from the European Research Council (ERC) under the European Union's Horizon 2020 research and innovation programme under grant agreement No 716155 (SACCRED). 
M. Ibrahimov and K. Grankin acknowledge an active participation of their colleagues: Drs. S.Yu. Melnikov, S.D. Yakubov, O.V. Ezhkova, S.A. Artemenko and some others - to obtain UBVRcIc photoelectric and BVRI CCD obsrevations of V1515 Cyg at Maidanak Astronomical Observatory (UBAI, Tashkent, Uzbekistan).
This article is partly based on observations made with the 1.5 m TCS and IAC-80 telescopes operated on the island of Tenerife by the Instituto de Astrofisica de Canarias in the Spanish Observatorio del Teide. 
B.K.~was supported by the National Research, Development and Innovation Fund of Hungary, financed under the 2020-1.2.1-GYAK-2020-00007 funding scheme.
The work was also supported by the Hungarian NKFIH grant KH-30526 and K-31508.
Our study is supported by the project ''Transient Astrophysical Objects'' GINOP 2.3.2-15-2016-00033 of the National Research, Development and Innovation Office (NKFIH), Hungary, funded by the European Union. 
This project has been supported by the Lendület grant LP2012-31 of the Hungarian Academy of Sciences.
J-E. Lee was supported by the Basic Science Research Program through the National Research Foundation of Korea (grant no. NRF-2018R1A2B6003423).
R.K.T. has been supported by the NKFIH/OTKA FK-134432 grant. 
K.CS. has been supported by the \'UNKP-21-2 New National Excellence Program of the Ministry for Innovation and Technology from the source of the National Research, Development and Innovation Fund.
This research was supported by the KKP-137523 `SeismoLab' \'Elvonal grant of the Hungarian Research, Development and Innovation Office (NKFIH).
A.B. is supported by the Lendület LP2014-17.
K.V. is supported by the Bolyai J\'anos Research Scholarship of the Hungarian Academy of Sciences. K.V. acknowledges  the support of the Hungarian National Research, Development and Innovation Office grant OTKA KH-30526 and K-3150. 
L.K. acknowledges the financial support of the Hungarian National Research, Development and Innovation Office grant NKFIH PD-134784 and K-131508. L.K. is a Bolyai János Research Fellow.
A.P. acknowledges the financial support of the Hungarian National Research, Development and Innovation Office (NKFIH) grant K-138962.
Á.S. is supported by the Élvonal grant KKP-137523. 

This paper includes data collected by the TESS mission. Funding for the TESS mission is provided by the NASA Explorer Program.. 
Based on observations made with the Nordic Optical Telescope, operated by the Nordic Optical Telescope Scientific Association at the Observatorio del Roque de los Muchachos, La Palma, Spain, of the Instituto de Astrofisica de Canarias. The data presented here were obtained (in part) with ALFOSC, which is provided by the Instituto de Astrofisica de Andalucia (IAA) under a joint agreement with the University of Copenhagen and NOTSA. 
The DASCH project at Harvard is partially supported by NSF grants AST-0407380, AST-0909073, and AST-1313370.
Based on observations obtained with the Samuel Oschin 48-inch Telescope at the Palomar Observatory as part of the Zwicky Transient Facility project. ZTF is supported by the National Science Foundation under Grant No. AST-1440341 and a collaboration including Caltech, IPAC, the Weizmann Institute for Science, the Oskar Klein Center at Stockholm University, the University of Maryland, the University of Washington, Deutsches Elektronen-Synchrotron and Humboldt University, Los Alamos National Laboratories, the TANGO Consortium of Taiwan, the University of Wisconsin at Milwaukee, and Lawrence Berkeley National Laboratories. Operations are conducted by COO, IPAC, and UW.
Based on observations obtained with the Samuel Oschin Telescope 48-inch and the 60-inch Telescope at the Palomar Observatory as part of the Zwicky Transient Facility project. ZTF is supported by the National Science Foundation under Grant No. AST-2034437 and a collaboration including Caltech, IPAC, the Weizmann Institute for Science, the Oskar Klein Center at Stockholm University, the University of Maryland, Deutsches Elektronen-Synchrotron and Humboldt University, the TANGO Consortium of Taiwan, the University of Wisconsin at Milwaukee, Trinity College Dublin, Lawrence Livermore National Laboratories, and IN2P3, France. Operations are conducted by COO, IPAC, and UW.
This paper also uses observations made at the Mount Suhora Astronomical Observatory (MSO) in Poland.

\vspace {5mm}

\facilities{NOT, BOAO, SOFIA, TESS}
\software{FITSH package \citep{pal2012}, Vartools \citep{hartman2016}, molecfit \citep{smette2015,kausch2015}, IRAF \citep{tody1986,tody1993}}

\appendix
\section{Photometry of V1515~Cyg}
Table~\ref{tab:phot} contains our original optical and near-infrared photometry of V1515~Cyg before the shifts discussed in Sec.~\ref{sec:obs}, while Table~\ref{tab:phot_wise} contains the saturation corrected WISE data that we use in Fig.~\ref{fig:lc}. Table~\ref{tab:shifts} contains shifts applied for the different data sets in the optical bands described in Sec.~\ref{sec:lc}

\startlongtable
\begin{deluxetable*}{ccccccccccccccc}
\tabletypesize{\footnotesize}
\tablecaption{Optical and near-IR photometry of V1515~Cyg.\label{tab:phot}}
\tablehead{%
\colhead{Date} & %
\colhead{MJD} &%
\colhead{$B$} &%
\colhead{$V$} &%
\colhead{$g'$} &%
\colhead{$R_{\rm C}$} &%
\colhead{$I_{\rm C}$} &%
\colhead{$R_{\rm J}$} &%
\colhead{$I_{\rm J}$} &%
\colhead{$r'$} &%
\colhead{$i'$} &%
\colhead{$J$} & \colhead{$H$} & \colhead{$K_{\rm s}$} & \colhead{Instrument}} 
\startdata
2005-11-18 & 53692.83 &\dots&$13.11$&\dots &$12.17$&$11.13$&\dots&\dots&\dots&\dots&\dots&\dots&\dots& RCC \\
2006-07-12 & 53928.15 & \dots & \dots &\dots & \dots & \dots &\dots&\dots&\dots&\dots&$9.04$&$8.18$&$7.55$& TCS \\
2006-07-19 & 53935.01 & $14.85$ & $12.97$ &\dots & $12.13$ & \dots &\dots&\dots&\dots&\dots&\dots&\dots&\dots& IAC80 \\
2011-08-05 & 55778.97 & $15.20$ & $13.35$ &\dots & $12.45$ & $11.39$ &\dots&\dots&\dots&\dots&\dots&\dots&\dots& Schmidt \\
2020-05-18 & 58988.06 & $15.91$ & $14.03$ &\dots&\dots&\dots&\dots&\dots& $13.42$ & $12.63$ &\dots&\dots&\dots& RC80 \\
2021-04-25 & 59329.20 & \dots & \dots & \dots &\dots&\dots&\dots&\dots& \dots & \dots &$9.72$&$8.52$&$7.92$& NOT \\
2022-01-23 & 59603.20 & $15.90$ & $14.04$ & $14.99$ &\dots&\dots&\dots&\dots& $13.43$ & $12.63$ &\dots&\dots&\dots& RC80 \\
\enddata
\tablecomments{The table is published in its entirety in the machine-readable format. A portion is shown here for guidance regarding its form and content.}
\label{tab:photometry}
\end{deluxetable*}

\startlongtable
\begin{deluxetable*}{ccccc}
\tabletypesize{\footnotesize} 
\tablecaption{Saturation corrected WISE data of V1515 Cyg}
\tablewidth{0pt}
\tablehead{\colhead{Date} &\colhead{MJD} & \colhead{W1 mag} & \colhead{W2 mag} & \colhead{Survey Phase}}
\startdata
2010-05-09  & 55334.3 & 6.94$\pm$0.03 & 6.07$\pm$0.01   &  4--Band Cryo \\
2010-11-14  & 55514.6 & 7.18$\pm$0.05 & 6.21$\pm$0.02   &  4--Band Cryo \\ 
2014-05-20  & 56797.8 & 7.04$\pm$0.02 & 6.27$\pm$0.01   &  Post--Cryo \\
2014-11-17  & 56978.0 & 7.03$\pm$0.03 & 6.23$\pm$0.01   &  Post--Cryo \\
2015-05-19  & 57161.1 & 7.10$\pm$0.02 & 6.29$\pm$0.01   &  Reactivation \\
2015-11-11  & 57337.7 & 7.18$\pm$0.02 & 6.39$\pm$0.01   &  Reactivation \\
2016-05-17  & 57525.2 & 7.11$\pm$0.02 & 6.39$\pm$0.02   &  Reactivation \\
2016-11-04  & 57696.3 & 7.19$\pm$0.02 & 6.44$\pm$0.01   &  Reactivation \\
2017-05-18  & 57891.1 & 7.17$\pm$0.02 & 6.41$\pm$0.01   &  Reactivation \\
2017-11-02  & 58059.7 & 7.21$\pm$0.02 & 6.46$\pm$0.02   &  Reactivation \\
2018-05-18  & 58256.6 & 7.26$\pm$0.02 & 6.49$\pm$0.01   &  Reactivation \\
2018-10-29  & 58420.2 & 7.25$\pm$0.03 & 6.45$\pm$0.02   &  Reactivation \\
2019-05-17  & 58620.7 & 7.23$\pm$0.02 & 6.46$\pm$0.01   &  Reactivation \\
2019-10-28  & 58784.7 & 7.23$\pm$0.02 & 6.43$\pm$0.01   &  Reactivation \\
2020-05-18  & 58987.8 & 7.35$\pm$0.02 & 6.52$\pm$0.01   &  Reactivation \\ 
2020-10-29  & 59151.3 & 7.26$\pm$0.02 & 6.49$\pm$0.02   &  Reactivation \\
2021-05-18  & 59352.6 & 7.26$\pm$0.02 & 6.52$\pm$0.02   &  Reactivation \\ 
2021-10-28  & 59516.1 & 7.27$\pm$0.02 & 6.51$\pm$0.02   &  Reactivation \\
\label{tab:phot_wise}
\enddata
\end{deluxetable*}

\startlongtable
\begin{deluxetable*}{ccccccccc}
\tabletypesize{\footnotesize} 
\tablecaption{Shifts applied to match overlapping optical data sets.}
\tablewidth{0pt}
\tablehead{\colhead{telescope/filter} & \colhead{Schmidt} & \colhead{RC80} & \colhead{Mt. Suhora} & \colhead{ZTF}  & \colhead{ASAS-SN} & \colhead{INTEGRAL} & \colhead{IAC} & \colhead{DASCH} }
\startdata
$B$              &  0.00   &  0.00  & +0.03      & +0.266 &  --     &  --      & -0.15 & -0.28 \\    
$V$              &  0.00   &  0.00  & +0.05      & +0.115 & -0.03   & -0.53    & +0.02 &  --   \\
$R_C$            &  0.00   &  0.00  & +0.07      & +0.16  & --      &  --      & --    &  --   \\ 
$I_C$            &  0.00   &  0.00  & +0.115     &  --    & --      &  --      & --    &  --   \\       
$g$              &   --    &  0.00  & +0.03      & +0.09  & +0.02   &  --      & --    &  --   \\   
$r$              &   --    &  0.00  & +0.07      & +0.15  & --      &  --      & --    &  --   \\      
$i$              &   --    &  0.00  & +0.11      &  --    & --      &  --      & --    &  --   \\ 
\enddata
\label{tab:shifts}
\end{deluxetable*}

\bibliography{paper}

\begin{thebibliography}{}
\expandafter\ifx\csname natexlab\endcsname\relax\def\natexlab#1{#1}\fi
\providecommand{\url}[1]{\href{#1}{#1}}
\providecommand{\dodoi}[1]{doi:~\href{http://doi.org/#1}{\nolinkurl{#1}}}
\providecommand{\doeprint}[1]{\href{http://ascl.net/#1}{\nolinkurl{http://ascl.net/#1}}}
\providecommand{\doarXiv}[1]{\href{https://arxiv.org/abs/#1}{\nolinkurl{https://arxiv.org/abs/#1}}}

\bibitem[{{{\'A}brah{\'a}m} {et~al.}(2018){{\'A}brah{\'a}m}, {K{\'o}sp{\'a}l},
  {Kun}, {Feh{\'e}r}, {Zsidi}, {Acosta-Pulido}, {Carnerero},
  {Garc{\'\i}a-{\'A}lvarez}, {Mo{\'o}r}, {Cseh}, {Hajdu}, {Hanyecz}, {Kelemen},
  {Kriskovics}, {Marton}, {Mez{\H{o}}}, {Moln{\'a}r}, {Ordasi},
  {Rodr{\'\i}guez-Coira}, {S{\'a}rneczky}, {S{\'o}dor}, {Szak{\'a}ts},
  {Szegedi-Elek}, {Szing}, {Farkas-Tak{\'a}cs}, {Vida}, \&
  {Vink{\'o}}}]{abraham2018}
{{\'A}brah{\'a}m}, P., {K{\'o}sp{\'a}l}, {\'A}., {Kun}, M., {et~al.} 2018,
  \apj, 853, 28, \dodoi{10.3847/1538-4357/aaa242}

\bibitem[{{Agra-Amboage} \& {Garcia}(2014)}]{agra-amboage2014}
{Agra-Amboage}, V., \& {Garcia}, P.~J.~V. 2014, \aap, 565, A92,
  \dodoi{10.1051/0004-6361/201323327}

\bibitem[{{Alcal{\'a}} {et~al.}(2017){Alcal{\'a}}, {Manara}, {Natta}, {Frasca},
  {Testi}, {Nisini}, {Stelzer}, {Williams}, {Antoniucci}, {Biazzo}, {Covino},
  {Esposito}, {Getman}, \& {Rigliaco}}]{alcala2017}
{Alcal{\'a}}, J.~M., {Manara}, C.~F., {Natta}, A., {et~al.} 2017, \aap, 600,
  A20, \dodoi{10.1051/0004-6361/201629929}

\bibitem[{{Audard} {et~al.}(2014){Audard}, {{\'A}brah{\'a}m}, {Dunham},
  {Green}, {Grosso}, {Hamaguchi}, {Kastner}, {K{\'o}sp{\'a}l}, {Lodato},
  {Romanova}, {Skinner}, {Vorobyov}, \& {Zhu}}]{audard2014}
{Audard}, M., {{\'A}brah{\'a}m}, P., {Dunham}, M.~M., {et~al.} 2014, in
  Protostars and Planets VI, ed. H.~{Beuther}, R.~S. {Klessen}, C.~P.
  {Dullemond}, \& T.~{Henning}, 387,
  \dodoi{10.2458/azu_uapress_9780816531240-ch017}

\bibitem[{{Azevedo} {et~al.}(2006){Azevedo}, {Calvet}, {Hartmann}, {Folha},
  {Gameiro}, \& {Muzerolle}}]{azevedo2006}
{Azevedo}, R., {Calvet}, N., {Hartmann}, L., {et~al.} 2006, \aap, 456, 225,
  \dodoi{10.1051/0004-6361:20054315}

\bibitem[{{Baek} {et~al.}(2015){Baek}, {Pak}, {Green}, {Meschiari}, {Lee},
  {Jeon}, {Choi}, {Im}, {Sung}, \& {Park}}]{baek2015}
{Baek}, G., {Pak}, S., {Green}, J.~D., {et~al.} 2015, \aj, 149, 11,
  \dodoi{10.1088/0004-6256/149/2/73}

\bibitem[{{Bell} {et~al.}(1995){Bell}, {Lin}, {Hartmann}, \&
  {Kenyon}}]{bell1995}
{Bell}, K.~R., {Lin}, D.~N.~C., {Hartmann}, L.~W., \& {Kenyon}, S.~J. 1995,
  \apj, 444, 376, \dodoi{10.1086/175612}

\bibitem[{{Bieging} {et~al.}(2002){Bieging}, {Rieke}, \& {Rieke}}]{bieging2002}
{Bieging}, J.~H., {Rieke}, M.~J., \& {Rieke}, G.~H. 2002, \aap, 384, 965,
  \dodoi{10.1051/0004-6361:20020063}

\bibitem[{{Blinova} {et~al.}(2016){Blinova}, {Romanova}, \&
  {Lovelace}}]{blinova2016}
{Blinova}, A.~A., {Romanova}, M.~M., \& {Lovelace}, R.~V.~E. 2016, \mnras, 459,
  2354, \dodoi{10.1093/mnras/stw786}

\bibitem[{{Cardelli} {et~al.}(1989){Cardelli}, {Clayton}, \&
  {Mathis}}]{cardelli1989}
{Cardelli}, J.~A., {Clayton}, G.~C., \& {Mathis}, J.~S. 1989, \apj, 345, 245,
  \dodoi{10.1086/167900}

\bibitem[{{Clarke} {et~al.}(2005){Clarke}, {Lodato}, {Melnikov}, \&
  {Ibrahimov}}]{clarke2005}
{Clarke}, C.~J., {Lodato}, G., {Melnikov}, S.~Y., \& {Ibrahimov}, M.~A. 2005,
  \mnras, 361, 942, \dodoi{10.1111/j.1365-2966.2005.09231.x}

\bibitem[{{Connelley} \& {Reipurth}(2018)}]{connelley2018}
{Connelley}, M.~S., \& {Reipurth}, B. 2018, \apj, 861, 145,
  \dodoi{10.3847/1538-4357/aaba7b}

\bibitem[{{Egan} {et~al.}(2003){Egan}, {Price}, {Kraemer}, {Mizuno}, {Carey},
  {Wright}, {Engelke}, {Cohen}, \& {Gugliotti}}]{egan2003}
{Egan}, M.~P., {Price}, S.~D., {Kraemer}, K.~E., {et~al.} 2003, VizieR Online
  Data Catalog, V/114

\bibitem[{{Facchini} {et~al.}(2016){Facchini}, {Manara}, {Schneider}, {Clarke},
  {Bouvier}, {Rosotti}, {Booth}, \& {Haworth}}]{facchini2016}
{Facchini}, S., {Manara}, C.~F., {Schneider}, P.~C., {et~al.} 2016, \aap, 596,
  A38, \dodoi{10.1051/0004-6361/201629607}

\bibitem[{{Feh{\'e}r} {et~al.}(2017){Feh{\'e}r}, {K{\'o}sp{\'a}l},
  {{\'A}brah{\'a}m}, {Hogerheijde}, \& {Brinch}}]{feher2017}
{Feh{\'e}r}, O., {K{\'o}sp{\'a}l}, {\'A}., {{\'A}brah{\'a}m}, P.,
  {Hogerheijde}, M.~R., \& {Brinch}, C. 2017, \aap, 607, A39,
  \dodoi{10.1051/0004-6361/201731446}

\bibitem[{{Fiorellino} {et~al.}(2022){Fiorellino}, {Park}, {K{\'o}sp{\'a}l}, \&
  {{\'A}brah{\'a}m}}]{Fiorellino2022}
{Fiorellino}, E., {Park}, S., {K{\'o}sp{\'a}l}, {\'A}., \& {{\'A}brah{\'a}m},
  P. 2022, arXiv e-prints, arXiv:2201.00784.
\newblock \doarXiv{2201.00784}

\bibitem[{{Fischer} {et~al.}(2022){Fischer}, {Hillenbrand}, {Herczeg},
  {Johnstone}, {K{\'o}sp{\'a}l}, \& {Dunham}}]{fischer2022}
{Fischer}, W.~J., {Hillenbrand}, L.~A., {Herczeg}, G.~J., {et~al.} 2022, arXiv
  e-prints, arXiv:2203.11257.
\newblock \doarXiv{2203.11257}

\bibitem[{{Foster}(1996)}]{foster1996}
{Foster}, G. 1996, \aj, 112, 1709, \dodoi{10.1086/118137}

\bibitem[{{Gagn{\'e}} {et~al.}(2015){Gagn{\'e}}, {Faherty}, {Cruz},
  {Lafreni{\'e}re}, {Doyon}, {Malo}, {Burgasser}, {Naud}, {Artigau},
  {Bouchard}, {Gizis}, \& {Albert}}]{gagne2015}
{Gagn{\'e}}, J., {Faherty}, J.~K., {Cruz}, K.~L., {et~al.} 2015, \apjs, 219,
  33, \dodoi{10.1088/0067-0049/219/2/33}

\bibitem[{{Gramajo} {et~al.}(2014){Gramajo}, {Rod{\'o}n}, \&
  {G{\'o}mez}}]{gramajo2014}
{Gramajo}, L.~V., {Rod{\'o}n}, J.~A., \& {G{\'o}mez}, M. 2014, \aj, 147, 140,
  \dodoi{10.1088/0004-6256/147/6/140}

\bibitem[{{Green} {et~al.}(2006){Green}, {Hartmann}, {Calvet}, {Watson},
  {Ibrahimov}, {Furlan}, {Sargent}, \& {Forrest}}]{green2006}
{Green}, J.~D., {Hartmann}, L., {Calvet}, N., {et~al.} 2006, \apj, 648, 1099,
  \dodoi{10.1086/505932}

\bibitem[{{Green} {et~al.}(2013{\natexlab{a}}){Green}, {Robertson}, {Baek},
  {Pooley}, {Pak}, {Im}, {Lee}, {Jeon}, {Choi}, \& {Meschiari}}]{green2013b}
{Green}, J.~D., {Robertson}, P., {Baek}, G., {et~al.} 2013{\natexlab{a}}, \apj,
  764, \dodoi{10.1088/0004-637X/764/1/22}

\bibitem[{{Green} {et~al.}(2013{\natexlab{b}}){Green}, {Evans},
  {K{\'o}sp{\'a}l}, {Herczeg}, {Quanz}, {Henning}, {van Kempen}, {Lee},
  {Dunham}, {Meeus}, {Bouwman}, {Chen}, {G{\"u}del}, {Skinner}, {Liebhart}, \&
  {Merello}}]{green2013}
{Green}, J.~D., {Evans}, Neal~J., I., {K{\'o}sp{\'a}l}, {\'A}., {et~al.}
  2013{\natexlab{b}}, \apj, 772, 117, \dodoi{10.1088/0004-637X/772/2/117}

\bibitem[{{Grindlay} {et~al.}(2012){Grindlay}, {Tang}, {Los}, \&
  {Servillat}}]{dasch2012}
{Grindlay}, J., {Tang}, S., {Los}, E., \& {Servillat}, M. 2012, in New Horizons
  in Time Domain Astronomy, ed. E.~{Griffin}, R.~{Hanisch}, \& R.~{Seaman},
  Vol. 285, 29--34, \dodoi{10.1017/S1743921312000166}

\bibitem[{{Hamann} \& {Persson}(1989)}]{hamman-persson1989}
{Hamann}, F., \& {Persson}, S.~E. 1989, \apj, 339, 1078, \dodoi{10.1086/167362}

\bibitem[{{Hamann} \& {Persson}(1990)}]{hamann-persson1990}
{Hamann}, F., \& {Persson}, S.~E. 1990, in Astronomical Society of the Pacific
  Conference Series, Vol.~9, Cool Stars, Stellar Systems, and the Sun, ed.
  G.~{Wallerstein}, 304--306

\bibitem[{Hargreaves {et~al.}(2010)Hargreaves, Hinkle, Bauschlicher, Wende,
  Seifahrt, \& Bernath}]{hargreaves2010}
Hargreaves, R.~J., Hinkle, K.~H., Bauschlicher, C.~W., {et~al.} 2010, 140, 919,
  \dodoi{10.1088/0004-6256/140/4/919}

\bibitem[{{Hartman} \& {Bakos}(2016)}]{hartman2016}
{Hartman}, J.~D., \& {Bakos}, G.~A. 2016, Astronomy and Computing, 17, 1,
  \dodoi{10.1016/j.ascom.2016.05.006}

\bibitem[{{Hartmann} {et~al.}(2016){Hartmann}, {Herczeg}, \&
  {Calvet}}]{hartmann2016}
{Hartmann}, L., {Herczeg}, G., \& {Calvet}, N. 2016, \araa, 54, 135,
  \dodoi{10.1146/annurev-astro-081915-023347}

\bibitem[{{Hartmann} {et~al.}(2004){Hartmann}, {Hinkle}, \&
  {Calvet}}]{hartmann2004}
{Hartmann}, L., {Hinkle}, K., \& {Calvet}, N. 2004, \apj, 609, 906,
  \dodoi{10.1086/421317}

\bibitem[{{Hartmann} \& {Kenyon}(1985)}]{hartmann1985}
{Hartmann}, L., \& {Kenyon}, S.~J. 1985, \apj, 299, 462, \dodoi{10.1086/163713}

\bibitem[{{Hartmann} \& {Kenyon}(1996)}]{kenyon&hartmann1996}
---. 1996, \araa, 34, 207, \dodoi{10.1146/annurev.astro.34.1.207}

\bibitem[{{Herbig}(1977)}]{herbig1977}
{Herbig}, G.~H. 1977, \apj, 217, 693, \dodoi{10.1086/155615}

\bibitem[{{Herbig} {et~al.}(2003){Herbig}, {Petrov}, \&
  {Duemmler}}]{herbig2003}
{Herbig}, G.~H., {Petrov}, P.~P., \& {Duemmler}, R. 2003, \apj, 595, 384,
  \dodoi{10.1086/377194}

\bibitem[{{Herbig} \& {Soderblom}(1980)}]{herbig1980}
{Herbig}, G.~H., \& {Soderblom}, D.~R. 1980, \apj, 242, 628,
  \dodoi{10.1086/158499}

\bibitem[{{Herbst} \& {Shevchenko}(1999)}]{herbst1999}
{Herbst}, W., \& {Shevchenko}, V.~S. 1999, \aj, 118, 1043,
  \dodoi{10.1086/300966}

\bibitem[{{Herter} {et~al.}(2013){Herter}, {Vacca}, {Adams}, {Keller},
  {Schoenwald}, {Hirsch}, {Wang}, {De Buizer}, {Helton}, \&
  {Llorens}}]{herter13}
{Herter}, T.~L., {Vacca}, W.~D., {Adams}, J.~D., {et~al.} 2013, \pasp, 125,
  1393, \dodoi{10.1086/674144}

\bibitem[{{Hillenbrand} {et~al.}(2019){Hillenbrand}, {Miller}, {Carpenter},
  {Kasliwal}, {Isaacson}, {Tang}, {Joshi}, {Banerjee}, \&
  {Cutri}}]{hillenbrand2019}
{Hillenbrand}, L.~A., {Miller}, A.~A., {Carpenter}, J.~M., {et~al.} 2019, \apj,
  874, 82, \dodoi{10.3847/1538-4357/ab06c8}

\bibitem[{{Hora} {et~al.}(2007){Hora}, {Bontemps}, {Megeath}, {Schneider},
  {Motte}, {Carey}, {Simon}, {Keto}, {Smith}, {Allen}, {Gutermuth}, {Fazio},
  {Kraemer}, {Mizuno}, {Price}, \& {Adams}}]{spitzer-cygnus2007}
{Hora}, J., {Bontemps}, S., {Megeath}, T., {et~al.} 2007, {A Spitzer Legacy
  Survey of the Cygnus-X Complex}, Spitzer Proposal ID \#40184

\bibitem[{{Joint Iras Science}(1994)}]{iras-catalogue1994}
{Joint Iras Science}, W.~G. 1994, VizieR Online Data Catalog, II/125

\bibitem[{{Jordi} {et~al.}(2006){Jordi}, {Grebel}, \& {Ammon}}]{jordi2006}
{Jordi}, K., {Grebel}, E.~K., \& {Ammon}, K. 2006, \aap, 460, 339,
  \dodoi{10.1051/0004-6361:20066082}

\bibitem[{{Kadam} {et~al.}(2020){Kadam}, {Vorobyov}, {Reg{\'a}ly},
  {K{\'o}sp{\'a}l}, \& {{\'A}brah{\'a}m}}]{kadam2020}
{Kadam}, K., {Vorobyov}, E., {Reg{\'a}ly}, Z., {K{\'o}sp{\'a}l}, {\'A}., \&
  {{\'A}brah{\'a}m}, P. 2020, \apj, 895, 41, \dodoi{10.3847/1538-4357/ab8bd8}

\bibitem[{{Kausch} {et~al.}(2015){Kausch}, {Noll}, {Smette}, {Kimeswenger},
  {Barden}, {Szyszka}, {Jones}, {Sana}, {Horst}, \& {Kerber}}]{kausch2015}
{Kausch}, W., {Noll}, S., {Smette}, A., {et~al.} 2015, \aap, 576, A78,
  \dodoi{10.1051/0004-6361/201423909}

\bibitem[{{Kenyon} {et~al.}(1988){Kenyon}, {Hartmann}, \&
  {Hewett}}]{kenyon-and-hartmann1988}
{Kenyon}, S.~J., {Hartmann}, L., \& {Hewett}, R. 1988, \apj, 325, 231,
  \dodoi{10.1086/165999}

\bibitem[{{Kenyon} \& {Hartmann}(1991)}]{kenyon-and-hartmann1991}
{Kenyon}, S.~J., \& {Hartmann}, L.~W. 1991, \apj, 383, 664,
  \dodoi{10.1086/170823}

\bibitem[{{Kenyon} {et~al.}(1991){Kenyon}, {Hartmann}, \& {Kolotilov}}]{KH1991}
{Kenyon}, S.~J., {Hartmann}, L.~W., \& {Kolotilov}, E.~A. 1991, \pasp, 103,
  1069, \dodoi{10.1086/132926}

\bibitem[{{Kim} {et~al.}(2002){Kim}, {Jang}, {Han}, {Jang}, {Sung}, {Chun},
  {Hyung}, {Yoon}, \& {Vogt}}]{kim2002}
{Kim}, K.-M., {Jang}, B.-H., {Han}, I., {et~al.} 2002, Journal of Korean
  Astronomical Society, 35, 221, \dodoi{10.5303/JKAS.2002.35.4.221}

\bibitem[{{Kochanek} {et~al.}(2017){Kochanek}, {Shappee}, {Stanek}, {Holoien},
  {Thompson}, {Prieto}, {Dong}, {Shields}, {Will}, {Britt}, {Perzanowski}, \&
  {Pojma{\'n}ski}}]{kochanek2017}
{Kochanek}, C.~S., {Shappee}, B.~J., {Stanek}, K.~Z., {et~al.} 2017, \pasp,
  129, 104502, \dodoi{10.1088/1538-3873/aa80d9}

\bibitem[{{Kolotilov}(1983)}]{kolotilov1983}
{Kolotilov}, E.~A. 1983, Pisma v Astronomicheskii Zhurnal, 9, 622

\bibitem[{{Kolotilov}(1990)}]{kolotilov1990}
---. 1990, Soviet Astronomy Letters, 16, 12

\bibitem[{{Kolotilov} \& {Kenyon}(1997{\natexlab{a}})}]{kolotilov1997}
{Kolotilov}, E.~A., \& {Kenyon}, S.~J. 1997{\natexlab{a}}, Information Bulletin
  on Variable Stars, 4494, 1

\bibitem[{{Kolotilov} \& {Kenyon}(1997{\natexlab{b}})}]{kolotilov&kenyon1997}
---. 1997{\natexlab{b}}, Information Bulletin on Variable Stars, 4494, 1

\bibitem[{{Kolotilov} \& {Petrov}(1981)}]{kolotilov1981}
{Kolotilov}, E.~A., \& {Petrov}, P.~P. 1981, Astronomicheskij Tsirkulyar, 1167,
  1

\bibitem[{{K{\'o}sp{\'a}l} {et~al.}(2017){K{\'o}sp{\'a}l}, {{\'A}brah{\'a}m},
  {Westhues}, \& {Haas}}]{kospal2017a}
{K{\'o}sp{\'a}l}, {\'A}., {{\'A}brah{\'a}m}, P., {Westhues}, C., \& {Haas}, M.
  2017, \aap, 597, L10, \dodoi{10.1051/0004-6361/201629447}

\bibitem[{{Kosp{\'a}l} {et~al.}(2018){Kosp{\'a}l}, {{\'A}br{\'a}ham}, {Zsidi},
  {Vida}, {Szab{\'o}}, {Mo{\'o}r}, \& {P{\'a}l}}]{kospal2018}
{Kosp{\'a}l}, {\'A}., {{\'A}br{\'a}ham}, P., {Zsidi}, G., {et~al.} 2018, \apj,
  862, 16, \dodoi{10.3847/1538-4357/aacafa}

\bibitem[{{K{\'o}sp{\'a}l} {et~al.}(2016){K{\'o}sp{\'a}l}, {{\'A}brah{\'a}m},
  {Acosta-Pulido}, {Dunham}, {Garc{\'\i}a-{\'A}lvarez}, {Hogerheijde}, {Kun},
  {Mo{\'o}r}, {Farkas}, {Hajdu}, {Hodos{\'a}n}, {Kov{\'a}cs}, {Kriskovics},
  {Marton}, {Moln{\'a}r}, {P{\'a}l}, {S{\'a}rneczky}, {S{\'o}dor},
  {Szak{\'a}ts}, {Szalai}, {Szegedi-Elek}, {Szing}, {T{\'o}th}, {Vida}, \&
  {Vink{\'o}}}]{kospal2016}
{K{\'o}sp{\'a}l}, {\'A}., {{\'A}brah{\'a}m}, P., {Acosta-Pulido}, J.~A.,
  {et~al.} 2016, \aap, 596, A52, \dodoi{10.1051/0004-6361/201528061}

\bibitem[{{Koutoulaki} {et~al.}(2019){Koutoulaki}, {Facchini}, {Manara},
  {Natta}, {Garcia Lopez}, {Fedriani}, {Caratti o Garatti}, {Coffey}, \&
  {Ray}}]{koutoulaki2019}
{Koutoulaki}, M., {Facchini}, S., {Manara}, C.~F., {et~al.} 2019, \aap, 625,
  A49, \dodoi{10.1051/0004-6361/201834713}

\bibitem[{{Kun} {et~al.}(2019){Kun}, {{\'A}brah{\'a}m}, {Acosta Pulido},
  {Mo{\'o}r}, \& {Prusti}}]{kun2019}
{Kun}, M., {{\'A}brah{\'a}m}, P., {Acosta Pulido}, J.~A., {Mo{\'o}r}, A., \&
  {Prusti}, T. 2019, \mnras, 483, 4424, \dodoi{10.1093/mnras/sty3425}

\bibitem[{{Kuznetsov} {et~al.}(2019){Kuznetsov}, {del Burgo}, {Pavlenko}, \&
  {Frith}}]{kuznetsov2019}
{Kuznetsov}, M.~K., {del Burgo}, C., {Pavlenko}, Y.~V., \& {Frith}, J. 2019,
  \apj, 878, 134, \dodoi{10.3847/1538-4357/ab1fe9}

\bibitem[{{Landolt}(1975)}]{landolt1975}
{Landolt}, A.~U. 1975, \pasp, 87, 379, \dodoi{10.1086/129778}

\bibitem[{{Landolt}(1977)}]{landolt1977}
---. 1977, \pasp, 89, 704, \dodoi{10.1086/130213}

\bibitem[{{Lin} \& {Papaloizou}(1985)}]{lin1985}
{Lin}, D.~N.~C., \& {Papaloizou}, J. 1985, in Protostars and Planets II, ed.
  D.~C. {Black} \& M.~S. {Matthews}, 981--1072

\bibitem[{Masci {et~al.}(2018)Masci, Laher, Rusholme, Shupe, Groom, Surace,
  Jackson, Monkewitz, Beck, Flynn, Terek, Landry, Hacopians, Desai, Howell,
  Brooke, Imel, Wachter, Ye, Lin, Cenko, Cunningham, Rebbapragada, Bue, Miller,
  Mahabal, Bellm, Patterson, Juri{\'{c}}, Golkhou, Ofek, Walters, Graham,
  Kasliwal, Dekany, Kupfer, Burdge, Cannella, Barlow, Sistine, Giomi, Fremling,
  Blagorodnova, Levitan, Riddle, Smith, Helou, Prince, \& Kulkarni}]{masci2018}
Masci, F.~J., Laher, R.~R., Rusholme, B., {et~al.} 2018, Publications of the
  Astronomical Society of the Pacific, 131, 018003,
  \dodoi{10.1088/1538-3873/aae8ac}

\bibitem[{{Milliner} {et~al.}(2019){Milliner}, {Matthews}, {Long}, \&
  {Hartmann}}]{milliner2019}
{Milliner}, K., {Matthews}, J.~H., {Long}, K.~S., \& {Hartmann}, L. 2019,
  \mnras, 483, 1663–1673, \dodoi{10.1093/mnras/sty3197}

\bibitem[{{Molinari} {et~al.}(1993){Molinari}, {Liseau}, \&
  {Lorenzetti}}]{molinari1993}
{Molinari}, S., {Liseau}, R., \& {Lorenzetti}, D. 1993, \aaps, 101, 59

\bibitem[{{Neugebauer} {et~al.}(1984){Neugebauer}, {Habing}, {van Duinen},
  {Aumann}, {Baud}, {Beichman}, {Beintema}, {Boggess}, {Clegg}, {de Jong},
  {Emerson}, {Gautier}, {Gillett}, {Harris}, {Hauser}, {Houck}, {Jennings},
  {Low}, {Marsden}, {Miley}, {Olnon}, {Pottasch}, {Raimond}, {Rowan-Robinson},
  {Soifer}, {Walker}, {Wesselius}, \& {Young}}]{iras1984}
{Neugebauer}, G., {Habing}, H.~J., {van Duinen}, R., {et~al.} 1984, \apjl, 278,
  L1, \dodoi{10.1086/184209}

\bibitem[{{Paczynski}(1976)}]{paczynski1976}
{Paczynski}, B. 1976, in IAU Symposium, Vol.~73, Structure and Evolution of
  Close Binary Systems, ed. P.~{Eggleton}, S.~{Mitton}, \& J.~{Whelan}, 75

\bibitem[{{P{\'a}l}(2012)}]{pal2012}
{P{\'a}l}, A. 2012, \mnras, 421, 1825, \dodoi{10.1111/j.1365-2966.2011.19813.x}

\bibitem[{{Park} {et~al.}(2020){Park}, {Lee}, {Pyo}, {Jaffe}, {Mace}, {Sung},
  {Lee}, {Kang}, {Oh}, {Yoon}, {Yoon}, \& {Green}}]{park2020}
{Park}, S., {Lee}, J.-E., {Pyo}, T.-S., {et~al.} 2020, \apj, 900, 36,
  \dodoi{10.3847/1538-4357/aba532}

\bibitem[{{Petrov} {et~al.}(2015){Petrov}, {Gahm}, {Djupvik}, {Babina},
  {Artemenko}, \& {Grankin}}]{petrov2015}
{Petrov}, P.~P., {Gahm}, G.~F., {Djupvik}, A.~A., {et~al.} 2015, \aap, 577,
  A73, \dodoi{10.1051/0004-6361/201525845}

\bibitem[{{Powell} {et~al.}(2012){Powell}, {Irwin}, {Bouvier}, \&
  {Clarke}}]{powell2012}
{Powell}, S.~L., {Irwin}, M., {Bouvier}, J., \& {Clarke}, C.~J. 2012, \mnras,
  426, 3315, \dodoi{10.1111/j.1365-2966.2012.21898.x}

\bibitem[{{Press}(1978)}]{press1978}
{Press}, W.~H. 1978, Comments on Modern Physics, Part C - Comments on
  Astrophysics, 7, 103

\bibitem[{{Rayner} {et~al.}(2009){Rayner}, {Cushing}, \& {Vacca}}]{rayner2009}
{Rayner}, J.~T., {Cushing}, M.~C., \& {Vacca}, W.~D. 2009, \apjs, 185, 289,
  \dodoi{10.1088/0067-0049/185/2/289}

\bibitem[{{Ricker} {et~al.}(2015){Ricker}, {Winn}, {Vanderspek}, {Latham},
  {Bakos}, {Bean}, {Berta-Thompson}, {Brown}, {Buchhave}, {Butler}, {Butler},
  {Chaplin}, {Charbonneau}, {Christensen-Dalsgaard}, {Clampin}, {Deming},
  {Doty}, {De Lee}, {Dressing}, {Dunham}, {Endl}, {Fressin}, {Ge}, {Henning},
  {Holman}, {Howard}, {Ida}, {Jenkins}, {Jernigan}, {Johnson}, {Kaltenegger},
  {Kawai}, {Kjeldsen}, {Laughlin}, {Levine}, {Lin}, {Lissauer}, {MacQueen},
  {Marcy}, {McCullough}, {Morton}, {Narita}, {Paegert}, {Palle}, {Pepe},
  {Pepper}, {Quirrenbach}, {Rinehart}, {Sasselov}, {Sato}, {Seager},
  {Sozzetti}, {Stassun}, {Sullivan}, {Szentgyorgyi}, {Torres}, {Udry}, \&
  {Villasenor}}]{ricker2015}
{Ricker}, G.~R., {Winn}, J.~N., {Vanderspek}, R., {et~al.} 2015, Journal of
  Astronomical Telescopes, Instruments, and Systems, 1, 014003,
  \dodoi{10.1117/1.JATIS.1.1.014003}

\bibitem[{{Rucinski} {et~al.}(2008){Rucinski}, {Matthews}, {Kuschnig}, G.,
  {Rowe}, {Guenther}, {Moffat}, {Sasselov}, {Walker}, \& {Weiss}}]{ruc08}
{Rucinski}, S.~M., {Matthews}, J.~M., {Kuschnig}, R., {et~al.} 2008, \mnras,
  391, 11, \dodoi{doi:10.1111/j.1365-2966.2008.14014.x}

\bibitem[{{Rucinski} {et~al.}(2010){Rucinski}, {Zwintz}, {Hareter},
  {Pojmanski}, {Kuschnig}, {Matthews}, {Guenther}, {Moffat}, {Sasselov}, \&
  {Weiss}}]{Rucinski2010}
{Rucinski}, S.~M., {Zwintz}, K., {Hareter}, M., {et~al.} 2010, \aap, 522, 8,
  \dodoi{10.1051/0004-6361/201014856}

\bibitem[{{Sandell} \& {Weintraub}(2001{\natexlab{a}})}]{sandell2001}
{Sandell}, G., \& {Weintraub}, D.~A. 2001{\natexlab{a}}, \apjs, 134, 115,
  \dodoi{10.1086/320360}

\bibitem[{{Sandell} \& {Weintraub}(2001{\natexlab{b}})}]{sw2001}
---. 2001{\natexlab{b}}, \apjs, 134, 115, \dodoi{10.1086/320360}

\bibitem[{{Shappee} {et~al.}(2014){Shappee}, {Prieto}, {Grupe}, {Kochanek},
  {Stanek}, {De Rosa}, {Mathur}, {Zu}, {Peterson}, {Pogge}, {Komossa}, {Im},
  {Jencson}, {Holoien}, {Basu}, {Beacom}, {Szczygie{\l}}, {Brimacombe},
  {Adams}, {Campillay}, {Choi}, {Contreras}, {Dietrich}, {Dubberley},
  {Elphick}, {Foale}, {Giustini}, {Gonzalez}, {Hawkins}, {Howell}, {Hsiao},
  {Koss}, {Leighly}, {Morrell}, {Mudd}, {Mullins}, {Nugent}, {Parrent},
  {Phillips}, {Pojmanski}, {Rosing}, {Ross}, {Sand}, {Terndrup}, {Valenti},
  {Walker}, \& {Yoon}}]{shappee2014}
{Shappee}, B.~J., {Prieto}, J.~L., {Grupe}, D., {et~al.} 2014, \apj, 788, 48,
  \dodoi{10.1088/0004-637X/788/1/48}

\bibitem[{{Siwak} {et~al.}(2020){Siwak}, {Og{\l}oza}, \&
  {Krzesi{\'n}ski}}]{siwak2020}
{Siwak}, M., {Og{\l}oza}, W., \& {Krzesi{\'n}ski}, J. 2020, \aap, 644,
  \dodoi{10.1051/0004-6361/202037607}

\bibitem[{{Siwak} {et~al.}(2013){Siwak}, {Ruci\'nski}, {Matthews}, {Kuschnig},
  {Guenther}, {Moffat}, {Rowe}, {Sasselov}, \& {Weiss}}]{siwak2013}
{Siwak}, M., {Ruci\'nski}, S.~M., {Matthews}, J.~M., {et~al.} 2013, \mnras,
  432, 194, \dodoi{10.1093/mnras/stt441}

\bibitem[{{Siwak} {et~al.}(2018){Siwak}, {Winiarski}, {Og{\l}oza},
  {Dro{\.z}d{\.z}}, {Zo{\l}a}, {Moffat}, {Stachowski}, {Rucinski}, {Cameron},
  {Matthews}, {Weiss}, {Kuschnig}, {Rowe}, {Guenther}, \&
  {Sasselov}}]{siwak2018}
{Siwak}, M., {Winiarski}, M., {Og{\l}oza}, W., {et~al.} 2018, \aap, 618,
  \dodoi{10.1051/0004-6361/201833401}

\bibitem[{{Smette} {et~al.}(2015){Smette}, {Sana}, {Noll}, {Horst}, {Kausch},
  {Kimeswenger}, {Barden}, {Szyszka}, {Jones}, {Gallenne}, {Vinther},
  {Ballester}, \& {Taylor}}]{smette2015}
{Smette}, A., {Sana}, H., {Noll}, S., {et~al.} 2015, \aap, 576, A77,
  \dodoi{10.1051/0004-6361/201423932}

\bibitem[{{Szab{\'o}} {et~al.}(2021){Szab{\'o}}, {K{\'o}sp{\'a}l},
  {{\'A}brah{\'a}m}, {Park}, {Siwak}, {Green}, {Mo{\'o}r}, {P{\'a}l},
  {Acosta-Pulido}, {Lee}, {Cseh}, {Cs{\"o}rnyei}, {Hanyecz},
  {K{\"o}nyves-T{\'o}th}, {Krezinger}, {Kriskovics}, {Ordasi}, {S{\'a}rneczky},
  {Seli}, {Szak{\'a}ts}, {Szing}, \& {Vida}}]{szabo2021}
{Szab{\'o}}, Z.~M., {K{\'o}sp{\'a}l}, {\'A}., {{\'A}brah{\'a}m}, P., {et~al.}
  2021, \apj, 917, 80, \dodoi{10.3847/1538-4357/ac04b3}

\bibitem[{{Szegedi-Elek} {et~al.}(2020){Szegedi-Elek}, {{\'A}brah{\'a}m},
  {Wyrzykowski}, {Kun}, {K{\'o}sp{\'a}l}, {Chen}, {Marton}, {Mo{\'o}r}, {Kiss},
  {P{\'a}l}, {Szabados}, {Varga}, {Varga-Vereb{\'e}lyi}, {Andreas}, {Bachelet},
  {Bischoff}, {B{\'o}di}, {Breedt}, {Burgaz}, {Butterley}, {Carrasco},
  {{\v{C}}epas}, {Damljanovic}, {Gezer}, {Godunova}, {Gromadzki}, {Gurgul},
  {Hardy}, {Hildebrandt}, {Hoffmann}, {Hundertmark}, {Ihanec}, {Janulis},
  {Kalup}, {Kaczmarek}, {K{\"o}nyves-T{\'o}th}, {Krezinger}, {Kruszy{\'n}ska},
  {Littlefair}, {Maskoli{\={u}}nas}, {M{\'e}sz{\'a}ros}, {Miko{\l}ajczyk},
  {Mugrauer}, {Netzel}, {Ordasi}, {Pak{\v{s}}tien{\.{e}}}, {Rybicki},
  {S{\'a}rneczky}, {Seli}, {Simon}, {{\v{S}}i{\v{s}}kauskait{\.{e}}},
  {S{\'o}dor}, {Sokolovsky}, {Stenglein}, {Street}, {Szak{\'a}ts}, {Tomasella},
  {Tsapras}, {Vida}, {Zdanavi{\v{c}}ius}, {Zieli{\'n}ski}, {Zieli{\'n}ski}, \&
  {Zi{\'o}{\l}kowska}}]{szegedi-elek2020}
{Szegedi-Elek}, E., {{\'A}brah{\'a}m}, P., {Wyrzykowski}, {\L}., {et~al.} 2020,
  \apj, 899, 130, \dodoi{10.3847/1538-4357/aba129}

\bibitem[{{Tody}(1986)}]{tody1986}
{Tody}, D. 1986, in Society of Photo-Optical Instrumentation Engineers (SPIE)
  Conference Series, Vol. 627, Instrumentation in astronomy VI, ed. D.~L.
  {Crawford}, 733, \dodoi{10.1117/12.968154}

\bibitem[{{Tody}(1993)}]{tody1993}
{Tody}, D. 1993, in Astronomical Society of the Pacific Conference Series,
  Vol.~52, Astronomical Data Analysis Software and Systems II, ed. R.~J.
  {Hanisch}, R.~J.~V. {Brissenden}, \& J.~{Barnes}, 173

\bibitem[{{Tofflemire} {et~al.}(2017){Tofflemire}, {Mathieu}, {Ardila},
  {Akeson}, {Ciardi}, {Johns-Krull}, {Herczeg}, \&
  {Quijano-Vodniza}}]{tofflemire2017}
{Tofflemire}, B.~M., {Mathieu}, R.~D., {Ardila}, D.~A., {et~al.} 2017, \apj,
  835, 11, \dodoi{10.3847/1538-4357/835/1/8}

\bibitem[{{Tofflemire} {et~al.}(2019){Tofflemire}, {Mathieu}, \&
  {Johns-Krull}}]{tofflemire2019}
{Tofflemire}, B.~M., {Mathieu}, R.~D., \& {Johns-Krull}, C. 2019, \aj, 158, 24

\bibitem[{{Turner} {et~al.}(1997){Turner}, {Bodenheimer}, \&
  {Bell}}]{turner1997}
{Turner}, N.~J.~J., {Bodenheimer}, P., \& {Bell}, K.~R. 1997, \apj, 480, 754,
  \dodoi{10.1086/303983}

\bibitem[{{Wallace} \& {Hinkle}(2001)}]{wallace2001}
{Wallace}, L., \& {Hinkle}, K. 2001, \apj, 559, 424, \dodoi{10.1086/322382}

\bibitem[{{Wright} {et~al.}(2010){Wright}, {Eisenhardt}, {Mainzer}, {Ressler},
  {Cutri}, {Jarrett}, {Kirkpatrick}, {Padgett}, {McMillan}, {Skrutskie},
  {Stanford}, {Cohen}, {Walker}, {Mather}, {Leisawitz}, {Gautier}, {McLean},
  {Benford}, {Lonsdale}, {Blain}, {Mendez}, {Irace}, {Duval}, {Liu}, {Royer},
  {Heinrichsen}, {Howard}, {Shannon}, {Kendall}, {Walsh}, {Larsen}, {Cardon},
  {Schick}, {Schwalm}, {Abid}, {Fabinsky}, {Naes}, \& {Tsai}}]{wright2010}
{Wright}, E.~L., {Eisenhardt}, P. R.~M., {Mainzer}, A.~K., {et~al.} 2010, \aj,
  140, 1868, \dodoi{10.1088/0004-6256/140/6/1868}

\bibitem[{{Young} {et~al.}(2012){Young}, {Becklin}, {Marcum}, {Roellig}, {De
  Buizer}, {Herter}, {G{\"u}sten}, {Dunham}, {Temi}, {Andersson}, {Backman},
  {Burgdorf}, {Caroff}, {Casey}, {Davidson}, {Erickson}, {Gehrz}, {Harper},
  {Harvey}, {Helton}, {Horner}, {Howard}, {Klein}, {Krabbe}, {McLean}, {Meyer},
  {Miles}, {Morris}, {Reach}, {Rho}, {Richter}, {Roeser}, {Sandell}, {Sankrit},
  {Savage}, {Smith}, {Shuping}, {Vacca}, {Vaillancourt}, {Wolf}, \&
  {Zinnecker}}]{young12}
{Young}, E.~T., {Becklin}, E.~E., {Marcum}, P.~M., {et~al.} 2012, \apjl, 749,
  L17, \dodoi{10.1088/2041-8205/749/2/L17}

\bibitem[{{Zechmeister} \& {K{\"u}rster}(2009)}]{zechmeister2009}
{Zechmeister}, M., \& {K{\"u}rster}, M. 2009, \aap, 496, 577,
  \dodoi{10.1051/0004-6361:200811296}

\end{thebibliography}
\bibliographystyle{aasjournal}

\end{document}